\pdfoutput=1
\documentclass{JINST}
\let\ifpdf\relax
\usepackage{url}

\title{High Voltage in Noble Liquids for High Energy Physics}         
\author{Edited by B.~Rebel$^a$ and
C.~Hall$^b$
with contributions from E.~Bernard$^c$,
C.~H.~Faham$^d$,
T.~M.~Ito$^e$,
B.~Lundberg$^b$,
M.~Messina$^f$,
F.~Monrabal$^g$,
S.~P.~Pereverzev$^h$,
F.~Resnati$^i$,
P.~C.~Rowson$^j$,
M.~Soderberg$^{a,k}$, 
T.~Strauss$^l$,
A.~Tomas$^m$,
J.~Va'vra$^j$,
H.~Wang$^n$\\
\llap{$^a$}Fermi National Accelerator Laboratory, Batavia, IL 60510, USA\\
\llap{$^b$}University of Maryland, College Park, MD 20742, USA\\
\llap{$^c$}Yale University, New Haven, CT 06520, USA\\
\llap{$^d$}Lawrence Berkeley National Laboratory, Berkeley, CA 94720, USA\\
\llap{$^e$}Los Alamos National Laboratory, Los Alamos, NM 87545, USA\\
\llap{$^f$}Columbia University, New York, NY 10027, USA\\
\llap{$^g$}IFIC, E-46980 Paterna, Spain\\
\llap{$^h$}Lawerence Livermore National Laboratory, Livermore, CA 94550, USA\\
\llap{$^i$}ETH Zurich, 8093 Zurich, Switzerland \\
\llap{$^j$}SLAC National Accerlerator Laboratory, Menlo Park, CA 94025, USA\\
\llap{$^k$}Syracuse University, NY 13210, USA\\
\llap{$^l$}University of Bern, 3012 Bern, Switzerland\\
\llap{$^m$}Imperial College London, SW7 2AZ London, UK\\
\llap{$^n$}University of California Los Angeles, Los Angeles, CA 90095, USA}

\abstract{
A workshop was held at Fermilab November 8-9, 2013 to discuss the challenges of using high voltage in noble liquids.  The participants spanned the fields of neutrino, dark matter, and electric dipole moment physics.  All presentations at the workshop were made in plenary sessions.  This document summarizes the experiences and lessons learned from experiments in these fields at developing high voltage systems in noble liquids.
}

\begin{document}


\maketitle



\tableofcontents
\setcounter{secnumdepth}{5}

\section{Introduction}

Noble liquids provide ideal detection media for a wide range of particle physics applications ranging from dark matter searches, to neutrino interaction and oscillation experiments, to searches for electric dipole moments.  Three noble liquids are the primary media used, helium, argon and xenon. 

The liquid argon time projection chamber (TPC) is an advanced detector technology suitable for large-scale neutrino and dark matter detectors. Particle interactions in liquid argon produce both atomic excitation and ionization leading to the creation of scintillation light. With a uniform electric field in a liquid argon TPC, ionized electrons are drifted toward segmented readout planes. The liquid argon TPC technique provides efficient particle identification and background rejection via good d$E$/d$x$ resolution and millimeter scale 3D precision particle tracking. A key concern of experiments using liquid argon TPCs to track charged particles is the purity of the argon with respect to electronegative impurities.  These impurities attach to the ionization electrons before the electrons can drift to the readout causing diminished signals.

Experiments searching for dark matter make use of a two-phase liquid argon TPC where scintillation light (S1) is detected via photodetectors and ionization electrons are drifted toward a gas phase, where they are converted into luminescent scintillation light (S2) and the latter is detected by the same photodetectors. This technique offers millimeter scale 3D position reconstruction and efficient background rejection via the S1 pulse shape and ratio of the S2 to S1 signal. Liquid xenon TPCs use the same technique to search for dark matter.  All TPCs require electric fields on the order of $0.1-1$~kV/cm to drift the ionized electrons. For this reason, larger TPCs will require higher cathode voltages to maintain the same electric field. 

This document summarizes the contents of a workshop held at Fermilab November 8-9, 2013 to discuss the challenges of using high voltage in noble liquids.  The experiences of several experiments in the electric dipole moment, neutrino and dark matter fields at developing high voltage delivery mechanisms inside of noble liquid detectors were presented.  This summary first high-lights important considerations based on properties of noble liquids, and then details the experiences of experiments in each of the mentioned areas.  The contributors for each section are identified to allow readers to contact them for further information.  A section high-lighting the important lessons learned based on the collective experience is provided at the end of the document.

\section{Considerations of Fundamental Properties of Noble Liquids}
\subsection{Electron emission enhancement by positive ions on metal surface}
\label{sec:PEREVERZEV}
{\it Contributed by S.~P.~Pereverzev, Lawerence Livermore National Laboratory, Livermore, CA 94550, USA}
\newline

For large detector sizes, even a moderate field in the drift region of the order 500~V/cm results in a large potential difference between cathode and anode. Consequently, a high voltage source, high voltage cable and feed-through are required to operate these detectors. In these components and in some regions inside of the detector the electric field can be much higher than the drift field in the active region or the amplification field in the gas region. These high field regions and components are prone to high voltage breakdown. The detector and high voltage components must be designed to account for the mechanisms of secondary ion production inside of the detector because the enhancement of electron emission by positive ions is a strong effect. 


The Malter effect was described 1937~\cite{Malter}. An aluminum surface was first oxidized electrochemically and then covered with a monatomic layer of cesium oxide. When an electron beam bombarded the surface in vacuum, the surface was charged positively due to secondary electron emission. As a result the aluminum surface may emit an electron current three orders of magnitude larger than the initial electron beam current. This Òcold emissionÒ can persist for some time after the initial electron beam is turned off.  This example illustrates that the effects of positive ions can be very strong.  

Another effect can be observed with a thin natural oxide film on aluminum. Studying argon  gas ionization detector with a aluminized Mylar film cathode, the authors of Ref.~\cite{Zorin} observed sensitivity of the device to visible light when positive ions were present on the surface of the aluminum cathode. In the absence of positive ions, visible light cannot cause the photoelectric effect from aluminum. Positive ions can be trapped on a metal surface due to the presence of a solid physisorbed film of a rare gas at cryogenic temperatures. Physisorbtion occurs because polarizable atoms are attracted to the metal surface. For noble gases, this attraction is weakest for helium and strongest for xenon. Formation of up to 10 solid monolayers of physisorbed helium film on a solid hydrogen surface was observed in~\cite{Cieslikowski}. In this experiment, a mobility of electrons localized above the helium film was monitored by capacitance technique. As the helium gas pressure increased at low temperatures the thickness of equilibrium physisorbed helium film also increased. The mobility exhibited  a maximum each time the top-most film layer was completely full. The formation of several solid monolayers of physisorbed  argon, krypton  and xenon films  on cold  surfaces was observed by different methods. It is also well-studied that a monolayer of argon, krypton or xenon on metal surface can change the work function for the metal~\cite{Brush,Forster,Huuckstadt}.  Because positive ion  mobility in solid rare gases is much lower than ion mobility in the liquid or gas phases, positive ions will be trapped for some time in solid physisorbed layers before they reach the metal surface and recombine with an electron. This trapping time can be long, and the enhancement of the photoelectric effect can be significant.

In the above examples, the exact mechanisms leading to the electron emission enhancement are not known. The presence of positive ions as well as other impurities on a metal surface can change the energies of electron states in a metal close to the surface.   These changes can alter the work function. For a large ion surface density, an average electric field produced by ions works as an external electric field which facilitates a photoemission and field emission.  For a small ion concentration, the presence of a positive ion can lower a tunneling barrier for an electron, increasing the escape probability for the electron. In this way, the quantum efficiency of the photoelectric effect can be increased without lowering the work function. There could be other scenarios of the electron emission enhancement, like resonant tunneling. An ion produces localized states for an electron under the tunneling barrier, and the tunneling can be a two-stage process: from the metal to the localized state inside the barrier and then to an unbounded state outside of the metal and the dielectric film. 
 
\subsubsection{Possible effects in liquid argon and xenon}

The presence of positive ions on a metal surface should cause an enhancement of the photoelectric effect, either through a red shift of the spectrum or increase of the quantum efficiency in liquid argon and liquid xenon.  Positive ions produce sites where the threshold for an electron tunneling out of the metal is locally lowered. This lowering of the threshold should produce enhancements of other electron emission processes such as field emission, thermal emission or emission caused by positive ion recombination on metal surface.  One can expect a nonlinear increase of an electron emission with an increasing positive ion surface density. One reason for this increase is that the collective electric field of ions promotes the electron tunneling. The other reason is that with the rise  of an ion surface density not only does the number of recombination events increase, but also the probability for an energetic electron to escape from the surface due to the easy tunneling site created by neighboring  positive ions also increases. The ion recombination energy is sufficient for extraction of two electrons out of the metal, so chain events could be  possible with the emission of  several electrons at a time. Such events can be misinterpreted as low energy ionization events in the detector.  The combination of a strong electric field near the metal surface with the presence of positive ions can lead to a non-linear increase of the field emission and photoemission. This combination can cause electric breakdowns or other high voltage instabilities.

\subsubsection{Comparison with effects observable in detectors}

It is known that many of the existing large detectors are operating at lower voltages than initially targeted in detector design. Unfortunately,  in  most cases, a detailed description of the difficulties and high voltage instabilities were never presented to the general public.  Therefore, this discussion focuses on the Livermore dual-phase argon  detector. This detector has an approximately 200 ml active volume and was operated in a large range of gas amplification fields up to 11 kV/cm and drift fields up to 4 kV/cm, with the total voltage on cathode up to 30 kV and up to 24 kV voltage on an extraction grid 8 mm below the liquid-gas interface. Single electron sensitivity was achieved with amplification of 8 photons detected by photomultiplier tubes (PMT) per each electron injected from liquid into gas.  The system demonstrated $\sim 20$\% energy resolution for 280 eV ionization events, indicating the detection of the $^{37}$Ar line. The goal was to measure the ionization yield of low energy nuclear recoils directly in liquid argon. 

When a solid copper cathode is used instead of a grid cathode, a $10$\% echo of the S2 signals delayed by the time required for the electrons reaching the liquid surface from the cathode is observed. The high amplitude of the echo requires 2-3 orders of magnitude larger quantum efficiency of the photoelectric effect from copper into liquid argon than the known efficiency of the effect from copper into vacuum for the same photon energy as argon scintillation photons. This large efficiency can be due to the presence of positive ions.

The dependence of the breakdown voltage of the cathode voltage lead on the gas amplification voltage has also been observed. This problem can be overcome by covering the negative tip of the feed-through with a plastic shrinkable tubing. Likely, large UV production in the gas during strong events, such as from cosmic ray muons, was triggering the breakdown on the feed-through. 

A stable operation of the detector for a high amplification level was achieved only after covering a metal extraction grid frame with Teflon.   Doing so means only the grid wires were exposed to UV radiation from the gas region. Without this cover the detector was prone to developing a glowing discharge, when the voltage on the grids was stable and discharge current was below 0.1 mA. After discharge ignition, it was possible to remove the drift field and observe a glowing discharge. It was concluded that the electron emission from the extraction grid and the grid frame was playing the major role to provide secondary electron emission, not the cathode grid.  It is not known what process is limiting current in this discharge. Possibly, positive ions are eventually discharging on the metal surface, while the production rate of secondary positive ions in high field regions around anode wires is low. The ion surface concentration also could be limited by a strong recombination rate dependence on a surface ion concentration. 

At high amplification, that is high voltage on the extraction grid, a several milliseconds long photon train was observed after some strong ionization events. These photon trains were observable also without a drift field in liquid, which means they originated on the extraction grid. The electric field around the extraction grid wires in the detector is higher than the field around the cathode grid wires and a possible explanation could be that for certain strong events a high positive ion concentration on some wires is obtained when a thin liquid layer above extraction grid is ionized. This situation can result in a non-linear increase in electron field emission similar to the Malter effect, or a non-linear increase in efficiency of the electron emission due to  ion recombination on the metal surface. Similar photon trains were observed in the Russian Emission Detector (RED) xenon detector in Moscow~\cite{LLNLAr}. This detector does not have an extraction grid, but has a strong electric field on the cathode wires. 

The total ionization load results in an increase of rate of low energy background events. These low energy events can be due to simultaneous emission of several electrons in chain surface recombination events,  as discussed before. An alternative explanation of the few electron events could be resonance tunneling; while positive ions drift toward metal surface through a thin solid layer, the resonant tunneling conditions are produced at some ion position, and a burst of electron emission takes place. Large xenon detectors also demonstrate an increase in low-energy background events with fewer than $7$ primary  electrons in their spectra~\cite{Sorensen} with no obvious reason for the increase in the number of events in a low energy part of the spectrum. 

\subsubsection{Avoiding problems by detector design}

It should be possible to avoid or compensate for these adverse effects on high voltage stability, in the case of dark matter searches, and on the low energy background. The best approach for doing so is to investigate the mentioned phenomena in detail in direct experiments. Even without such experiments, some recommendations for detector design can be formulated. An experimental test of high voltage components stability in a liquid or gas in absence of positive ions and UV radiation does not guarantee a stable operation of those components in the detector. Ideally, one would avoid having a strong electric field, positive ions and UV radiation near a metal surface anywhere in a liquid or gas inside of the detector.  It is impossible to avoid a high field, positive ions and UV light on a cathode and on the extraction grid. Still, there are several possible ways to lower positive ion concentration on the wire surface. 

Positive ion mobility in a solid increases nonlinearly with an electric field, such that the positive ion trapping time on the thin solid film decreases nonlinearly with the field strength. Moreover, one should expect an electric breakdown in this thin solid film at some voltage. This phenomenon limits the maximum ion surface density on a solid helium film, for example. A solid helium film on a metal surface in liquid helium starts to conduct current for the voltage drop at about  50~V. Taking into account a small film thickness, this voltage corresponds to a rather high electric field strength that is far higher than 50 kV/cm. For argon and xenon detectors, there could be an optimum cathode wire thickness where the strong electric field decreases the lifetime of trapped positive ions,  while it is not large enough to cause strong field emission. Interestingly, the application of a strong AC electric field activates ion mobility more efficiently than application of a strong DC electric field.  The ion trapping time can be shortened by an application of a strong AC electric field between cathode wires in addition to a DC drift field. Application of this AC field will not change the ion and electron drift patterns in the regions away from the grid. 

\begin{figure}
\begin{center}
\includegraphics[width=3in]{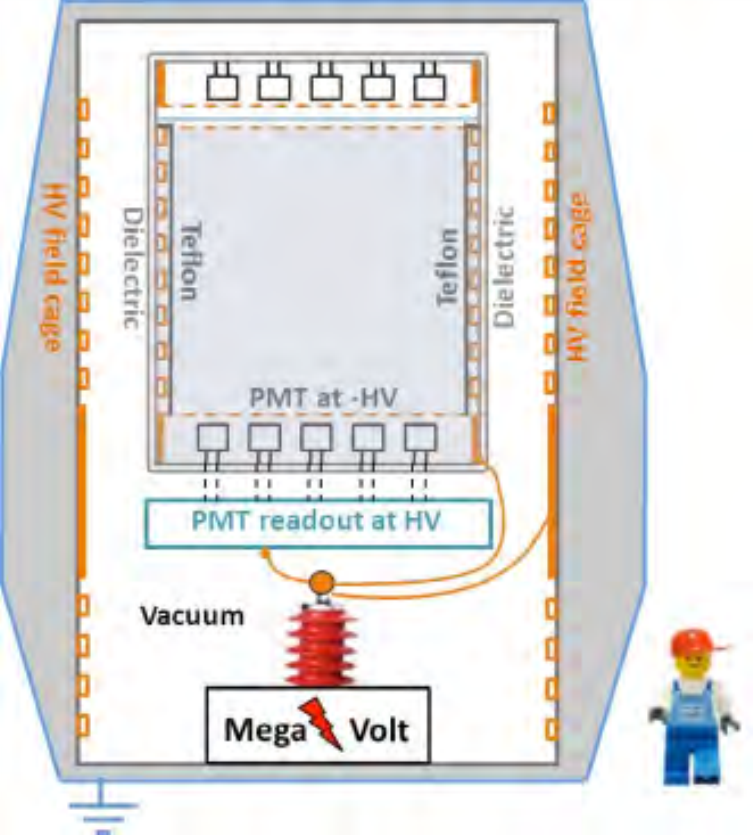}
\caption{Possible design for a large dielectric cryostat. }
\label{fig:sp3}
\end{center}
\end{figure}

It may be possible to avoid the problems mentioned above using a dielectric cryostat, such as that shown in Fig.~\ref{fig:sp3}, where the only high electric field inside the detector is contained in a small volume around the cathode, anode, and extraction grids.  Using such a design it should be possible to produce an amplification electric field with magnitude of 10 kV/cm in gas and between $0.1-10$~kV/cm in liquid, even for a large liquid volume of 10~m or more in height.  The total potential difference between the cathode and anode could reach up to 1~MV or more. This design does not require using a high voltage cable or feed-through to bring the high voltage to the cathode in the liquid.  It is likely that such a large dielectric cryostat for a DM detector will require development effort to find composite polymer structural materials that have low levels of radioactive impurities and sufficiently inert polymer material, such as Teflon, to line the inside of the liquid reservoir to avoid diffusion of organic impurities into the high purity liquid argon or xenon from the dielectric walls. The thick outer walls of the cryostat need to hold high voltage, but they do not need to be made of low out gassing materials or have extremely low levels of radioactivity.  Thus, having two electrostatic cages helps ease the requirements on the materials used for the cryostat.
The covering the inner field cage with Teflon would also stop ultraviolet radiation originating in the gas amplification region.  The Teflon surface would charge until the electric field component normal to the surface vanishes. The close potential distribution on the field cage resembles a linear potential drop, where a smaller charge surface density is required to nullify the normal field component.  The probability for charges to move along the surface due to the presence of a tangential electric field decreases with decreasing surface charge density and residual events caused by charge trapping and released by the Teflon walls can be geometrically rejected.

The effects of positive ions on metal surfaces is in need of theoretical models, and direct experimental determination of probabilities of electron emission and simultaneous emission of several  electrons as a function of electric field and ion surface density are required. High voltage stability of large detectors and  the non-radioactive part of low energy backgrounds are related to those effects. To understand the limits of operation of direct dark matter ionization detectors, knowledge of the effects caused positive ions is as important as knowledge of ionization efficiency by low energy nuclear recoils.

\subsection{Resistivity at Cold Temperatures}
{\it Contributed by J.~Va'vra, SLAC National Accerlerator Laboratory, Menlo Park, CA 94025, USA}
\newline
\subsubsection{Summary of measurements}
 
\begin{figure}[ht]
\centering
\includegraphics[width=4in]{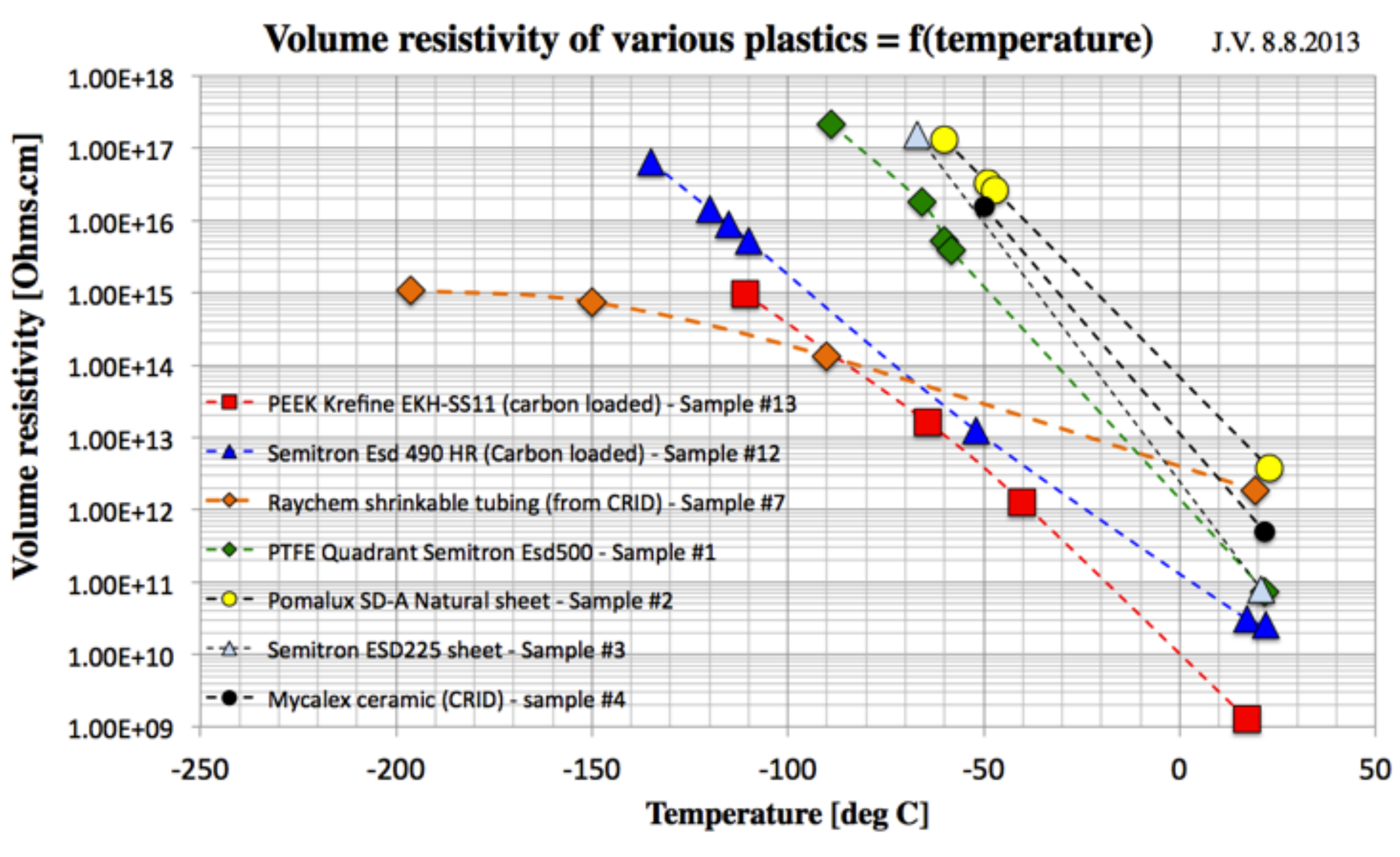}
\caption{Volume resistivity of insulator materials as a function of temperature. }
\label{fig:jv1}
\end{figure}

\begin{figure}[Htb]
\centering
\includegraphics[width=4in]{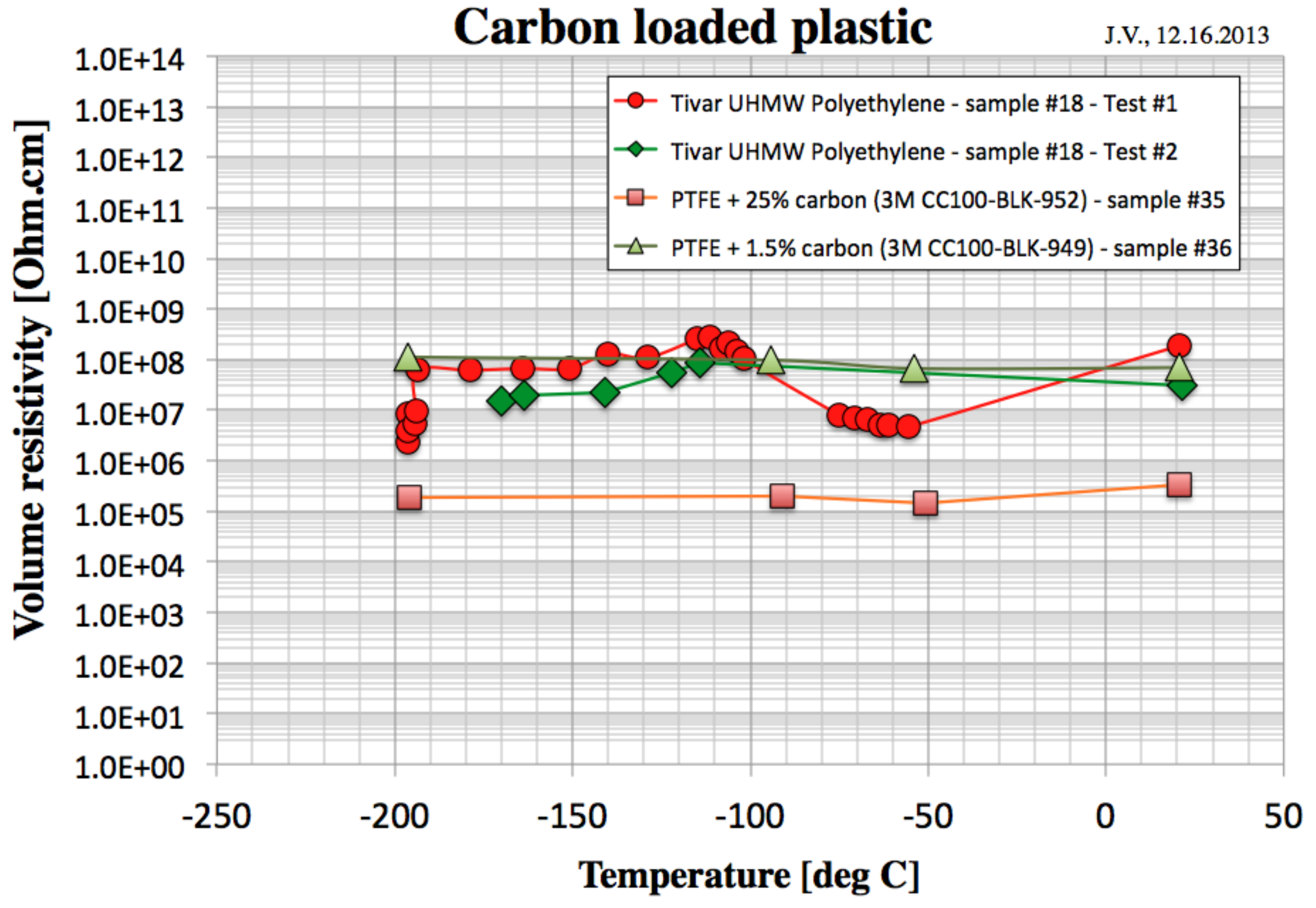}
\caption{The resistivity of carbon-loaded plastics are nearly constant between room temperature and liquid nitrogen temperatures.}
\label{fig:jv2}
\end{figure}

The resistivity of various insulators at very low temperatures has been measured~\cite{jv1} and is summarized in Table~\ref{table:jv1}. Many samples, which have a low resistivity at room temperature, show an exponential resistivity increase by approximately $9$ orders of magnitude as one goes from room to liquid nitrogen temperatures as shown in Fig.~\ref{fig:jv1}. Samples such as Raychem shrinkable tubing, Semitron ESd 490 and PEEK Krefine EKH-SS11 have a relatively low volume resistance at liquid xenon temperature ($-109^\circ$C). Many of the lower resistivity samples are loaded by carbon. Several of them show a constant resistivity between room temperature and liquid nitrogen temperature as seen in Fig.~\ref{fig:jv2}. Among lower resistivity samples, there is none, which would replace Teflon in terms of the photon reflectivity. Many samples, such as Teflon, polyethylene, and acrylic have a volume resistance of more than $4\times10^{18} \Omega$cm already at room temperature. Assuming a similar temperature dependence, their resistance may approach a value of $\rho \sim10^{27} \Omega$cm at liquid nitrogen temperature. More study is needed to find slightly more conducting material, which is as reflective as Teflon.
 \begin{table*}[tb]
\centering
\caption{\label{table:jv1} Volume resistivity measurements at room (20$^\circ$C), liquid nitrogen ($-196.4^\circ$C), and liquid xenon ($-109^\circ$C) temperatures.  All values are in units of $[\Omega$cm$]$.}
\resizebox{0.8\textwidth}{!}{\begin{tabular}{|c|l|c|c|c|}
\hline
Sample &	Material	                                                                             & 20$^\circ$C                     & $-196.4^\circ$C               & $-109^\circ$C    \\ \hline
1	     &   Semitron ESd 500 (PTFE) (0.29 inch-thick) 	                  & $7.5 \times 10^{10}$       & $\gg 10^{18}$                  & $\geq 10^{18}$  \\ \hline
2	     & Pomalux SD-A natural (0.092~inch-thick) 	                                   & $3.8 \times 10^{12}$	   & $\gg 10^{18}$                  & $\gg 10^{18}$    \\ \hline
3	     & Semitron ESd 225 (0.052~inch-thick) 	                                   & $8.0 \times 10^{10}$	   & $\gg 10^{18}$                  & $\gg 10^{18}$    \\ \hline
4	     & Mycalex sheet (0.130~inch-thick) 	                                           & $5 \times 10^{11}$	   & $\sim1.5 \times 10^{17}$ & -                         \\ \hline
5	     & Bakelite sheet (0.080~inch-thick)	                                           & $1.0 \times 10^{12}$        & $\sim1.5 \times 10^{17}$ & -                         \\ \hline
6	     & Raychem shrinkable tubing from CRID days (0.050~inch-thick) & $1.9 \times 10^{12}$	   & $\sim 1 \times 10^{15}$   & $\sim 3.5 \times 10^{14}$ \\ \hline
7	     & Teflon PTFE (0.030~inch-thick) - from EXO exp.                        & $> 4 \times 10^{18}$	   & $\gg 10^{18}$                  & $\gg 10^{18}$    \\ \hline
8	     & Fused silica sheet (0.125~inch-thick)	                                           & $\sim2.2 \times 10^{18}$  & $\gg 10^{18}$                 & $\gg 10^{18}$     \\ \hline
9	     & Mylar (0.005~inch-thick)	                                                           & $\sim2.1 \times 10^{18}$   & $\gg 10^{18}$                 & $\gg 10^{18}$     \\ \hline
10	     & Acrylic sheet used in EXO test (0.057~inch-thick)                      & $\sim4.0 \times 10^{18}$   & $\gg 10^{18}$                  & $\gg 10^{18}$      \\ \hline
11	     & Raychem, RNF-100-4-BK-STK (0.023~inch-thick)                     & $\sim7 \times 10^{18}$	    & $\gg 10^{18}$                 & $\gg 10^{18}$      \\ \hline
12	     & Semitron ESd 490HR (0.046~inch-thick) (carbon-loaded)	         & $2.6 \times 10^{10}$         & $> 10^{18}$                    & $\sim 5.2 \times 10^{15}$ \\ \hline
13	     & PEEK Krefine EKH-SS11 (0.042~inch-thick)  (carbon-loaded)	& $8.6 \times 10^9$              & $> 10^{18}$                    & $10^{15}$          \\ \hline
14	     & Raychem, MWTM-115/34-1500/U (0.04~inch-thick) 	                 & $> 6 \times 10^{17}$         & $\gg 10^{18}$                  &-                         \\ \hline
15	     & Rexolite (0.035~inch-thick) 	                                                   & $>3 \times 10^{18}$          & $\gg 10^{18}$                  & $\gg 10^{18}$    \\ \hline
16	     & PVC (0.043~inch-thick) 	                                                           & $>3 \times 10^{18}$	    & $\gg 10^{18}$                  & $\gg 10^{18}$    \\ \hline 
17	     & LDPE (0.039~inch-thick) 	                                                           & $>3 \times 10^{18}$	    & $\gg 10^{18}$                  & $\gg 10^{18}$    \\ \hline
18	     & TIVAR 1000 ESD \& EC (0.046~inch-thick)  (carbon-loaded)      & $10^7 - 10^8	$                   & $10^7 - 10^8$                 & $10^7 - 10^8$   \\ \hline
19	     & Delrin (0.041~inch-thick) 	                                                           & $\sim5.8 \times 10^{16}$   & $\gg 10^{18}$                  & $\gg 10^{18}$    \\ \hline
20	     & Polycarb (0.041~inch-thick) 	                                                   & $>3 \times 10^{18}$	    & $\gg 10^{18}$                   & $\gg 10^{18}$      \\ \hline
21	     & Polypro (0.044~inch-thick) 	                                                   & $>3 \times 10^{18}$	    & $\gg 10^{18}$                   & $\gg 10^{18}$     \\ \hline
22	     & Ultem (0.040~inch-thick) 	                                                           & $>3 \times 10^{18}$	    & $\gg 10^{18}$                  & $\gg 10^{18}$    \\ \hline
23	     & Peek (0.043~inch-thick) 	                                                           & $>3 \times 10^{18}$	    & $\gg 10^{18}$                   & $\gg 10^{18}$     \\ \hline
24	     & Teflon (0.041~inch-thick) 	                                                           & $>3 \times 10^{18}$	    & $\gg 10^{18}$                   & $\gg 10^{18}$     \\ \hline
25	     & Acrylic (0.040~inch-thick) 	                                                           & $>3 \times 10^{18}$	    & $\gg 10^{18}$                  & $\gg 10^{18}$    \\ \hline
26	     & Cast-33 glue (cure 26, TFE) (0.064~inch-thick)	                         & $7.8 \times 10^{12}$	    & $\gg 10^{18}$                  & $\gg 10^{18}$    \\ \hline
27	     & Insul Cast-502 (cure 26,TFE) (0.063~inch-thick)                        & $1.5 \times 10^{16}$	    & -                                      & -                         \\ \hline
28	     & Lord-340 glue cast (\#70, 100-8) 	                                         & $1.3 \times 10^{15}$	    & $\gg 10^{18}$                 & $\gg 10^{18}$    \\ \hline
29	     & CLR-1066/CLH 6330/TEE glue cast 	                                         & $7 \times 10^{15}$	             & -                                      & -                        \\ \hline
30	     & CAST-502 clear clue, Insulcure-26, BYK-A-500                         & $6.4 \times 10^{16}$	     & -                                      & -                        \\ \hline
31	     & Hysol Dexter glue cast (0.057~inch-thick) 	                                 & $2.5 \times 10^{15}$	     & $\gg 10^{18}$                  & $\gg 10^{18}$   \\ \hline
32	     & Sample 2 - unspecified glue cast (0.059~inch-thick) 	                 & $2.1 \times 10^{15}$	     & -                                      & -                        \\ \hline
33	     & Sample 3 - unspecified glue cast (0.062~inch-thick) 	                 & $1.0 \times 10^{16}$	     & -                                      & -                        \\ \hline
34	     & Sample 4 - unspecified glue cast (0.059~inch-thick) 	                 & $3.3 \times 10^{15}$	     & -                                      & -                        \\ \hline
35	     & PTFE + 25\% Carbon - 3M CC100-952 (0.032~inch-thick)	         & $3.4 \times 10^5$	              & $1.93 \times 10^5$         & -                       \\ \hline
36	     & PTFE + 1.5\% Carbon - 3M CC100-949 (0.067~inch-thick)        & $1.46 \times 10^7$	              & $1.09 \times 10^8$         & -                        \\ \hline
\end{tabular}}
\end{table*}

One should also investigate if Teflon is not shedding its polymer-chain fragments, coating cathode wire surfaces. In the past Teflon tubing was to be avoided as it would increase the wire aging rate~\cite{jv2}. It is assumed that Teflon coated cathode wires with an insulator triggered the Malter electron emission by approaching ions. This assumption needs to be studied in noble liquid TPCs as well.

\subsubsection{Electrodless TPC}

The Allison paper~\cite{ja} was the first to show the importance of positive ions in the TPC electric field uniformity. Their simulation shows that the TPC field has ripples initially when there are no free charges on the surface. Once the positive charge is introduced, ripples are removed, and the deposition of positive ions is self-limiting. Allison used fiber glass boards with volume resistivity of about $3\times10^{17} \Omega$cm and surface resistivity of approximately $10^{15} \Omega/$cm$^{2}$. The charging time constant was a few seconds. The question is what time constants are in a structure where there is no amplification to produce positive ions, it is located deeply underground, and uses Teflon with a volume resistivity of 8 orders of magnitude higher than G-10. The ion deposition will be driven by the radioactive source calibration. Will it distort the drift field? This question needs to be studied in detail with a laser.
    
However, one can ask a different question: could one use a carbon-loaded plastic to produce an electrodeless TPC operating at liquid xenon temperature? A TPC made of a 1 m-long, 1 cm-thick, 1 m-radius cylinder using material \#13 in Table~\ref{table:jv1}, would draw $\sim0.6$~nA at 100~kV; one would have to increase the carbon content to get the right current to prevent distortions, while preventing heating of the liquid. A benefit of such a design is that it would eliminate a resistor chain and the electrode grid, and provide a uniform non-localized heating.

     

\section{EDM Experiments}
\subsection{SNS nEDM}
\label{sec:snsnedm}
{\it Contributed by T.~M.~Ito, Los Alamos National Laboratory, Los Alamos, New Mexico 87545, USA in collaboration with \\
D.~H.~Beck, University of Illinois, Urbana, Illinois 61801, USA, \\
S.~M.~Clayton, S.~A.~Currie, W.~C.~Griffith, J.~C.~Ramsey, A.~L.~Roberts, Los Alamos National Laboratory, Los Alamos, New Mexico 87545, USA, \\
C.~Crawford, University of Kentucky, Lexington, Kentucky 40506, USA, \\
R.~Schmid, California Institute of Technology, Pasadena, California 91125, USA, \\
G.~M.~Seidel, Brown University, Providence, Rhode Island 02912, USA, \\
D. Wagner, University of Kentucky, Lexington, Kentucky 40506, USA, \\
W. Yao, Oak Ridge National Laboratory, Oak Ridge, Tennessee 37831, USA}

\subsubsection{Introduction}

\begin{figure}[h]
\centering
\includegraphics[width=4in]{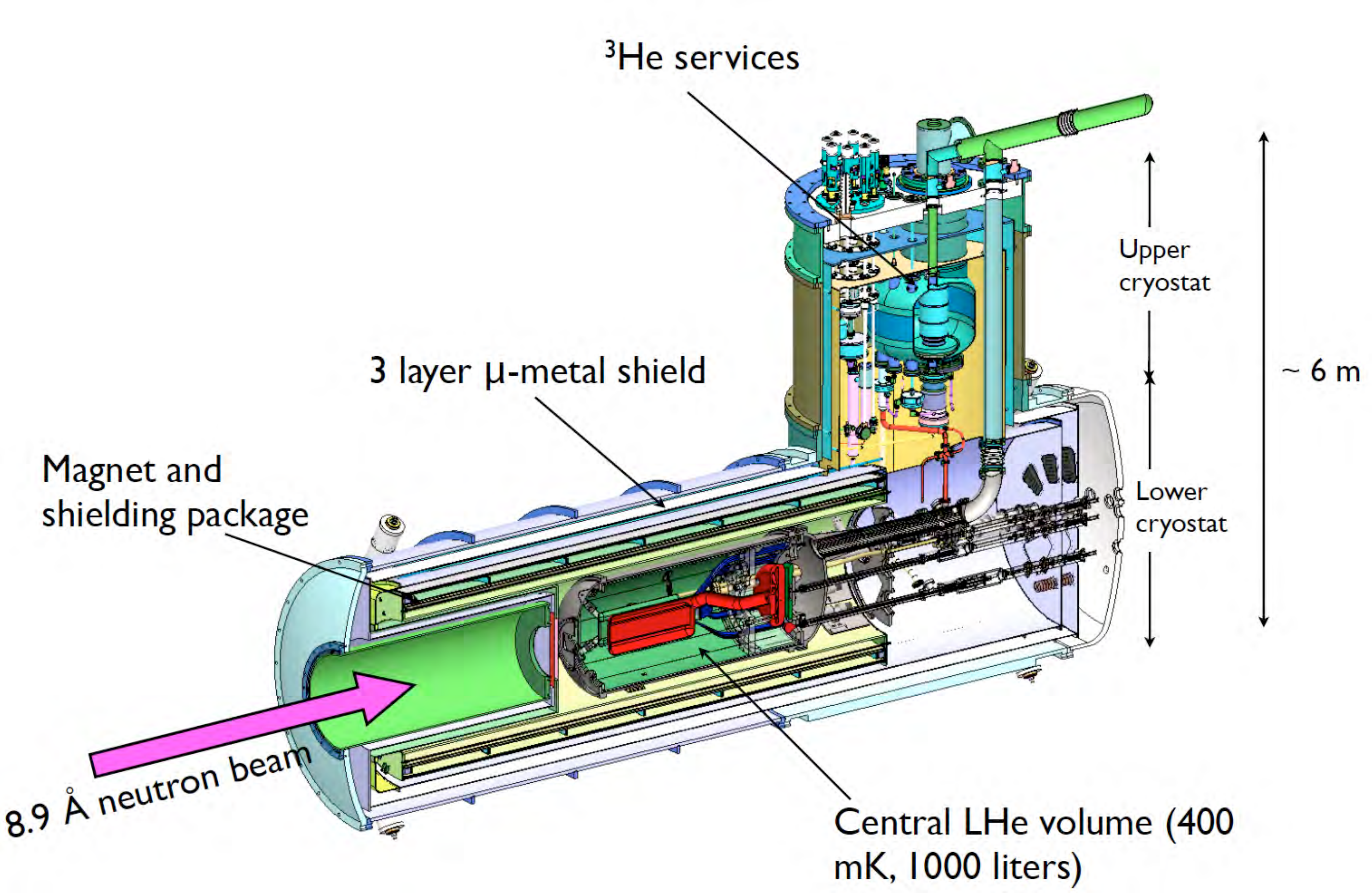}
\caption{A schematic of the apparatus for the SNS nEDM experiment, as is currently designed}
\label{fig:SNSnEDM}
\end{figure}

A nonzero permanent electric dipole moment (EDM) of a nondegenerate state of a system with spin $J \neq 0$ violates time reversal invariance as well as invariance under parity operation. The time reversal invariance violation implies a $CP$ violation through the $CPT$ theorem. Given the smallness of the standard model $CP$ violating contributions induced by quark mixing, an EDM is a sensitive probe of new physics. A new experiment to search for the permanent EDM of the neutron, SNS nEDM~\cite{ITO07}, based on the method proposed by Golub and Lamoreaux~\cite{GOL94}, is being developed to be mounted at the Spallation Neutron Source at Oak Ridge National Laboratory, with a sensitivity goal of $\sim 5\times 10^{-28}$~$e\cdot$cm, an improvement of roughy two orders of magnitude over the current limit~\cite{BAK06}. A schematic of the apparatus for the SNS nEDM experiment is shown in Fig.~\ref{fig:SNSnEDM}.

\subsubsection{High voltage requirements}

The SNS nEDM experiment requires that a high, stable electric field of approximately $75$~kV/cm be applied in the region inside the two measurement cells that are sandwiched between electrodes. The measurement cells and electrodes are immersed in 0.4~K liquid helium. The measurement cells, which store ultracold neutrons, are filled with isotopically pure liquid $^4$He at $0.4$~K with the relative $^3$He concentration of $10^{-10}$. The measurement cells are made of PMMA and are 10.16~cm$\times$ 12.70~cm $\times$ 42~cm in outer dimension with a wall thickness of 1.2~cm. The electrodes, roughly $10$~cm~$\times$~40~cm$\times$~80~cm in size, are made of PMMA coated with a material that needs to meet various requirements related to electrical resistivity, neutron activation properties, and magnetic properties. The leakage current along the cell walls need to be minimized.

\subsubsection{General remarks on electrical breakdown in liquid helium}
\label{sec:snsbreakdown}

Electrical breakdown in liquid helium, or more in general electrical breakdown in any dielectric liquid, is rather poorly understood. Data exist on electrical breakdown in liquid helium for temperatures of $1.2-4.2$~K, with the bulk of data being taken at 4.2~K, mostly at the saturated vapor pressure (SVP) for various electrode geometries, including sphere-to-sphere, sphere-to-plane, and plane-to-plane. However, there is little consistency among the data, and therefore there is no consistent theoretical interpretation.

However, a rather simple consideration of the mean free path of ions in liquid helium and the electric field strength necessary to accelerate them to an energy sufficiently high to generate subsequent ionization leads to a conclusion that the intrinsic dielectric strength of bulk liquid helium is greater than 10~MV/cm, a field much higher than breakdown fields experimentally observed. This leads to the following generally-accepted picture for the mechanism of generation of electrical breakdown in liquid helium:
\begin{enumerate}
\item A vapor bubble is formed on the surface of the electrode, e.g. by field emission from roughness on the cathode
\item The vapor bubble grows, presumably by heating of the gas by accelerated electrons and evaporation of the liquid as a result, and forms a column of gas reaching from one electrode to the other
\item Electrical breakdown occurs through the gas column
\end{enumerate}
It follows that the parameters that can affect the breakdown field strength include: (i) electrode material, in particular the surface properties,and (ii) liquid helium temperature and pressure. In addition, because electrical breakdown is a stochastic process, the size of the system affects the breakdown field strength and its distribution~\cite{WEB56}.

\subsubsection{Research \& Development (R\&D) Approach}

The consideration given above indicates that the R\&D for the SNS nEDM experiment requires a study of electrical breakdown in liquid helium in a condition (i.e. temperature, pressure, size) as close as possible to that expected for the SNS nEDM experiment, using suitable candidate materials.

It is also very important to study the effect of the presence of a dielectric insulator sandwiched between electrodes, as such will be the geometry for the SNS nEDM experiment. Note that even in a room temperature vacuum system, electric fields exceeding a few 100~kV/cm are possible when there is no insulator between the two electrodes.  In a study performed using a room temperature vacuum aparatus~\cite{GOL86} similar to those used in the previous nEDM experiments, such as Ref.~\cite{BAK06}, the electric field was limited to $\sim 30$~kV/cm due to the presence of the UCN confining wall that was sandwiched between the two electrodes~\cite{GOL86,nEDMnote}.  Field emission at the cathode-insulator junction is thought to be responsible for initiating breakdown~\cite{KOF60}, which is expected to be suppressed at cryogenic temperatures.

In order to study the relevant aspects of electrical breakdown in liquid helium with a goal of establishing the feasibility of the SNS nEDM experiment as well as guiding the design of the apparatus, an apparatus called the Medium Scale High Voltage (MSHV) Test Apparatus was constructed and is described below. 
\begin{figure}[h]
\centering
\includegraphics[width=4in]{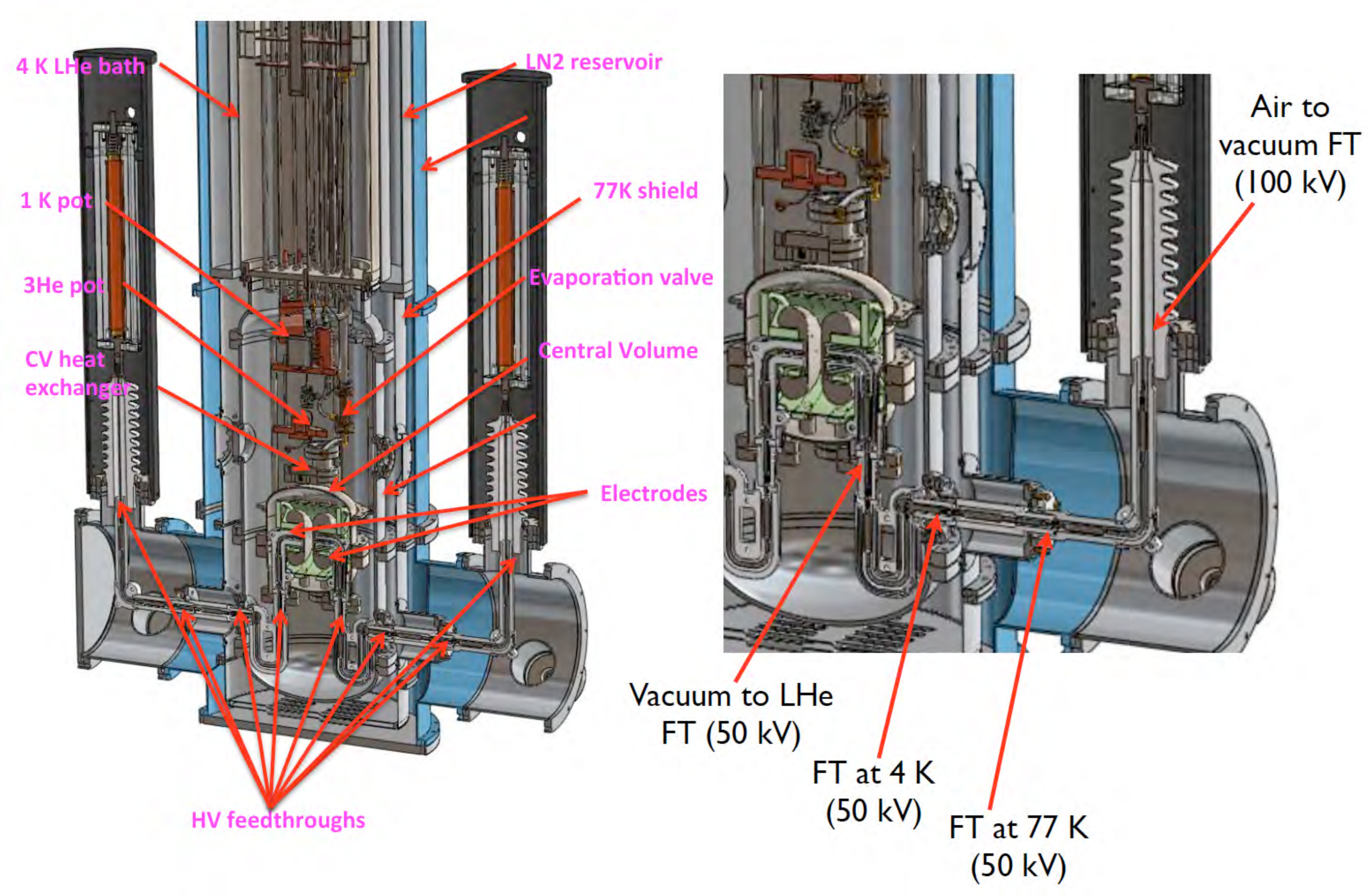}
\caption{A schematic of the MSHV system for the SNS nEDM experiment.}
\label{fig:schematic}
\end{figure}

\subsubsection{Medium scale high voltage test apparatus}

The purpose of the MSHV system is to study electrical breakdown in liquid helium in a condition approximating that of the SNS nEDM experiment using suitable electrode candidate materials. The lowest operating temperature of the MSHV system is designed to be 0.4~K, corresponding to the operating temperature of the SNS nEDM experiment. Since it is expected that the pressure is an important parameter affecting the breakdown field strength in liquid helium, the MSHV system is designed so that the pressure of the liquid helium volume in which the electrodes are placed can be varied between the SVP and 1~atm. Note that the SVP is approximately $10^{-6}$~torr at 0.4~K. The size of the liquid helium volume was determined as a compromise between the following two competing factors:
$i)$ short turnaround time of the system to allow multiple electrode material candidates to be tested, and $ii)$ size large enough to give information relevant for the SNS nEDM experiment.  The electrodes are 12~cm in diameter. The gap size is adjustable between 1 and 2~cm. Each dimension is within a factor of 10 of the SNS nEDM experiment's high voltage system. A schematic of the MSHV system is shown in Fig.~\ref{fig:schematic}.
The Central Volume (CV), a 6-liter liquid helium volume that houses the electrodes, is cooled by a $^3$He refrigerator. 

High voltage between $-50$~kV and $+50$~kV can be provided to each electrode through a high voltage feed line. The high voltage feed lines are made of thin wall stainless steel tubing and are thermally anchored at the liquid nitrogen heat shield and at the 4~K heat shield, in order to minimize the heat leak to the high voltage electrodes. Heat leak to the high voltage electrodes can cause vapor bubbles to be created on the surface of the electrodes, which in turn can initiate electrical breakdown, potentially leading to erroneous results. 

Commercially available models are used for all the high voltage feedthroughs. For the air-to-vacuum feedthroughs, CeramTec Model 6722-01-CF feedthroughs, rated for 100~kV, are used. For all other feedthroughs, including ones on the CV that need to be superfluid tight, CeramTec Model 21183-01-W, rated for 50~kV and for liquid helium temperature operation, were chosen because the spatial limitations did not allow larger sized feedthroughs. In an offline test of these 50~kV feedthroughs, we found that, after proper cleaning, they can withstand up to about $90$~kV with a leakage current of less than 1~nA in vacuum when cooled to 77~K.

Most of high voltage components were tested in a separate system for hold-off voltage before being installed into the MSHV system. Experience shows that a component that functions in a room temperature vacuum generally functions in liquid helium.

\begin{figure}[Htb]
\centering
\includegraphics[width=5in]{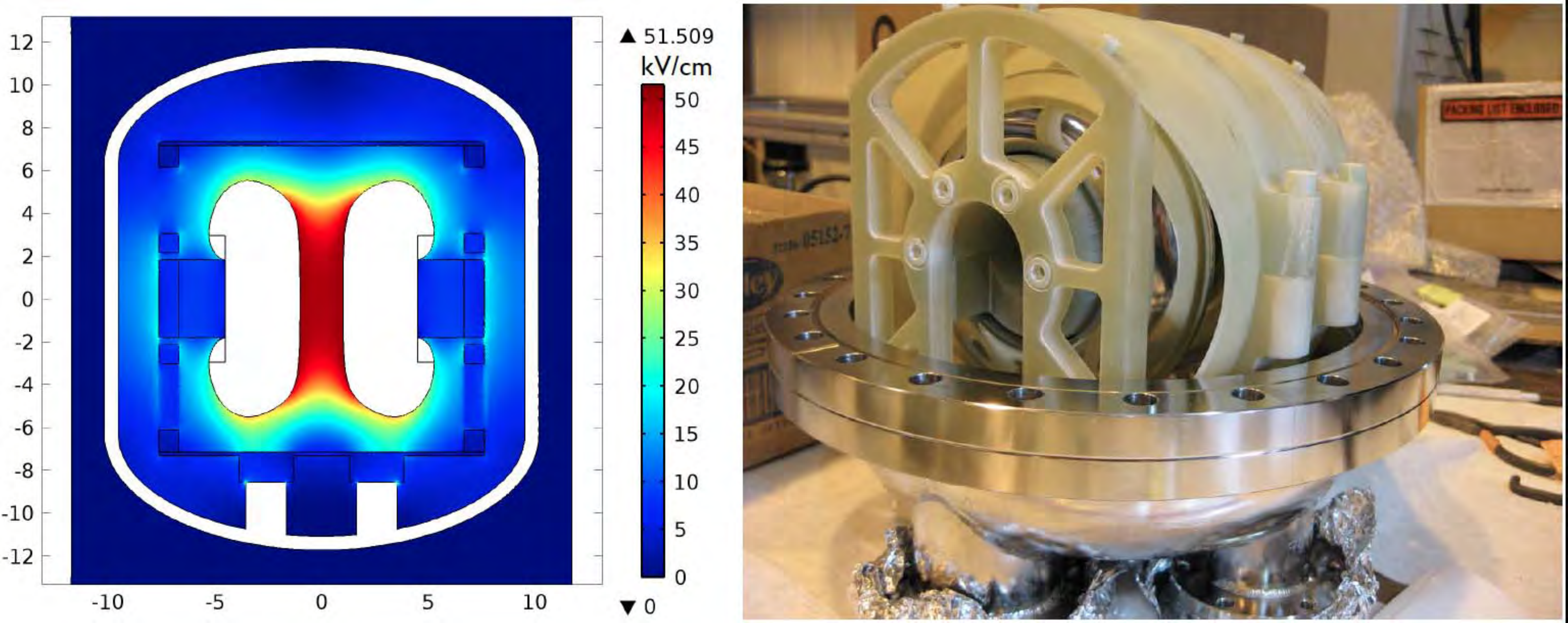}
\caption{Rogowski electrodes for the MSHV system of the SNS nEDM experiment. (left) FEM calculation of the electric field inside the CV showing a uniform field in the gap. (right) A photograph of the Rogowski electrodes installed in the CV.}
\label{fig:electrodes}
\end{figure}

The initial electrodes are made of electropolished stainless steel and have the so-called Rogowski profile~\cite{COB58}, which provides a uniform electric field in the gap and ensures that the gap has the highest field in the system, as seen in Fig.~\ref{fig:electrodes}.

\subsubsection{Progress to date}

The MSHV system was successfully constructed and commissioned. The CV can be cooled to 0.4~K with it filled with liquid helium, and the pressure inside the CV can be varied and controlled easily.
\begin{figure}[Htb]
\centering
\includegraphics[width=4in]{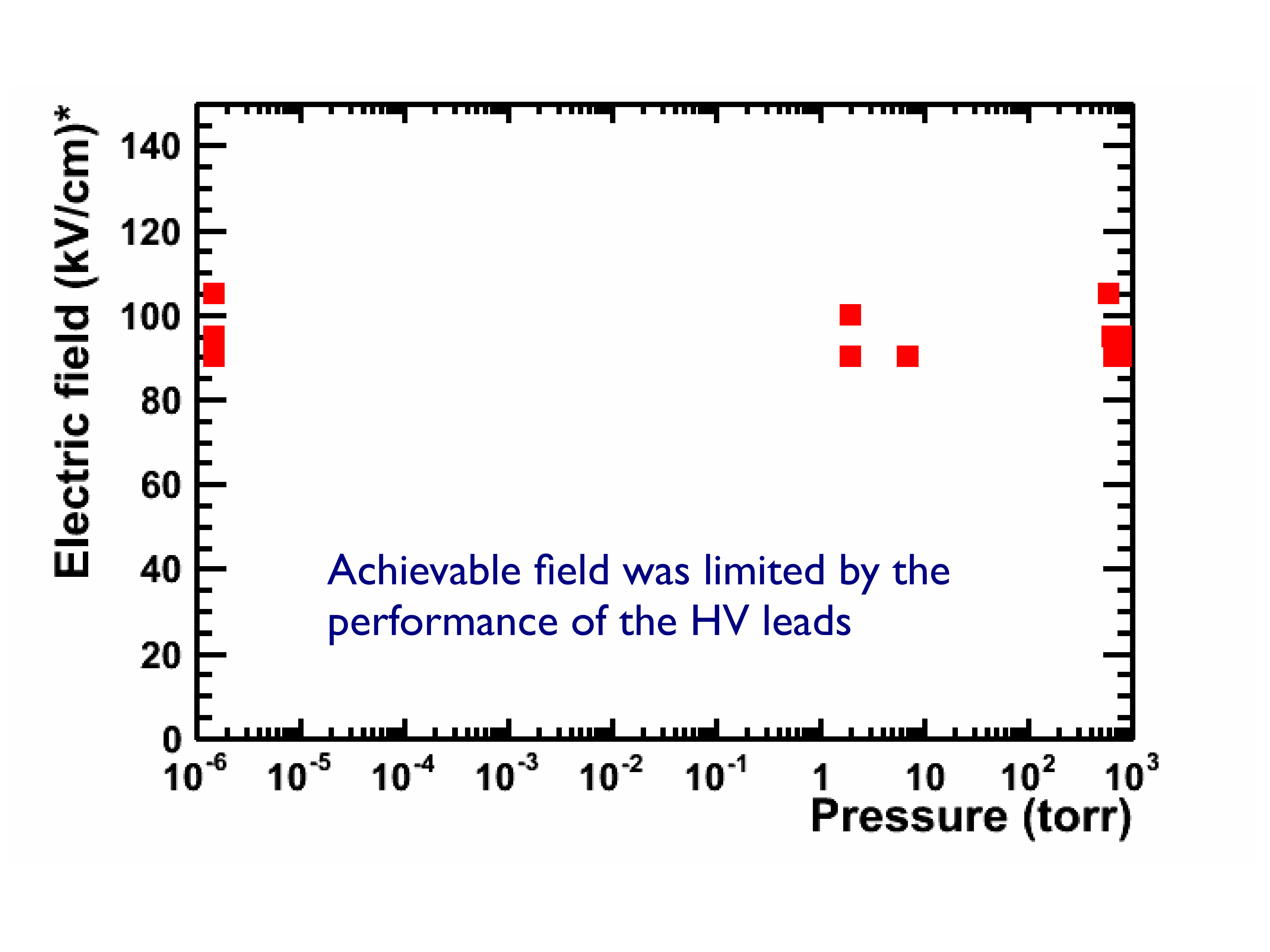}
\caption{Electric field strength achieved in liquid helium for a 1~cm gap between two electropolished stainless steel electrodes 12~cm in diameter. Note that the achievable field was limited by the performance of the high voltage leads. The hydrolic pressure of liquid helium is $\sim 0.1$~torr/cm}
\label{fig:results}
\end{figure}

In addition, an electric field exceeding 100~kV/cm can be stably applied in a 1-cm gap between the two 12-cm diameter electrodes made of electropolished stainless steel for a wide range of pressures, as seen in Fig.~\ref{fig:results}. The achievable field was limited by the performance of the high voltage leads. No breakdown was observed in the gap between the two electrodes. Also the leakage current between the two electrodes was measured to be less than 1~pA at a 50~kV potential difference. This current was measured with one of the high voltage electrodes set to the ground potential to avoid the leakage current in the high voltage cable and feedthroughs dominating the measurement.

%

\section{Liquid Argon Neutrino and Dark Matter Experiments}

In this section we review the high voltage experience of several working liquid argon experiments (MicroBooNE, ArgoNeuT, LBNE, DarkSide, and GLACIER), as well as R\&D in preparation for new large-scale experiments (ARGONTUBE and CAPTAIN). 

\subsection{MicroBooNE}
{\it Contributed by B.~Lundberg, Fermi National Accelerator Laboratory, Batavia, IL 60510, USA for the MicroBooNE Collaboration}
\newline

The MicroBooNE experiment at Fermilab is assembling a 170~ton, 70~ton fiducial, liquid argon TPC that will be installed at the Liquid Argon Test Facility (LArTF) site in the Booster Neutrino Beam.  This TPC must efficiently drift electrons created from particle tracks and collect them on wires up to 2.5~m away from their source. To do so requires a drift electric field with a magnitude of 300 to 500~V/cm generated by a high potential applied at one side of the TPC active volume. In MicroBooNE this cathode plane and the rectangular TPC assembly are located within the cylindrical volume of the cryostat. Hence there are locations at high potential, near the corners of the TPC framework, that are places of concern for operating at the highest applied voltages.  The literature on liquid dielectrics is extensive and provides general guidance for design. However, the wide ranging conditions with very specific intended uses renders predictions of exact behaviors in the MicroBooNE system ambiguous. The operating condition in MicroBooNE is {\it terra incognita} for particle physics detectors, and no guarantee exists for the successful operation with the highest voltages.
\begin{figure}[ht]
\centering
\includegraphics[height=3.in]{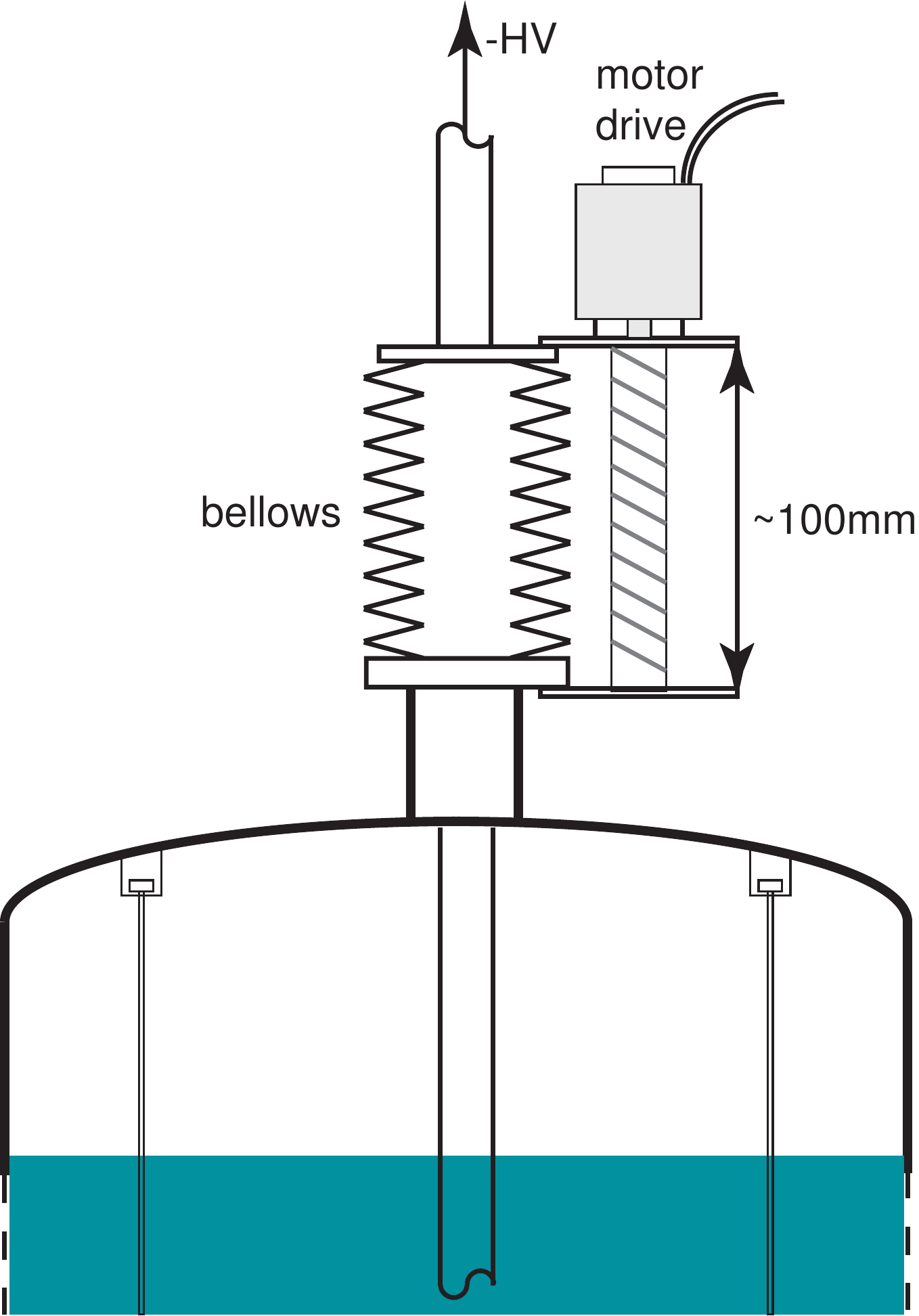}
\caption{A schematic of the top of the MicroBooNE liquid argon test cryostat showing the movable feedthrough, bellows and motor drive. The electrodes are 
mounted below and not shown here.}
\label{fig:top_assy}
\end{figure}

It is clearly not practical to study the effects of high voltage dielectric breakdown within the actual TPC and cryostat assemblies before filling the cryostat at the LArTF. Instead, a smaller, dedicated cryostat will be placed at the LArTF to contain a a few hundred liters of liquid argon, in which tests will be conducted to ascertain with some precision the properties of the dielectric cryogenic liquid. This dewar will be connected to the LArTF purification and circulation system and provided  with a purity monitor to determine the drift lifetime of the electrons. The proposed studies include understanding dielectric failure with respect to electrode surface quality, electrode geometry and liquid argon purity. The tests are designed to provide generic and useful information for future TPC design, but also to address MicroBooNE-specific design issues.

At the LArTF, the ``test cryostat'' will be installed and plumbed at the end of the full cryogenic system. The outlet of the test cryostat feeds into a purity monitor. The commercially-built cryostat has an 880 liter capacity. Inside the cryostat volume, a high voltage feedthrough charges interchangeable stainless spherical electrodes which are suspended above a flat stainless plate electrode at ground. The distance between electrodes can be varied accurately from 0~mm, or contact, to 100~mm. The high voltage power supply is capable of 150~kV.

The testing apparatus consists three main parts: $i)$ a free-standing cryostat to contain liquid argon, $ii)$ the high voltage feed-through and power supply to provide the high potential to the $iii)$ electrodes which have an accurately maintained distance between them.  There are two complications to an otherwise very simple design. First, the feed-through must be easily extracted and the electrode at its end changed-out then reinserted.  Secondly, the feed-through must be mounted to a carefully controlled, vertically moving assembly so that the distance between the two electrodes can be varied in a reproducible way. The feed-through and grounded plate electrode are mounted, aligned and tested prior to attaching the top section to the main part of the cryostat.   A schematic of the top of the test system is shown in Figure~\ref{fig:top_assy}.

The tests to be performed include $i)$ measuring breakdown voltage as a function of electrode size and shape as well as argon purity, $ii)$l ooking for corona from current/current pulses and onset of breakdown, and $iii)$ dielectric constant measurements.  These tests will be conducted before the large MicroBooNE TPC and cryostat are installed at LArTF in March 2014.

\subsection{ArgoNeuT}
{\it Contributed by M.~Soderberg, Fermi National Accelerator Laboratory, Batavia, IL 60510, USA and Syracuse University, Syracuse, NY 13210, USA}
\newline

\begin{figure}[h]
\centering
\includegraphics[height=1.6in]{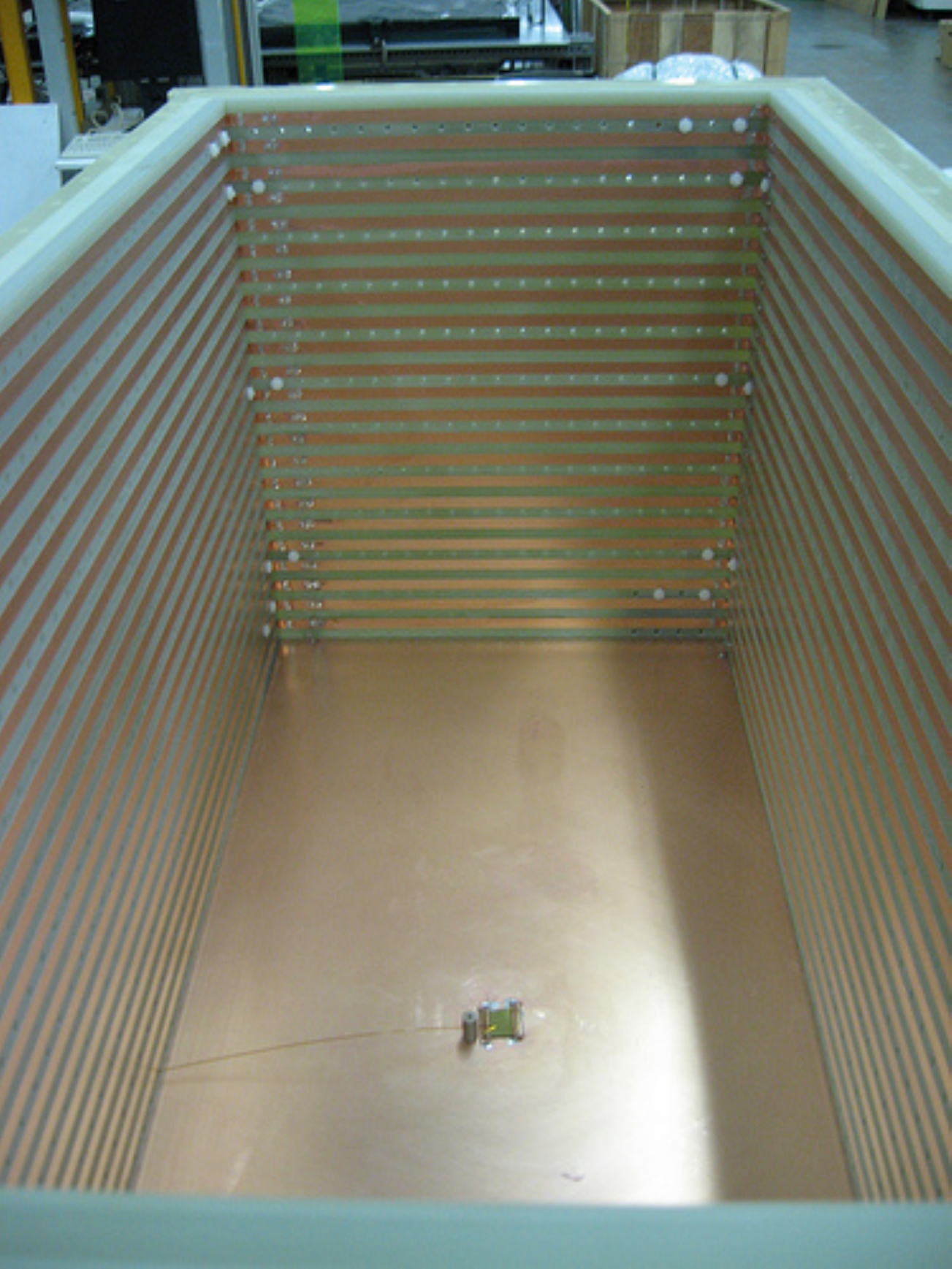} 
\includegraphics[height=1.6in]{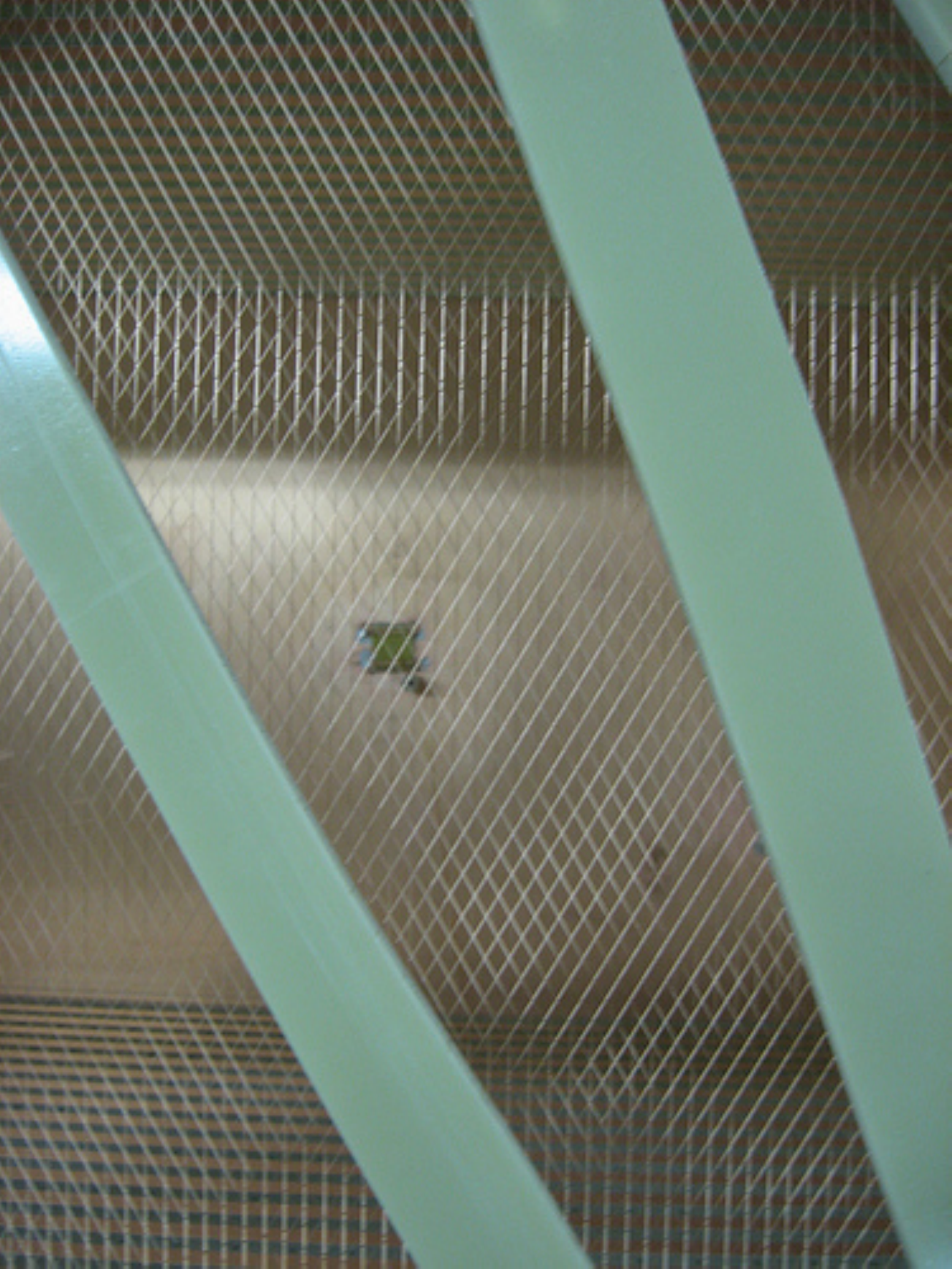} \\
\includegraphics[width=2in]{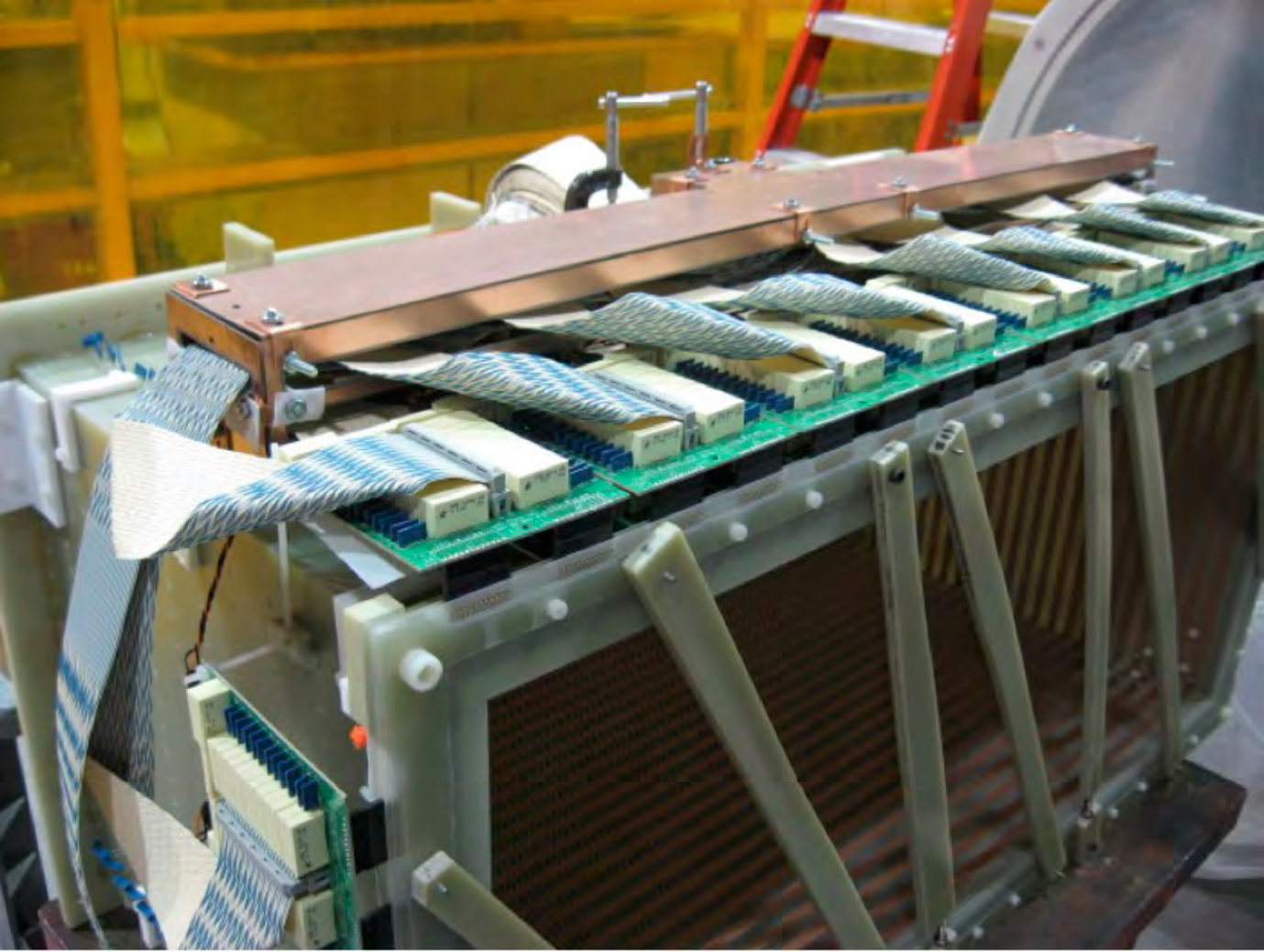}
\includegraphics[width=2in]{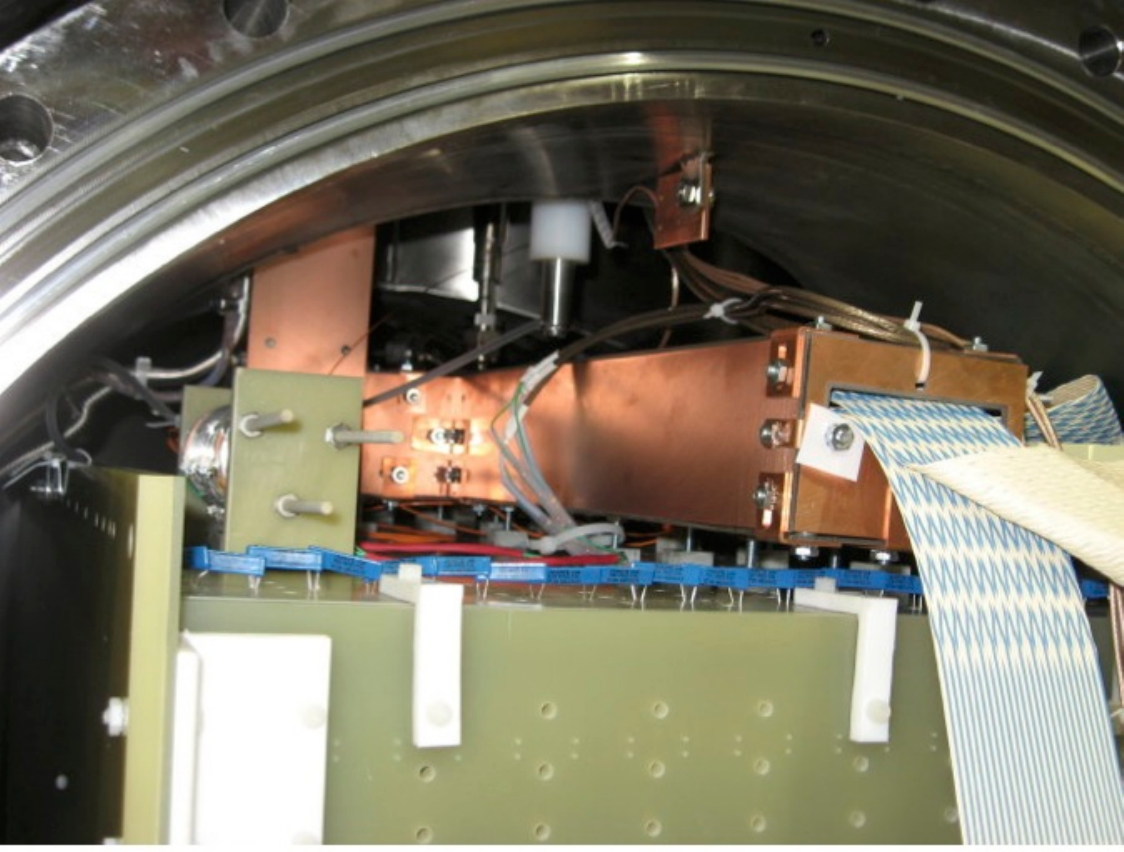}
\caption{Looking into the ArgoNeuT liquid argon TPC field cage before (top left) and after (top right) installation of the wire planes.  The ArgoNeuT TPC before (bottom left) and after (bottom right) installation in the cryostat.}
\label{fig:tpc}
\end{figure}

The ArgoNeuT experiment featured a LArTPC with inner dimensions of 40 cm (height) $\times$ 90 cm (length) $\times$ 47.5 cm (width).  Photographs of the TPC are shown in Fig.~\ref{fig:tpc}.  The maximum drift distance of 47.5 cm from the cathode to the innermost anode plane, and the desired drift electric-field of 500 V/cm, make the required high-voltage 23750 V, or $\approx$25 kV.  Neighboring stages within the TPC field cage were connected by four 100 M$\Omega$ resistors arranged in parallel.  The resistance of the TPC, and an external high voltage filter that was located in series with the TPC, was 670 M$\Omega$.

\begin{figure}[h]
\centering
\includegraphics[height=3in]{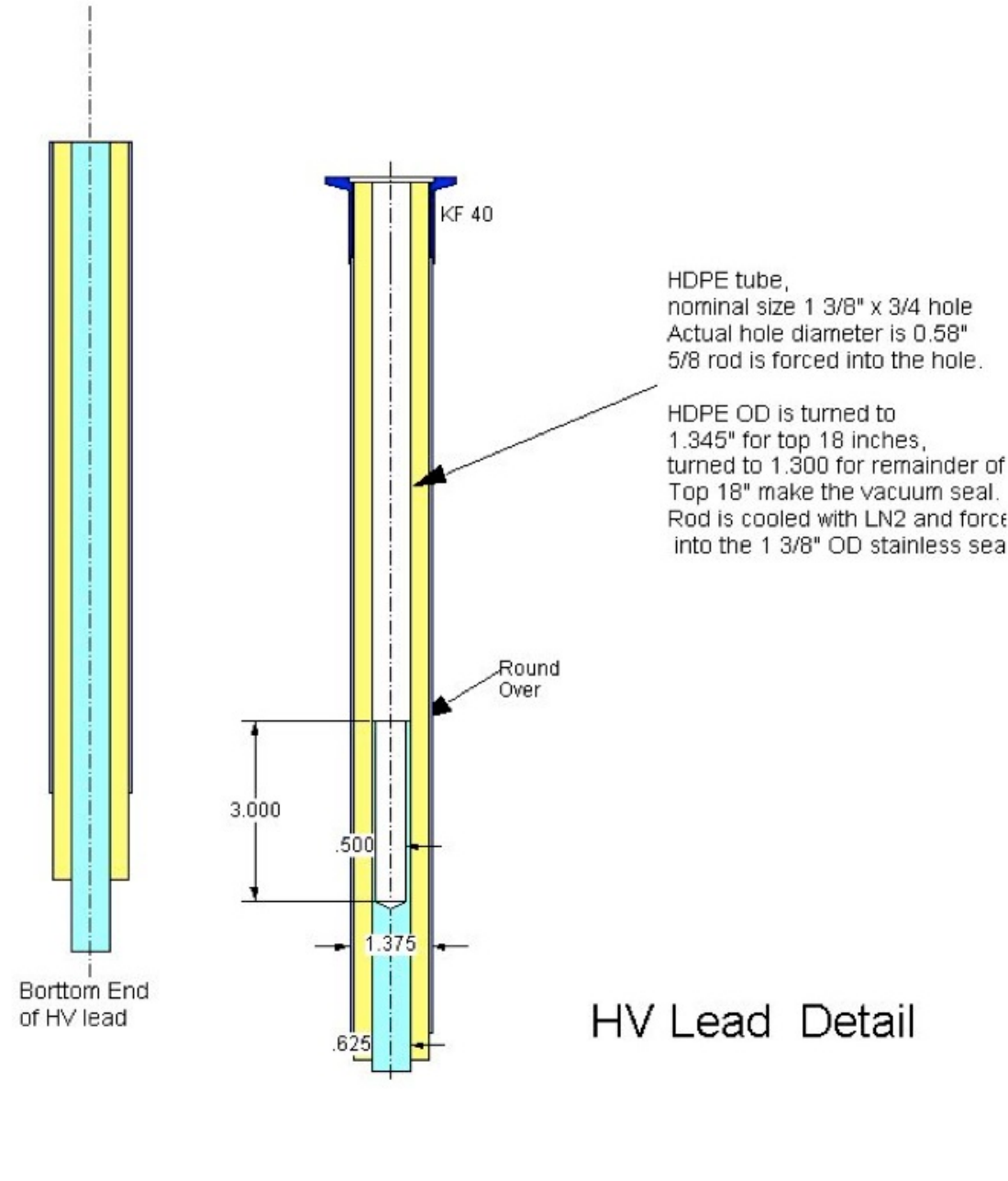}
\includegraphics[width=2in]{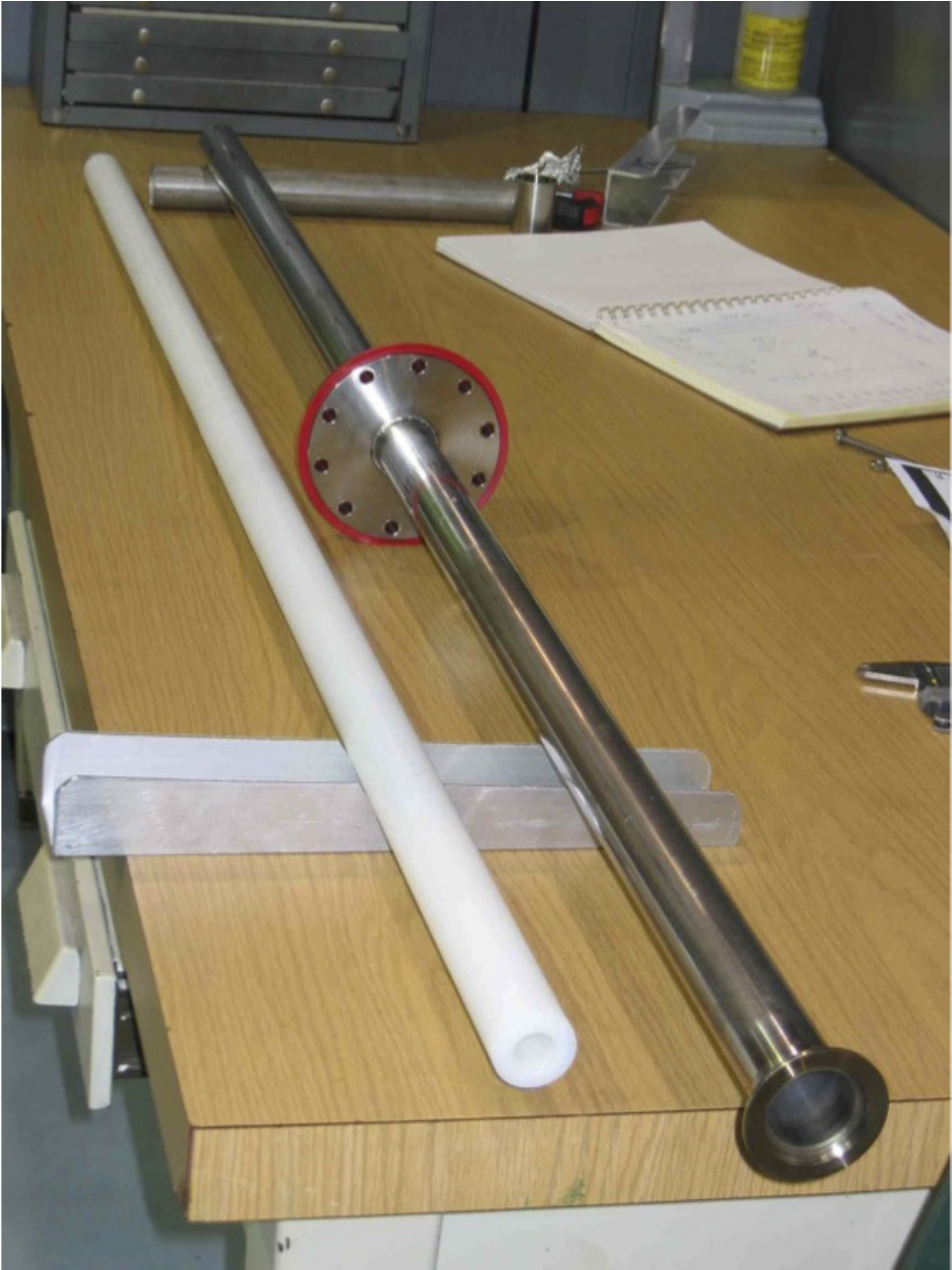}
\caption{High voltage feed-through for the ArgoNeuT liquid argon TPC.}
\label{fig:feed}
\end{figure}
The feed-through that brings high voltage from the outside to the inside of the cryostat was constructed by inserting a solid stainless steel (SS) rod with 5/8 inch outer diameter into a high density polyethylene (HDPE) tube that had been bored out to 0.58~inches.  The pressure required to complete this insertion forms a seal that is leak tight.  The outside face of the HDPE tube is machined to 1.345~inches in diameter at the upper, warm end, while the bottom, cold end is machined to 1.300~inches in diameter.  The HDPE+SS assembly is cooled using liquid nitrogen, temporarily causing it to contract, and then inserted into a stainless-steel tube with outer diameter of 1.375~inches.  Upon warming up, the HDPE+SS tube expands and forms a vacuum-tight seal against the outer SS tube on its upper, thicker, end.  The completed assembly was helium-leak checked and found to have a leak rate of $<10^{-10}$~torr/liter/sec. The outer SS tube is outfitted with a 4-5/8~inch conflat flange which is mounted on the ArgoNeuT cryostat.  The feed-through was tested for its ability to achieve the desired voltage by immersing the cold end in non-purified liquid argon and gradually increasing the voltage up to -50~kV, or twice the nominal operating voltage of -25 kV, and verifying the system was stable for a period of at least 1 hour.   Figure \ref{fig:feed} shows the high voltage feed-through design and components under construction.
\begin{figure}[htb]
\centering
\includegraphics[width=4in]{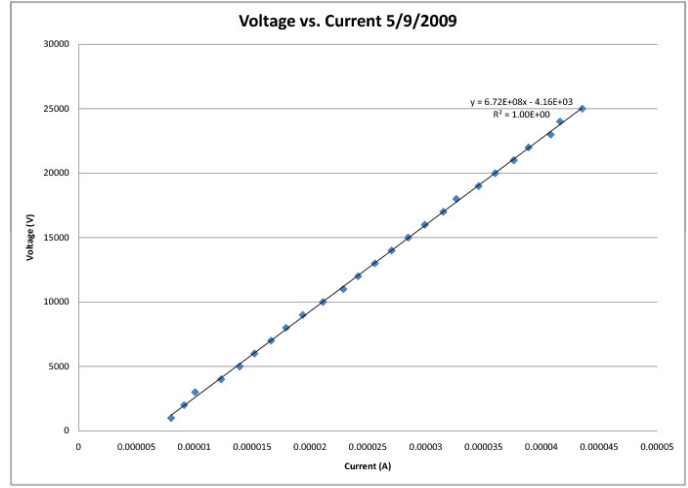}
\caption{Current-Voltage curve for ArgoNeuT liquid argon TPC.}
\label{fig:iv}
\end{figure}

After the ArgoNeuT cryostat was filled with liquid, the voltage on the TPC cathode was gradually increased and the resulting current through the TPC was recorded.  Figure \ref{fig:iv} shows the current-voltage curve as the voltage is gradually increased on the cathode.  The resistance of the TPC circuit is determined by fitting the slope of this curve, and the measured value of 672~M$\Omega$ is consistent with the expected value of 670~M$\Omega$.  This check was performed periodically throughout the physics run of ArgoNeuT to verify the performance of the TPC field cage.  The TPC and connected high voltage feed-through operated at -25 kV continuously in the NuMI tunnel for a period of about 9 months without incident.  

\subsection{ARGONTUBE}
\label{sec:strauss}
{\it Contributed by T.~Strauss, Laboratory for High Energy Physics, Albert Einstein Center for Fundamental Physics (AEC), University of Bern, 3012 Bern, Switzerland in collaboration with \\
A.Blatter, A.Ereditato, C.-C.Hsu, S.Janos, I.Kreslo, M.Luethi, C.Rudolf von Rohr, M.Schenk, M.S.Weber, M.Zeller, Laboratory for High Energy Physics, Albert Einstein Center for Fundamental Physics (AEC), University of Bern, 3012 Bern, Switzerland}
\newline

\subsubsection{Introduction}
\begin{figure}[h]
\begin{center}
\includegraphics[width=4in]{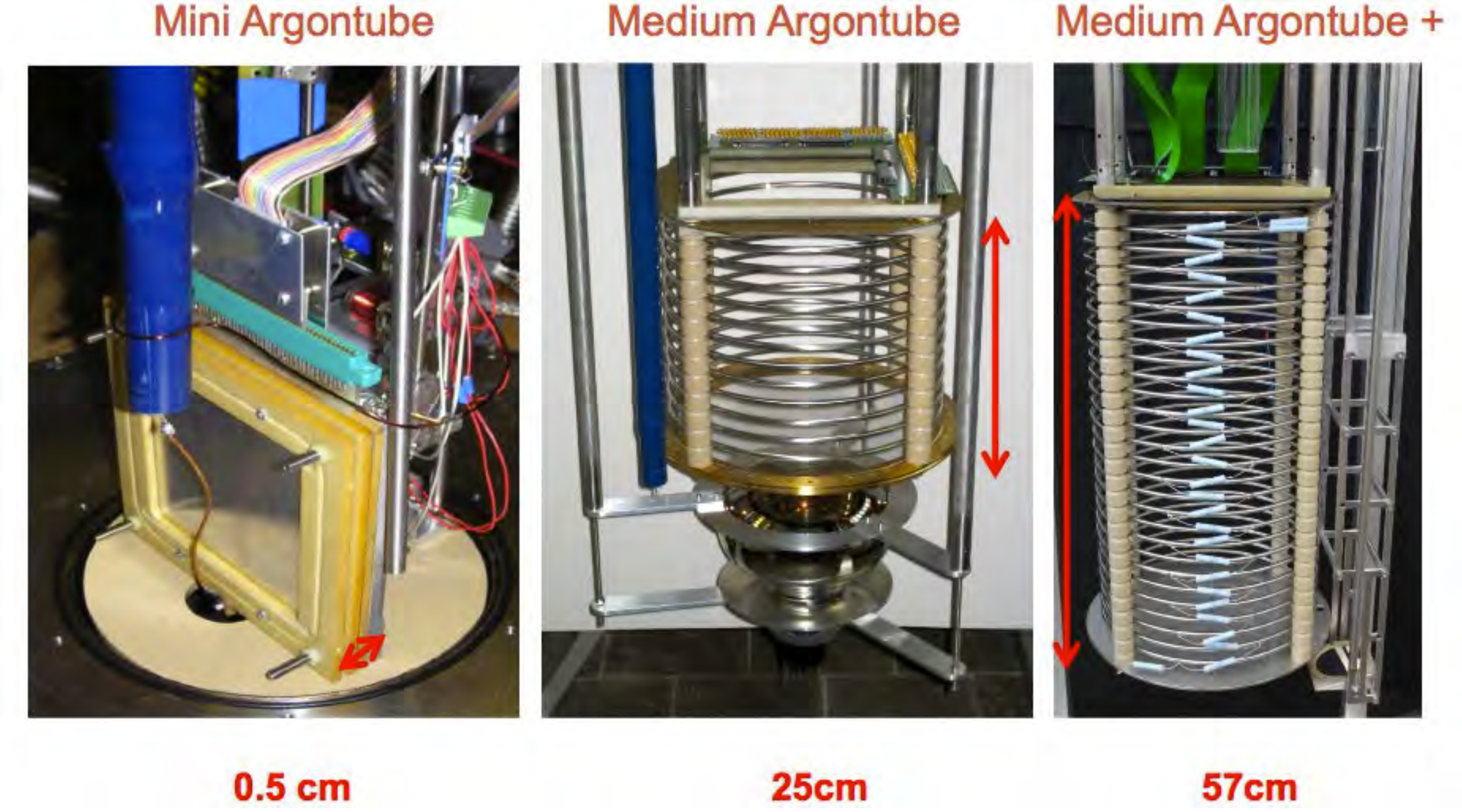}
\end{center}
\caption{Liquid argon TPC evolution at LHEP Bern. From very short drift of 0.5~cm distance TPCs the drift distance increased in size to a 25~cm drift TPC equipped with a Hamamatsu PMT, and a 57~cm version for UV laser ionization measurements.}
\label{tpcdevelopment}
\end{figure}

At the University of Bern Laboratory for High Energy Physics (LHEP), a rich R\&D program for future liquid argon detectors is on-going~\cite{futureargon1,futureargon2,futureargon3} . Over the years, several TPC's have been built and are in use~\cite{nitrogen1,nitrogen2,masterzeller}. The sizes of the TPCs grew over time with the gained experience, as shown in Fig.~\ref{tpcdevelopment}. The latest smaller sized TPC's had a drift length up to 57~cm and were used in UV ionization measurements and novel light readout tests~\cite{uvlaser1,uvlaser2,gapd}. For these early TPC's a ICARUS type high voltage feed-through~\cite{icarus} was used and provided up to 30~kV~\cite{30kvhv}, and a resistor chain supplied a uniform drift field. 

For the next generation of liquid argon TPCs, the main identified challenges investigated at LHEP Bern were the long drift and the high voltage supply. The drift depends on three main factors,  the electric field, drift distance and argon purity. The better the purity, the longer the possible drift distance; the same is true for the applied drift field. Given a fixed drift distance, the drift time can be varied with the applied field. The current magnitude limit of the possible drift fields with a ICARUS design feed-through is about 150~kV. 
\begin{figure}[tb]
\begin{center}
\includegraphics[height=3in]{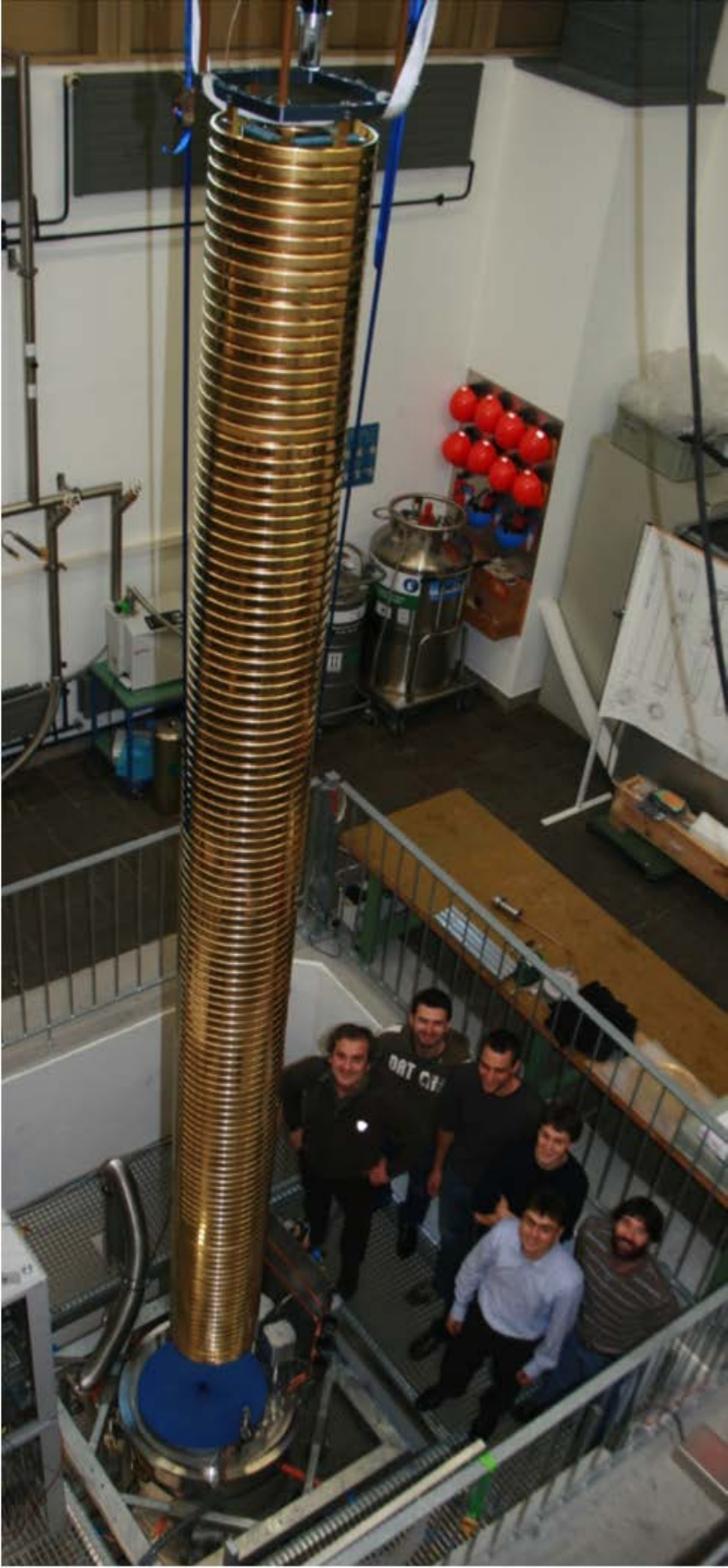}
\end{center}
\caption{TPC of the ARGONTUBE at LHEP Bern. Gold plated field shaping rings to reduce high voltage breakdowns cover 5~m drift length.}
\label{argontube1}
\end{figure}
\begin{figure}[h]
\begin{center}
\includegraphics[width=3.5in]{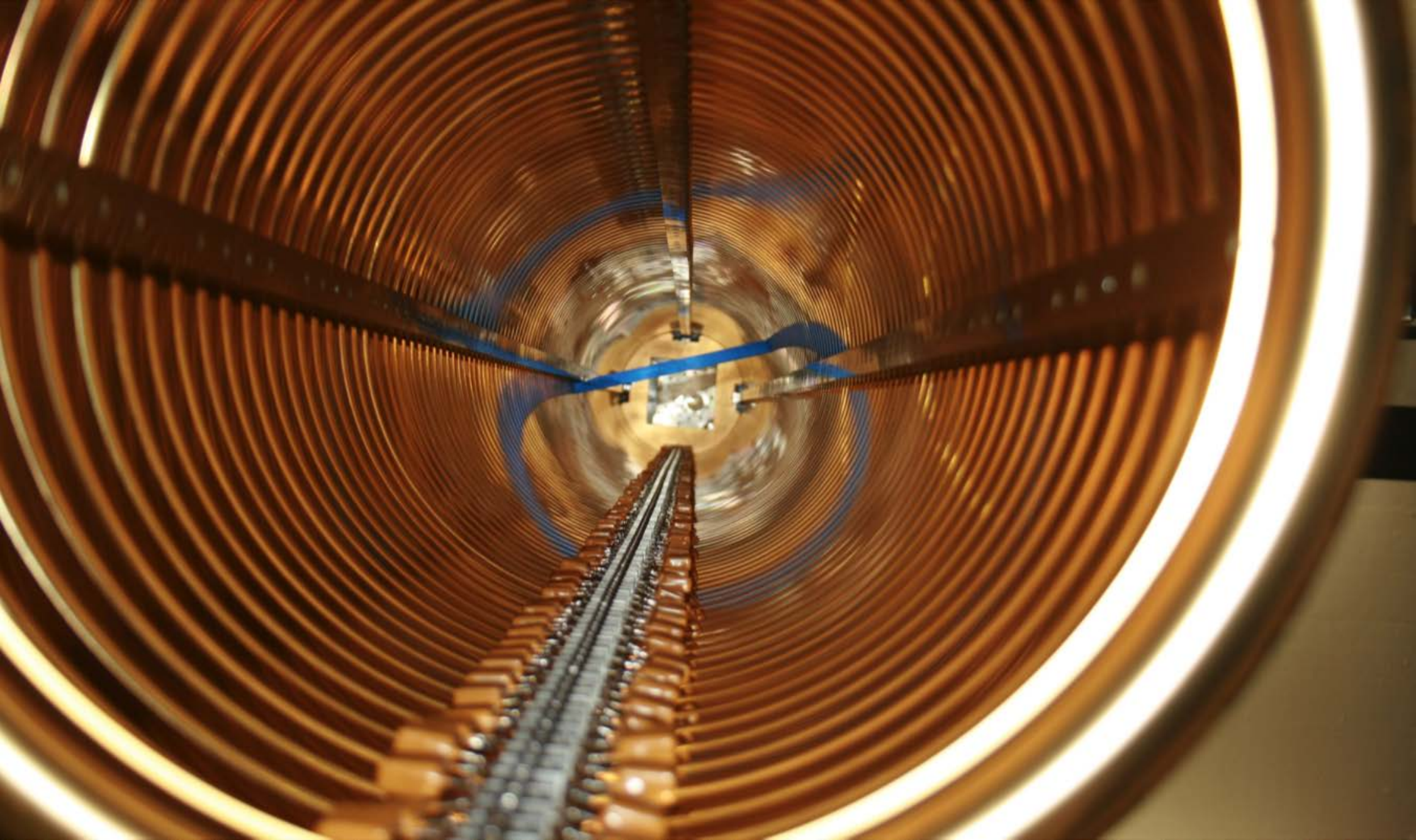}
\end{center}
\caption{View inside the ARGONTUBE TPC at LHEP Bern. The Cockcroft-Walton, or Greinacher, circuit is supplying the high voltage to the field shaping rings.}
\label{argontube2}
\end{figure}
\subsubsection{ARGONTUBE detector}
\begin{figure}[h]
\begin{center}
\includegraphics[width=2in]{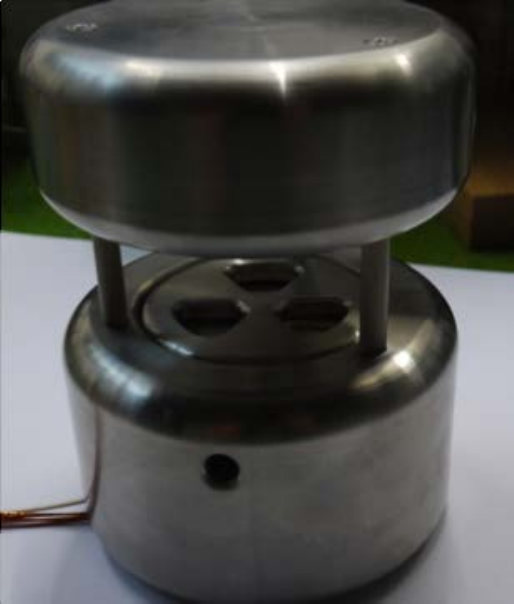}
\includegraphics[height=2.5in]{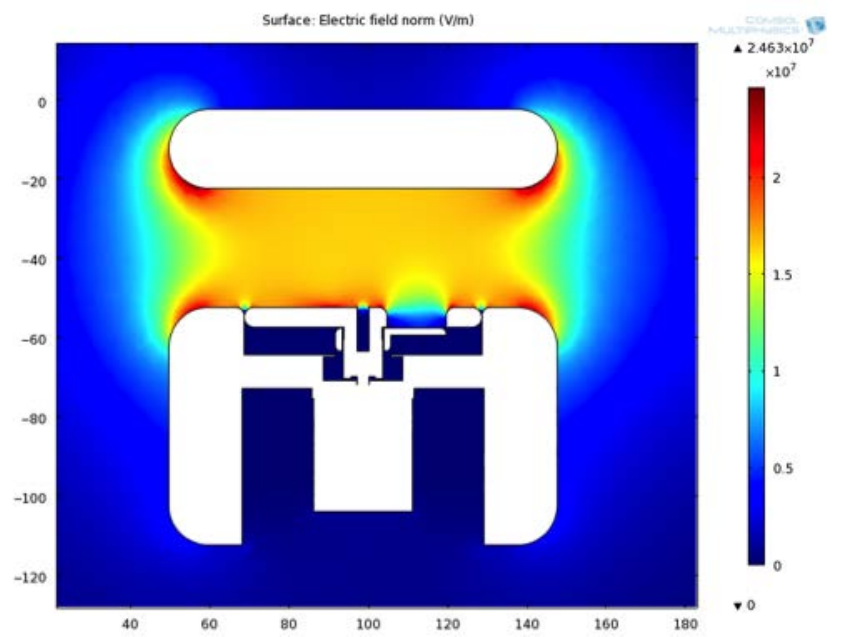}
\end{center}
\caption{Field mill used to measure the applied high voltage at the Cathode of the ARGONTUBE detector.}
\label{fieldmill}
\end{figure}

To address these challenges, a 5~m long TPC was designed at the LHEP, called ARGONTUBE and shown in Fig.~\ref{argontube1}, which uses a novel idea for applying the high voltage~\cite{argontuberef}.  A Cockcroft-Walton, or Greinacher, circuit is connected to the field shaping rings of the detector, with a supply current of 4~VAC, and a total of 125 stages which allow a design high voltage of up to 500~kV. In Fig.~\ref{argontube2} a view inside the TPC shows the circuit. A purification system allows the recirculation of the liquid argon and a cryocooler is used to close the system and allow long-term runs. An outer bath of liquid argon allows the thermal insulation. The readout is performed with warm pre-amplifiers and CAEN ADC, however a recent test used the MicroBooNE ASIC boards as pre-amplifiers in the liquid argon~\cite{microboone}. A field mill shown in Fig.~\ref{fieldmill} is used to cross-check the applied high voltage at the cathode. The operating of the field mill is as follows: a rotator with windows spins above sensed electrodes; charges are induced and lost when the electrodes are shielded along this capacitance, $C$. This periodic charge and discharge voltage, $U$, can be measured and depends on the rotation frequency; the induced charge is proportional to the area exposed to the electric field, $A(t)$.  This allows one to extract the high voltage information from the signal, $S$,  using $S\propto Q=\frac{U}{C}\epsilon_0\epsilon_rA(t)$, where $\epsilon_0$ is the dielectric constant and $\epsilon_r$ is the dielectric strength of argon.

\begin{figure}[h]
\begin{center}
\includegraphics[width=1.75in]{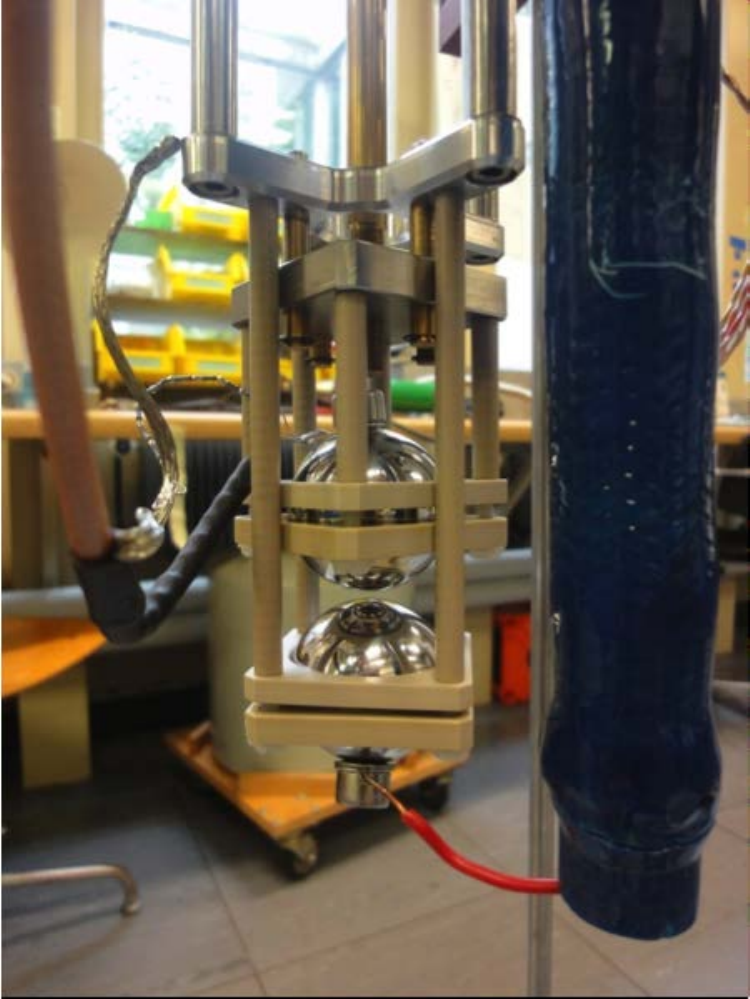}
\includegraphics[width=3.5in]{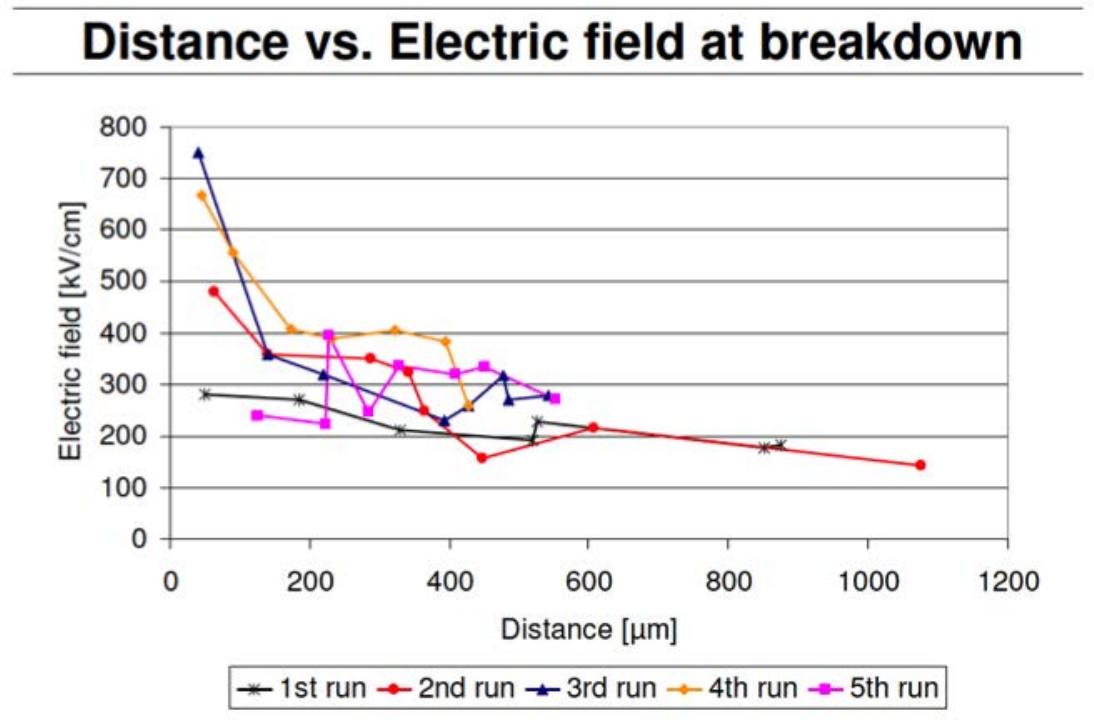}
\end{center}
\caption{Breakdown measurements in liquid argon at the LHEP Bern using a maximum electric field of 30~kV and a maximum distance of 1~cm distance measured with micrometric precision.}
\label{breakdown}
\end{figure}

\begin{figure}
\begin{center}
\includegraphics[width=4in]{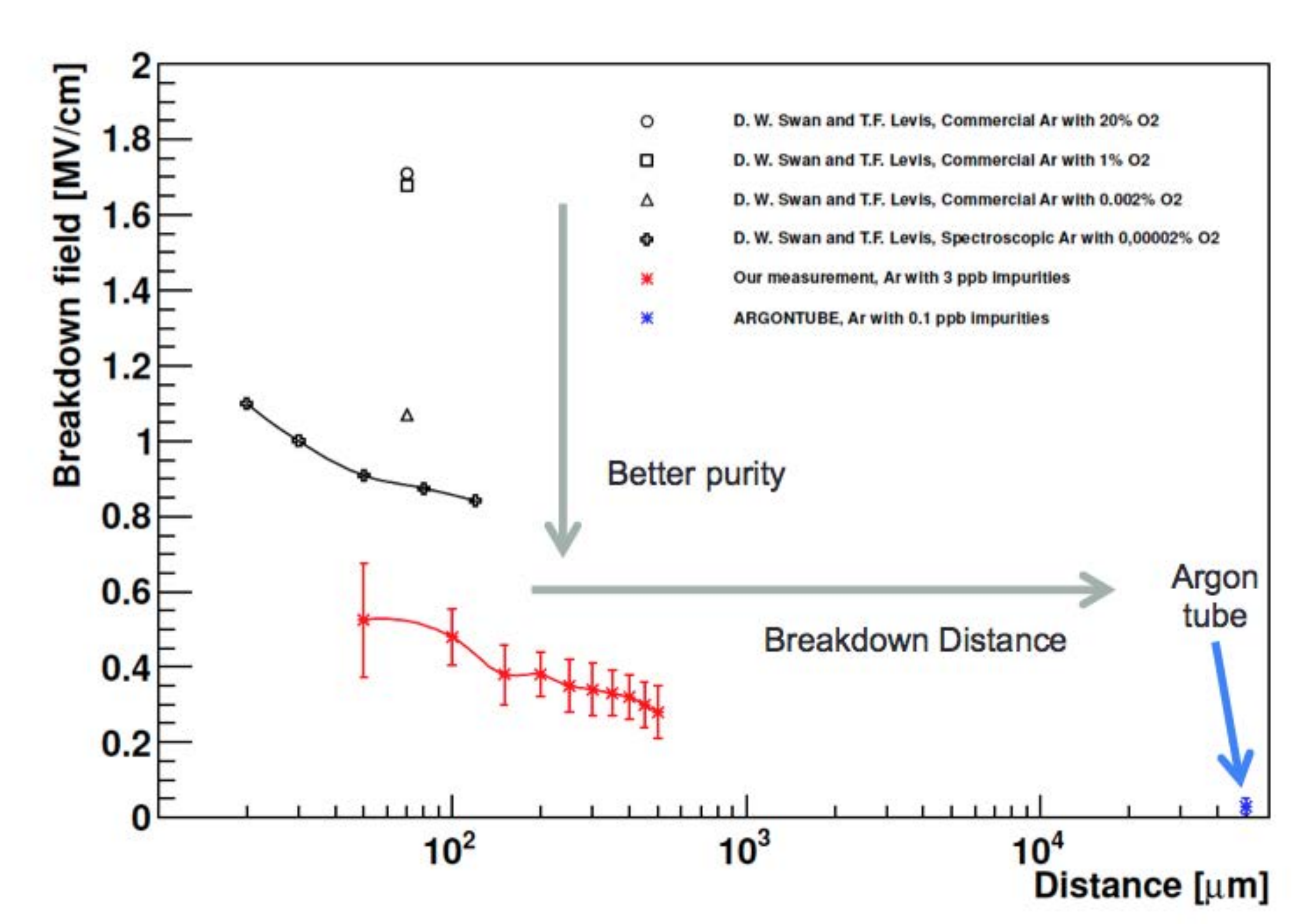}
\end{center}
\caption{Literature versus recent measurements of high voltage breakdowns in liquid argon at LHEP Bern. The higher the argon purity, the lower the breakdown field. The effect on the cathode ground distance needs to be studied further.}
\label{zeller}
\end{figure}

During the operation a maximum stable high voltage of 100~kV was achieved. Around 100~kV spontaneous discharges of the circuit took place, which corresponds to about 30~kV/cm for the highest field to ground. Several runs investigated the breakdowns and argon bubbles, possible contamination with copper dust from the purity filter and detector misalignment due to electrostatic forces were all excluded as possible causes. The conclusion of these tests was that the value of the breakdown voltage is less than the published value of 1MV/cm~\cite{swan1,swan2,swan3} .

\begin{figure}[ht]
\begin{center}
\includegraphics[width=1.5in]{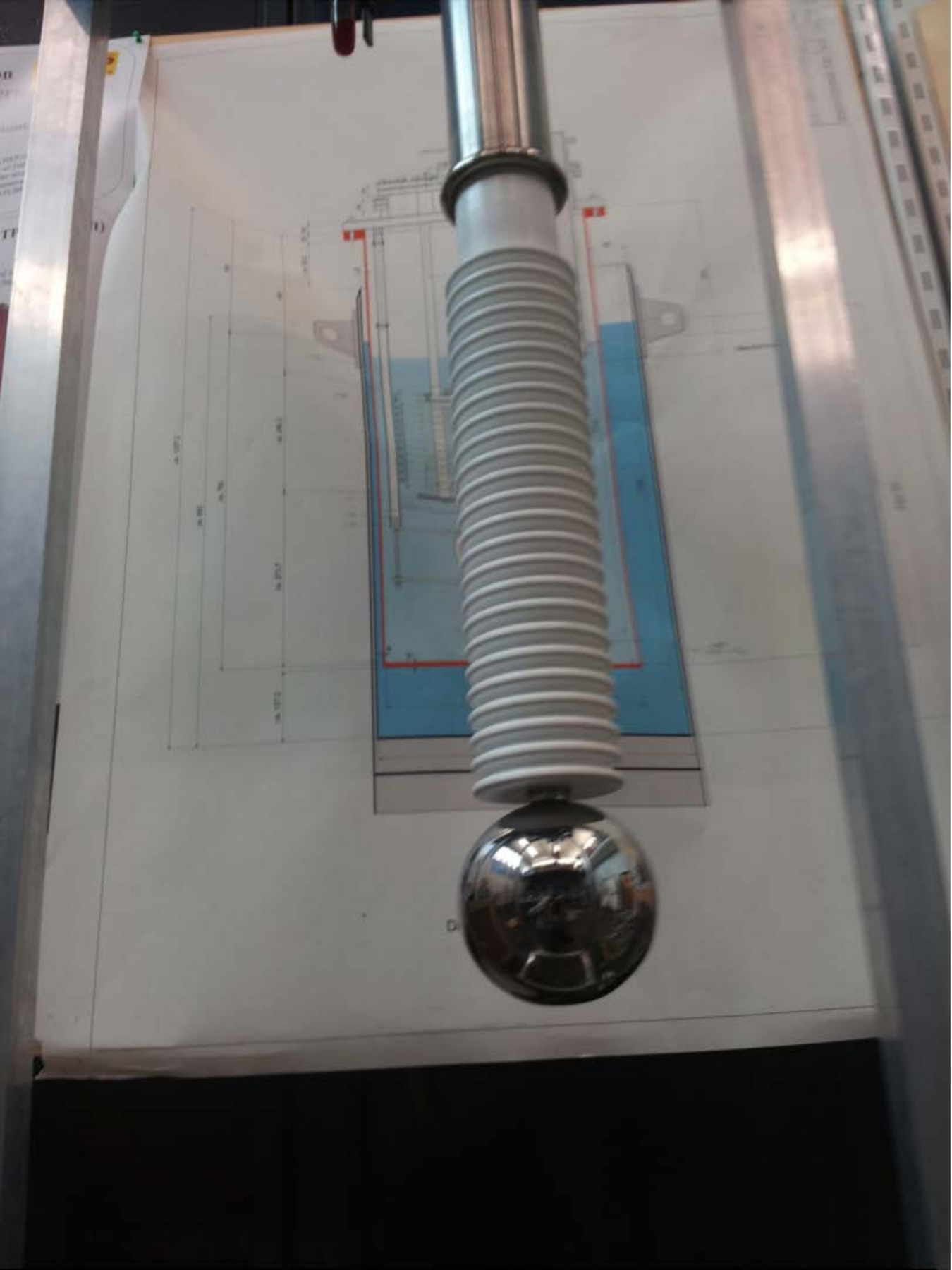}
\includegraphics[width=1.25in]{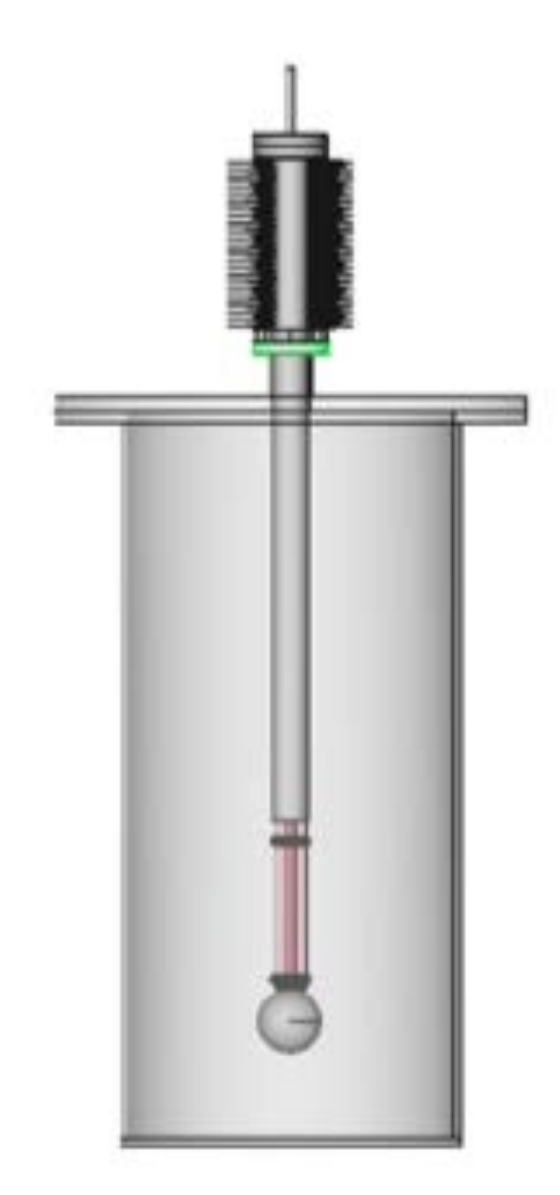}
\end{center}
\caption{New test stand for liquid argon breakdown measurements at the LHEP Bern using a maximum electric field of 1.2~MV/cm, a maximum voltage difference of 130~kVcm, and a distance range of 1~mm up to 10~cm measured with 100 microns accuracy.}
\label{breakdown2}
\end{figure}

\subsubsection{High voltage breakdown measurements}
A dedicated test stand was used at LHEP Bern to cross-check the literature value of the breakdown voltage in liquid argon of 1MV/cm with the results shown in Fig.~\ref{breakdown}.

As summarized in~\cite{phdzeller} and shown in Fig.~\ref{zeller} the breakdown voltage seems to be much lower at large distance and high purities compared to the published measurements~\cite{swan1,swan2,swan3}.

A new test stand at LHEP Bern is currently used to measure the distance dependence in the range from 1~mm to 10~cm with 100~microns accuracy. This discharge will be to a flat surface using a new  ICARUS style high voltage feed through that is protected with a 20~M$\Omega$ resistor. The test stand is shown in Fig.~\ref{breakdown2}. 

\clearpage
\subsection{LBNE, CAPTAIN, and DarkSide}
\label{sec:hanguo}
{\it Contributed by H.~Wang, University of California, Los Angeles, CA 90095, USA in collaboration with \\
A. Fan, Y. Meng, Y. Suvorov, A. Teymourian, University of California, Los Angeles, CA 90095, USA \\
E. Pantic, University of California, Davis, CA 95616, USA \\
Yi Wang, Institute of High Energy Physics, Beijing, China}
\newline

\subsubsection{Physics requirements}
\label{sec:intro}


\paragraph{LBNE}
\label{sec:lbne}

The Long Baseline Neutrino Experiment (LBNE) plans to utilize a 34 kton liquid argon TPC detector located at the Sanford Underground Research Facility (SURF), receiving neutrinos from a wide-band muon-neutrino beamline facility located 1300~km away at Fermilab to precisely measure the neutrino mixing matrix parameters. A near detector on site at Fermilab will provide measurements of various beamline parameters necessary to control systematic uncertainties. The LBNE program has many other physics goals, which are outlined in~\cite{bib1} such as a proton decay search or study of the neutrino flux from a core-collapse supernova within our galaxy. The conceptual design of the far detector consists of two large detectors, which are constructed from an array of TPC modules with a maximal drift length of 3.4~m. The nominal drift electric field of LBNE is 500~V/cm. A high voltage feedthrough capable of delivering 170 kV is needed for LBNE.
\begin{figure}[ht]
\centering
\includegraphics[height=1.5in]{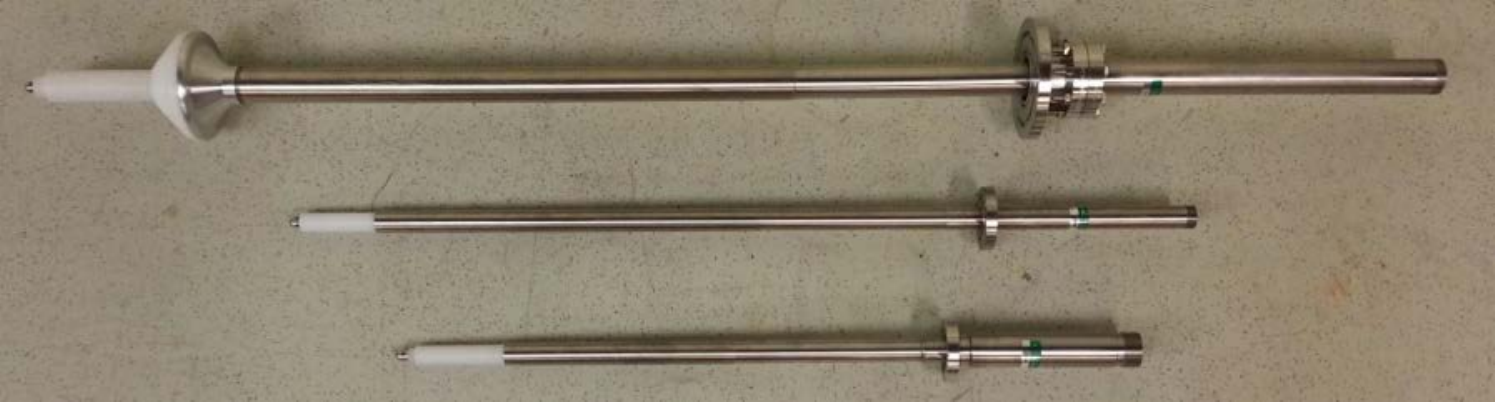}
\caption{High voltage feed-throughs designed for the LBNE, DarkSide, and CAPTAIN liquid argon TPCs, from top to bottom.}
\label{fig:hvFT1} 
\end{figure}

\paragraph{CAPTAIN}
\label{sec:captain}

The Cryogenic Apparatus for Precision Tests of Argon Interactions with Neutrinos (CAPTAIN) program aims to make precision measurements of neutron interactions in argon, which are of scientific importance to the LBNE project and investigate the atmospheric neutrino and supernova neutrino program \cite{bib2}. The program employs two liquid argon hexagonal shape TPC detectors with photon detection systems: the CAPTAIN detector, with a 5 ton active target, 2,000 wire charge readout channels and 1 m electron maximum drift length, and the mini-CAPTAIN prototype detector, with 1,000 wire charge readout channels and a 32 cm total drift length in a 1700 L cryostat. The prototype detector will allow an end-to-end test of all components such as the laser calibration system and development of analysis techniques. The CAPTAIN detector will require a high voltage feed-through capable of delivering 50 kV while the mini-CAPTAIN detector will require a 16 kV high voltage feed-through. 

\paragraph{DarkSide}
\label{sec:darkside}

The DarkSide staged program searches for interactions of cold dark matter in two-phase liquid argon cylindrical TPC detectors at Laboratori Nazionali del Gran Sasso (LNGS). The first physics detector, DarkSide-50, is in operation. The TPC with 35 cm drift length and 50 kg active underground liquid argon, strongly depleted in the $^{39}$Ar radioisotope, is coupled within a 4 m borated liquid scintillator neutron veto detector, which is itself within a 10 m water muon veto detector, resulting in a detector system capable of achieving background-free operation in a 0.1 ton-year exposure \cite{bib3}. The DarkSide-G2, a second-generation detector with a 3.8 ton active mass of underground argon  in a 1.5 m long TPC, is proposed and currently in the R\&D phase. If approved, the DarkSide-G2 detector will be hosted in the existing muon veto and neutron veto system.  The DarkSide-50 high voltage feed-through is designed and built to deliver 43 kV. The planned DarkSide-G2 detector will require a high voltage feed-through capable of 75 kV.

\subsubsection{High Voltage Feedthroughs}
\label{sec:hvfeed-through}

The high voltage feed-through designed and constructed at UCLA for the LBNE, CAPTAIN, and DarkSide projects all consist of an outer stainless steel grounding tube, an inner stainless steel high voltage conductor, and an Ultrahigh Molecular Weight Polyethylene (UHMWPE) dielectric between the two.  This design creates a coaxial design capable of withstanding very high voltages.  Depending on the ultimate design goals, the diameters of the outer grounding tube and inner conductor are tuned to provide the highest voltage with a given outer diameter of the ground tube, with an appropriate safety factor. UHMWPE can withstand an electric field up to 2,500 kV/cm, and so the dimensions of the high voltage feed-through keep the field within the polyethylene well below its limit.  The length of the high voltage feed-through is set to accommodate the geometry of the experiment in which it is used.  A flange welded onto the outer ground tube allows for the high voltage feed-through to be coupled to the cryostat with an ultra high vacuum seal. At the bottom tip of the high voltage feed-through, the UHMWPE dielectric and inner conductor extend further than the grounding tube.  The design at this tip is critical to ensure safe operation up to the voltage needed.  Although liquid argon has a breakdown strength of $1.1 - 1.42$~MV/cm, however, because insulating materials are used near the tip, charge buildup can cause sparking even if the field is below the liquid argon breakdown limit. Ionizing radiation is constantly entering the liquid argon creating charge carriers.  The electric field generated by the high voltage feed-through separates the charge carriers and creates a charge buildup on the insulating surfaces when the liquid argon has a reasonably good purity.  If the total surface charge density exceeds the holding, the charges will move causing a gas bubble to form.  The breakdown strength of gaseous argon is very low, 6.2 kV/cm~\cite{bib4}, and so if the bubble forms in a region with an electric field greater than this value, breakdown will occur.  This breakdown initiates a spark, and so it is important to minimize the electric field in the regions around the tip and to prevent the charges from moving.

Any exposed insulators in the region around the tip are grooved in order to trap charges and prevent any movement, including the UHMWPE dielectric that extends below the outer grounding conductor.  The inner conductor extends slightly beyond the dielectric and provides a means to supply the voltage to the cathode.  The inner and outer tubes are cut to length, connectors for the air-side and liquid-side of the inner conductor are machined and welded onto the inner conductor, and the vacuum flange is welded onto the outer ground tube.  The dielectric is purchased as an extruded UHMWPE tube with the correct inner diameter, and the outer diameter is machined down to be slightly larger than the inner diameter of the outer ground tube.  Grooves are then machined into the bottom end of the dielectric.  The inner conductor is pressed into the dielectric, and the dielectric/conductor combination is cryofitted into the outer tube.  This simple procedure has been used several times for ICARUS, ZEPLIN II, DarkSide-10, DarkSide-50, XENON1T, LBNE and CAPTAIN with ultra high vacuum seal and operated at the designed voltage.

\begin{figure}
\centering
\includegraphics[height=2in]{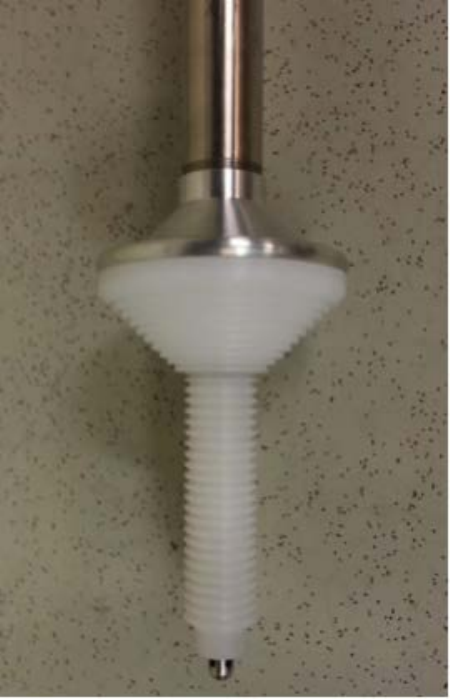}
\includegraphics[width=2.2in]{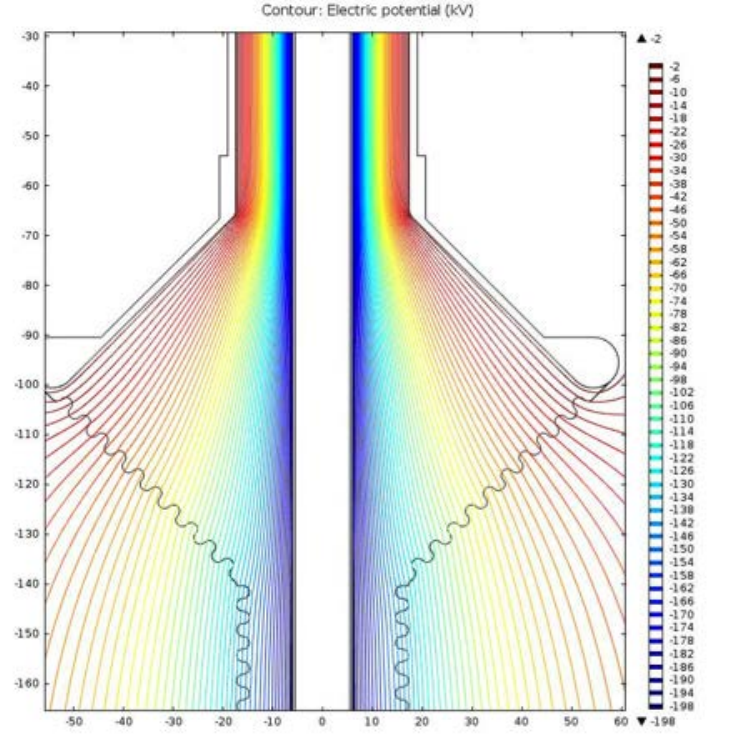}
\caption{(left) A close-up of the tip of the LBNE liquid argon high voltage feed-through. (right) Electric potential of the LBNE high voltage feed-through near the tip.}
\label{fig:hvFTtip}
\end{figure}

The high voltage feed-through design is shown in Fig.~\ref{fig:hvFT1} and is used for the CAPTAIN and DarkSide experiments, which require a voltage of $\sim$50 kV.  The LBNE experiment however requires a much higher voltage, up to 200 kV. The original design created a high electric field at the end of the outer ground tube, which did not allow the voltage to reach above about $150$~kV.  A modification of the tip of the high voltage feed-through is required.  The end of the outer ground tube is flared out to a diameter of more than 4~inches, and a ring shape is machined onto the end.   This design lowers the electric field at the end of the ground tube as seen in Fig.~\ref{fig:hvFTtip}.  A second UHMWPE piece is used to fill the region between the ground tube and the dielectric, and grooves are machined into this piece to prevent any charge migration.  With this design, the voltage can be held at 200~kV without any sparking at zero psig pressure on the liquid argon surface.

\begin{figure}
\centering
\includegraphics[width=2in]{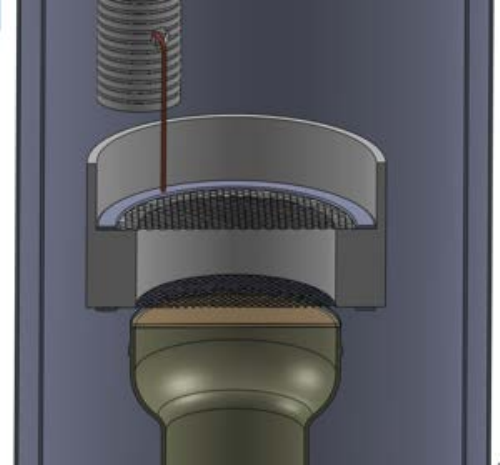}
\includegraphics[width=3.5in]{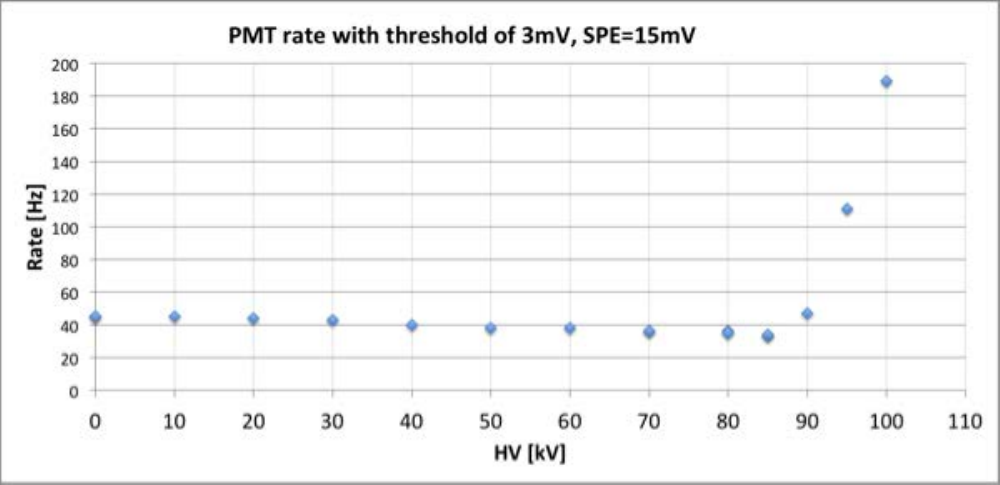}
\caption{(left)The test system for XENON1T including the high voltage feed-through, two grids, and a PMT. (right) PMT rate versus applied cathode voltage.}
\label{fig:hvFTtestgrid} 
\end{figure}

The high voltage feed-throughs were tested by submerging them in liquid argon in a dewar. A polyethylene cup was placed inside the dewar to shield the walls of the dewar from the high voltage feed-through tip.  The voltage was gradually increased up to the design limit of the specific high voltage feed-through being tested.  In some cases, when the surfaces were unpolished, a few sparks occurred during the ramping up of the voltage.  This sparking was attributed to sharp points on the conductors creating very high electric fields.  These initial sparks cause the sharp points to be {\it burned} off and allowed the voltage to increase.  In all cases, the high voltage feed-through was able to run at the maximum designed voltage with no sparking.  The tests were run at both atmospheric pressure and also at a slight overpressure, up to 9~psi.

In addition to high voltage feed-through testing in liquid argon, application of such high voltage feed-through in liquid xenon is planned for XENON1T and its upgrade, see \S~\ref{sec:messina}. In the case of liquid xenon TPCs for dark matter search applications, due to high light yield demands, a high transparency fine mesh cathode is generally used. The ability of running the fine mesh cathode on the designed voltage while close to ground is essential to minimize the amount of liquid xenon used between the cathode and the ground plane. A detailed set-up to study the high voltage on a fine cathode mesh was constructed with care taken on all insulating surfaces, as seen in Fig.~\ref{fig:hvFTtestgrid}. The fine cathode mesh, 2 mm pitch hexagonal mesh with 95\% optical transparency, was placed 3 cm from another of the same mesh, which was grounded. A 3-inch PMT was placed below the grounded mesh to look for any sparking while the voltage was raised up to -100 kV. The voltage was maintained at this level with no sparking. While increasing the voltage to this level, the count rate at 1/3 pe threshold of the PMT was monitored for any possible electroluminescence of the liquid xenon.  Beginning at $-90$~kV, the count rate from the PMT began to steadily increase with the voltage, indicating the presence of electroluminesence.




\subsection{GLACIER/LBNO}
{\it Contributed by F.~Resnati, ETH Zurich - Institute for Particle Physics, 8093 Zurich, Switzerland in collaboration with \\
F.~Bay, C. Cantini, S.~Murphy, A.~Rubbia, F.~Sergiampietri, S.~Wu, ETH Zurich - Institute for Particle Physics, 8093 Zurich, Switzerland}
\newline

\label{sec:resnati}
The LBNO experiment~\cite{Stahl:2012} aims at very large statistics, excellent background rejection and good energy resolution in order to measure precisely the oscillation pattern as a function of the neutrino energy. The Giant Liquid Argon Charge Imaging ExpeRiment (GLACIER) is a very large liquid argon TPC that will perform neutrinos physics as well as look for proton decay.
GLACIER plans to use the innovative double phase, liquid-vapor, operation, which permits the amplification of the signal by means of charge avalanche in the vapor, yielding a larger signal to noise ratio and an overall better image quality than single phase operation~\cite{Cantini:2013, Badertscher:2013, Badertscher:2011a, Badertscher:2011b, Badertscher:2010, Badertscher:2008}. Several technological challenges need to be addressed to achieve the successful operation of this experiment, with the high voltage being a critical one. The assumed electron drift length is 20~m, with an electric field of 1~kV/cm. These design parameters require a voltage at the cathode of $-2$~MV. In the current design~\cite{LAGUNA_LBNO}, a distance of 1.5~m is foreseen between the cathode structure and the bottom of the tank at ground, giving an average electric field of 13.3~kV/cm. In the vicinity of the cathode tubes the electric field reaches 50~kV/cm over distances of the order of centimeter. Therefore, GLACIER performed a dedicated test to measure the maximum electric field, up to 100~kV over 1~cm, that the liquid argon can sustain. The setup was operated for the first time in December 2013. The description is given in \S~\ref{sec:apparatus} and the results obtained in \S~\ref{sec:test}.

\subsubsection{The apparatus}
\label{sec:apparatus}
\begin{figure}
\centering
\includegraphics[width=4in]{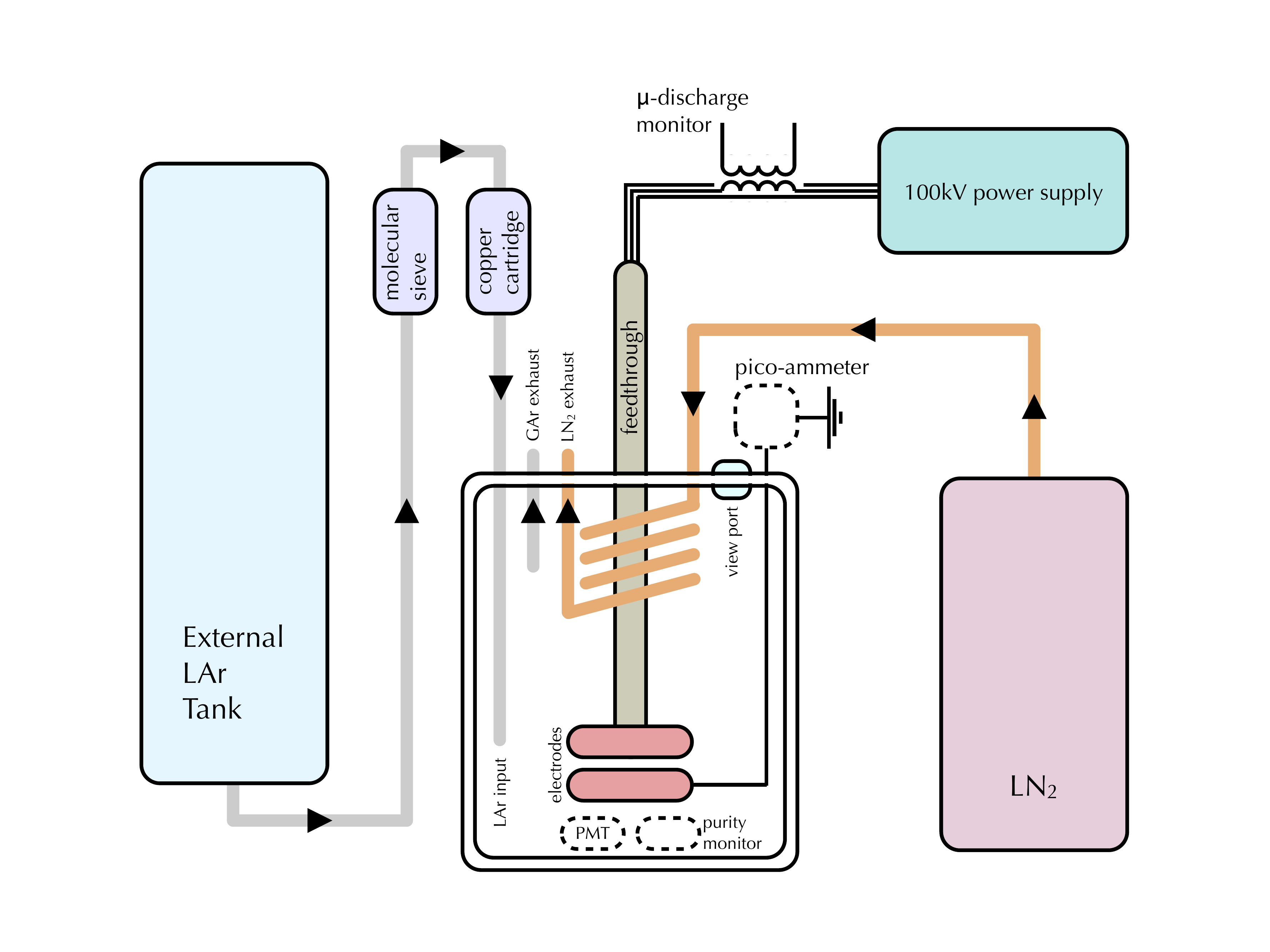}
\caption{Schematic representation of the GLACIER liquid argon test apparatus.}
\label{fig:scheme}
\end{figure}

Figure~\ref{fig:scheme} shows the schematic representation of the test setup, which consists of a vacuum insulated dewar hosting a high voltage feed-through and a couple of electrodes, in between which the high electric field is generated.  During operation, the dewar is filled with liquid argon purified through a molecular sieve, ZEOCHEM Z3-06, which blocks the water molecules, and a custom-made copper cartridge, which absorbs oxygen molecules. Before the filling, the vessel is evacuated to residual pressure lower than $10^{-4}$~mbar in order to remove air traces, favor the outgassing of the materials in the dewar, and check the absence of leaks towards the atmosphere.  During the filling, the boiled-off argon gas is exhausted to control the pressure in the dewar. The pressure is always kept at least 100~mbar above the atmospheric pressure, so that air contaminations are minimized. Once the detector is completely full, the exhaust is closed, and the liquid argon is kept cold by means of liquid nitrogen flowing into a serpentine, which acts as heat exchanger that condenses the boiled-off argon gas. The liquid argon level is visually checked through a vacuum sealed viewport.

\begin{figure}[htbp]
\centering
\includegraphics[width=4in]{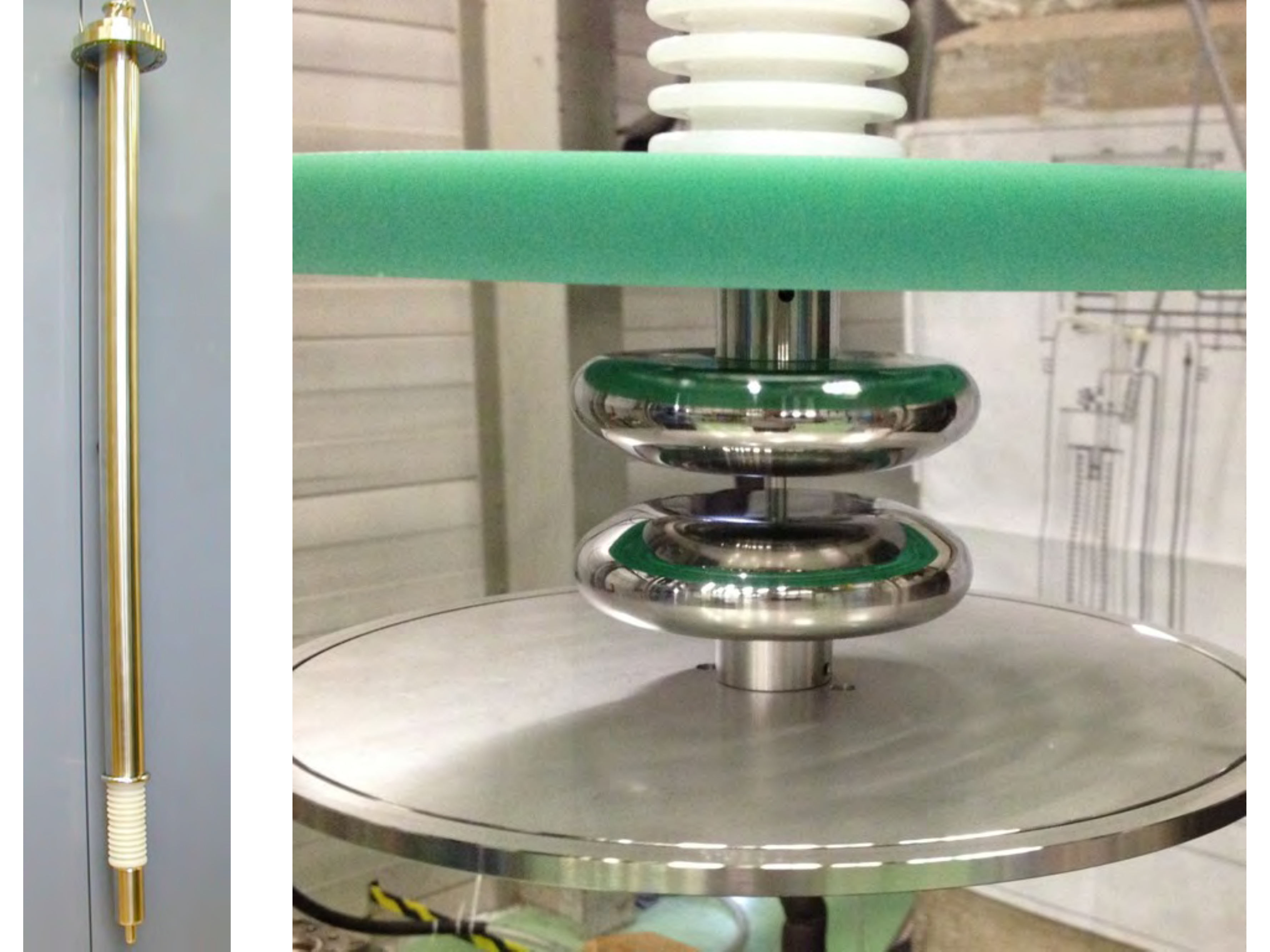}
\caption{(left) Image of the GLACIER test apparatus high voltage feed-through. (right) The electrodes structure.}
\label{fig:feedthroughg}
\end{figure}

The voltage is provided by a 100~kV power supply~\cite{Heinzinger} though a high voltage coaxial cable modified to inductively couple the high voltage wire to an oscilloscope via a 1:1 transformer. When a pulsed current flows through the cable, it is detected on the scope. This sensor is used to monitor the frequency of small discharges that are not necessarily happening between the electrodes, but rather the feedthrough and the cable may have leakage currents. It is also used to monitor the current delivered by the power supply during the charging up. The cable enters into a custom-made high voltage feedthrough, as seen in Fig.~\ref{fig:feedthroughg}, analogous to the one developed by ICARUS~\cite{icarus}. It is vacuum tight and designed to sustain voltages larger than 150~kV.

\begin{figure}
\centering
\includegraphics[width=2.2in]{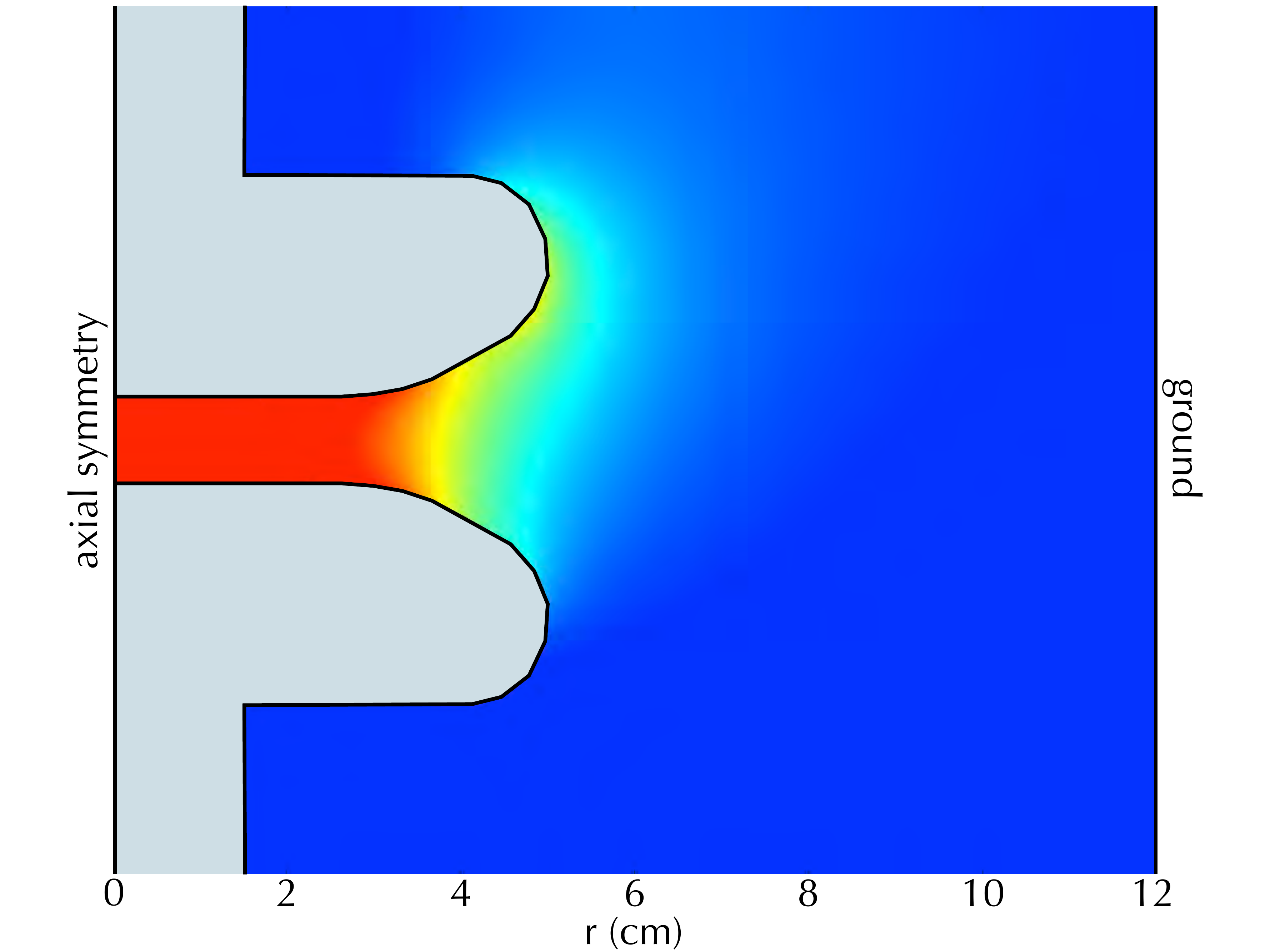}
\includegraphics[width=2.2in]{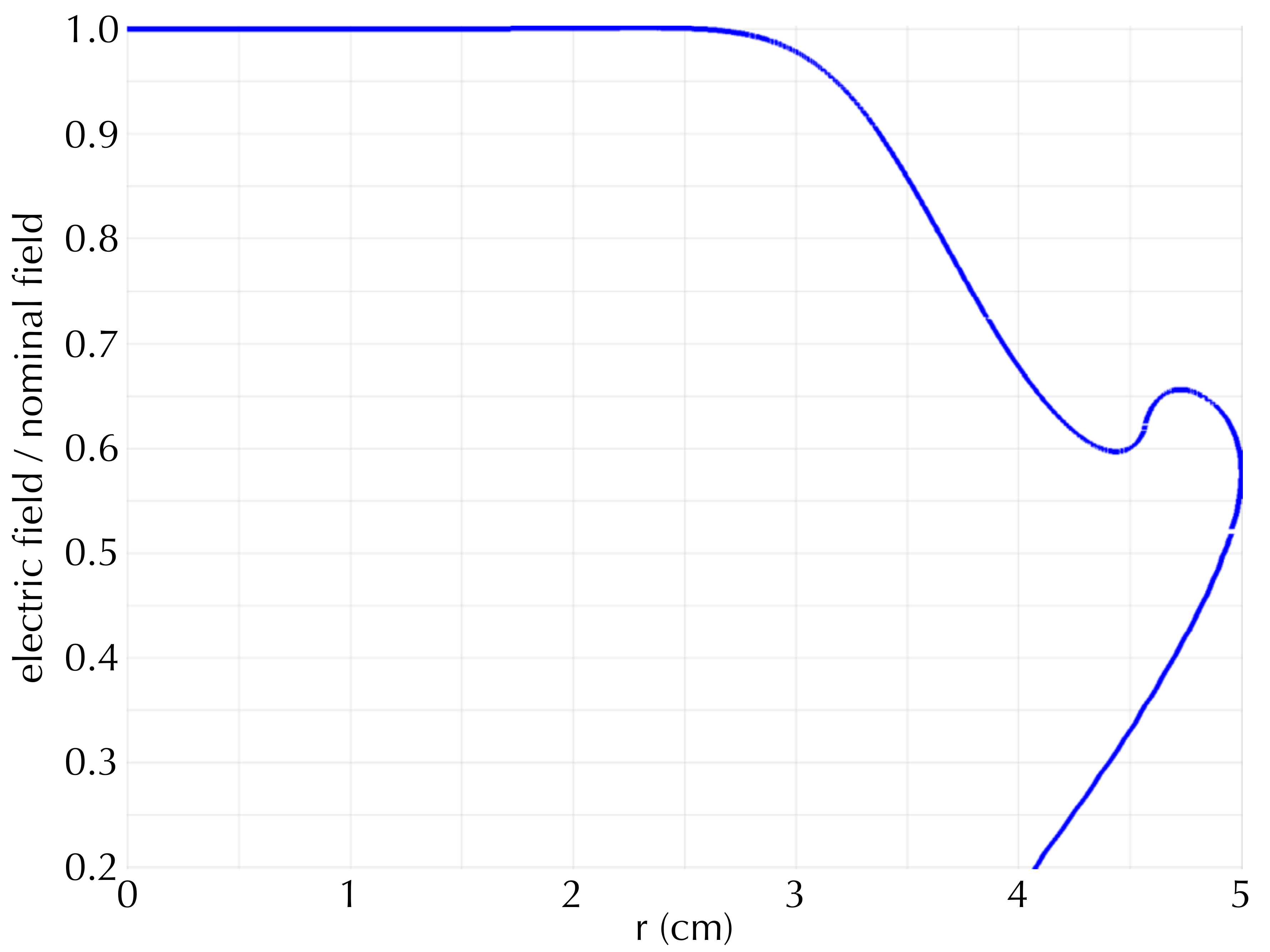}
\caption{(left) Computed electric field for GLACIER test electrodes. The cross sections of the electrodes are shown in grey, and the color pattern is proportional to the absolute value of the electric field. The electric field is essentially uniform in the central region. (right) Computed electric field on the profile of the top electrode as a function of the radius. The largest field is attained between the electrodes.}
\label{fig:field}
\end{figure}

High electric fields can be achieved with low potentials and electrodes with small radii of curvature, but since the breakdown is a random process, it is important to test a sizable region of the electrodes. For these reasons, a system that provides a uniform electric field over 20~cm$^2$ area was designed. A picture of the electrodes structure is shown in Fig.~\ref{fig:feedthroughg}. The two 10~cm diameter electrodes have the same shape and are facing each other at a distance of 1~cm. The top electrode is connected to the live contact of the high voltage feedthrough, and the bottom one is connected to ground through the vessel. The electrodes, made out of mechanically polished stainless steel, are shaped according to the Rogowski profile~\cite{Rogowski:1923} that guarantees that the highest electric field is almost uniform in a region of about 5~cm in diameter, and confined between the two electrodes, as shown in Fig.~\ref{fig:field}. The left image shows, in cylindrical coordinates, the absolute value of the electric field in the vicinity of the electrodes, computed with COMSOL~\cite{COMSOL}. On the right the electric field along the profile of the top electrode as a function of the radius is shown.

The two electrodes form a standalone structure, that is assembled first and then mounted. By construction, the structure ensures the parallelism of the electrodes when cooled down to the liquid argon temperature. The shrinkage of the materials in cold is computed to affect the distance between the electrodes less than 1\%.

\subsubsection{Results}
\label{sec:test}
In December 2013 the setup was operated for the first time. The goal of the first test was to commission the setup and to check if a field of 100~kV/cm can be reached in stable conditions. The operation with the high voltage power supply switched on lasted about 4~hours.

With the liquid argon temperature below the boiling point at a given pressure, a voltage of -100~kV was applied to the top electrode. This value was limited by the maximum voltage of the power supply. This configuration corresponds to a uniform electric field of 100~kV/cm in a region of about 20~cm$^2$ area between the electrodes. Several cycles of discharging and charging up of the power supply were performed. The system could also be stressed several times by ramping up the voltage from 0~V to -100~kV in about 20~s without provoking any breakdown.

A completely different behavior was observed with boiling argon.  Several breakdowns occurred between the electrodes at fields as low as 40~kV/cm. The stillness of the liquid argon was controlled by varying the pressure of the argon vapor and was monitored visually by looking through the viewport. The pressure was regulated by acting on the flow of the liquid nitrogen passing through the heat exchanger. In fact, the thermal inertia of the liquid argon bulk makes temperature variations very slow, hence increasing the cooling power translates in an rapid decrease of the vapor pressure. When the pressure is above the boiling one, the liquid argon surface becomes flat and still. On the contrary, the more the pressure decreases below the boiling point, the more the liquid argon boils.

Above about 1000~mbar the argon was not boiling, and 100~kV/cm could be achieved. Around 930~mbar breakdowns occurred at 70~kV/cm, and below 880~mbar at 40kV/cm. Since argon gas has a much lower dielectric rigidity compared to the liquid, this behavior is interpreted as the evidence that breakdown can be triggered by bubbles in the liquid.

\subsubsection{Conclusions From GLACIER}
\label{sec:glacieroutlook}

The dielectric rigidity of liquid argon with temperature below the boiling point is larger than 100~kV/cm. The occurrence of breakdowns was found to be significantly sensitive to the presence of bubbles: in boiling liquid argon the breakdown field is as low as 40~kV/cm. The breakdown is thought to be due to the bubbles themselves that convey the breakdown forming a preferred channel for the discharge development, because of the much lower dielectric rigidity of the argon gas. The GLACIER group intends to repeat the test to quantitatively measure the correlation between the bubble formation and the breakdowns. They also plan to test larger electrode distances with larger voltages and measure the dielectric rigidity of liquid argon as a function of the argon purity and electrode distance. The setup will be upgraded installing a PMT, a liquid argon purity monitor and a pico-ammeter, represented with dashed lines in Fig.~\ref{fig:scheme}. The main purpose of the PMT is to establish the presence of luminescent precursors of the breakdown in liquid. The purity monitor will consist of a small TPC to measure, with cosmic muons, the lifetime of the drifting ionization electrons. Since this setup will also be used to characterize dielectric materials in presence of high electric fields and in cryogenic conditions, a pico-ammeter will be used to measure the current flowing through the materials placed between the two electrodes to measure their volumetric impedance and the surface currents.

\section{Liquid Xenon Experiments}

In this section we review the high voltage experience of several working liquid xenon experiments (EXO, LUX, and XENON100), as well as R\&D in preparation for new large-scale experiments (nEXO, XENON1T and LZ), and one gas-phase xenon TPC (NEXT). 
\subsection{The EXO-200 experiment}
{\it Contributed by P.~C.~Rowson, SLAC National Accelerator Laboratory, Menlo Park, CA 94025, USA}
\newline
\subsubsection{Introduction}

The EXO-200 detector is a liquid xenon TPC and is described elsewhere \cite{EXOdet}.  The device is contained within a copper vessel with a horizontal right-cylindrical layout of $\sim$40 cm length and diameter, the TPC itself consisting of a central cathode plane at negative high voltage, and two anode readout planes for charge and light collection spaced at $\sim$20 cm on either side.  The readout planes contain two orthogonal closely spaced wire planes for charge collection and position determination in front of a support plane holding a LAAPD array for light readout.  The central cathode is operated a high negative potential, with a design maximum of -75 k), the front wire plane for induction signals at a much lower negative potential with a design maximum of -4 kV, and the second wire plane for charge collection at virtual ground.  The LAAPD plane is set to $\sim$ -1.4 kV to bias the diodes.  The two drift regions on both sides of the cathode include field shaping rings at large radius connected by voltage divider resistors, in front of which are placed thin teflon panels used to reflect the VUV scintillation light.

EXO-200 has encountered instabilities, in the form of small discharge-like pulses or `glitches' observed on an oscilloscope capacitively coupled to the cathode, for cathode voltages $\sim$-10 kV. Presently EXO-200 operates with a cathode voltage of -8 kV, and a induction plane voltage of -780 V.  While there is no clear understanding of this phenomenon, there is evidence that light production can be associated with a glitch, and that some, but not all, of these events appear to occur at large radii behind our teflon reflectors.  

A test apparatus was assembled at SLAC to make controlled measurements of electric fields in liquid xenon, and is described in what follows.

\begin{figure}
\begin{center}
\includegraphics[height=2in]{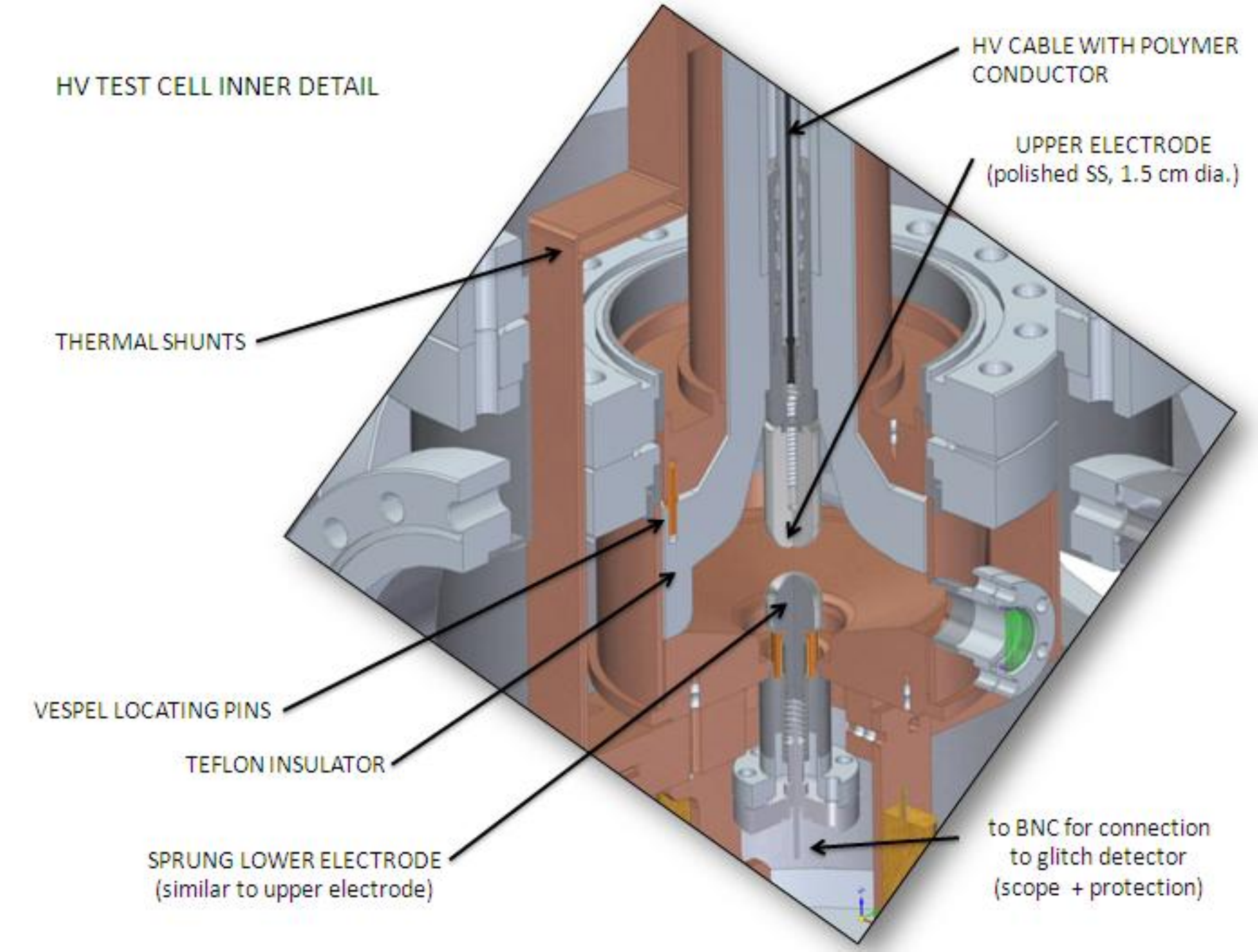}
\caption{The EXO-200 high voltage test cell detail : The moveable upper and fixed lower electrodes, teflon filler material, and view port with sapphire windows are all visible.  Shown here is the hemispherical upper electrode.  The view ports are miniConflat flanges to set the scale. Note that the copper xenon vessel is cooled from below by a cold finger immersed in liquid nitrogen with trim heaters for temperature control.}
\label{fig:testcell}
\end{center}
\end{figure}

\subsubsection{The Test Apparatus}

The R\&D setup used to design the xenon handling, xenon purification and cryosystems for EXO-200 has been adapted for high voltage studies.  A small test cell with a capacity of $\sim$400 cc's for liquid xenon has been built as seen in Fig.~\ref{fig:testcell}.  The cell includes a SS smooth hemispherical electrode of 1.5 cm diameter at virtual ground located at the bottom, and a precisely locatable opposing electrode that can be brought to negative high voltage.  Various upper electrode designs have been used. The cell includes teflon space-filling parts to reduce the required liquid xenon volume.  Xenon gas from a SS storage bottle is purified by a single pass through a SAES 3000 series purifier \cite{SAES}, and then into the test cell for condensation.  

A limitation of this experiment is that the xenon purity in the cell is not known.  With the cell filled to a level visible in an upper viewport, the fill is ended and high voltage testing begins.  Typically, a few hours of testing are followed by xenon recovery.  Another limitation of these tests is that they are all short term experiments.  Mitigating these shortcomings are the facts that the high voltage issues observed in EXO-200 do not clearly correlate with xenon purity, effects are similar for a wide range of purity levels, and that they appeared immediately after high-voltage turn-on, typically on time scales of a few minutes.

\subsubsection{Measurements}

Possible causes for the high voltage issues in EXO-200 might fall into several categories: $i)$ mundane causes such as debris and sharp points, $ii)$ bubble formation from heat leaks or electrostrictive bubble formation, $iii)$ static charge on dielectrics or charge build up including flow related effects, and $iv)$ inherent liquid xenon effects such as UV transparency and enhanced photoelectric or field emission.  Most of these possibilities have been investigated in the high voltage test cell, as described below, although examining EXO-200 specific effects requires testing the TPC itself. In what follows, the program MAXWELL \cite{MAXWELL}, in 3D format, is used to determine the surface electric fields.

\subsubsection{Liquid Xenon Breakdown Field}

With a hemispherical electrode installed in the upper position, breakdown is observed with a 1 mm electrode separation at a voltage corresponding to a 250-300 kV/cm electric field, which determines a high dielectric strength, as expected, for liquid xenon.  For electrode separations greater than 3 mm ,the maximum voltage of the power supply, -75~kV can be maintained without instability.

\subsubsection{Surface Field Studies}

A series of measurements have been made to look for surface field limitations in liquid xenon, and to replicate the glitch phenomenon observed in EXO-200.  The lower electrode is connected through a protection circuit to an oscilloscope.  An upper electrode holding a small wire loop of 5mm diameter is employed for testing; first using a sample of EXO-200 125 micron etched phosphor bronze identical to the material used for the wire planes, and then an off-the-shelf round phosphor bronze wire of identical diameter.  Gold-plated tungsten wires of 50 and 20 microns have also been tested.  Breakdown and glitch instability behavior are observed at surface electric fields of $\>$600 kV/cm and $\>$350 kV/cm respectively, much higher than those nominally encountered in EXO-200 where the highest surface fields are the $\sim$4 kV/cm at the induction wires.  Relatively small differences , $<20\%$ lower breakdown voltages, are seen for the etched wires compared to the round wires.  

In addition, the test cell is sometimes instrumented with a Hamamatsu R8520-406 PMT that views the electrodes through two sapphire windows, one on the cell, and one on the cryostat, inside of an evacuated vessel so that UV light can be detected.  It is seen that each glitch observed on the oscilloscope is accompanied by a PMT pulse, and by venting the evacuated chamber the light is confirmed via absorption in air to be in the ultraviolet region of the spectrum.

\begin{figure}[h]
\begin{center}
\includegraphics[width=4in]{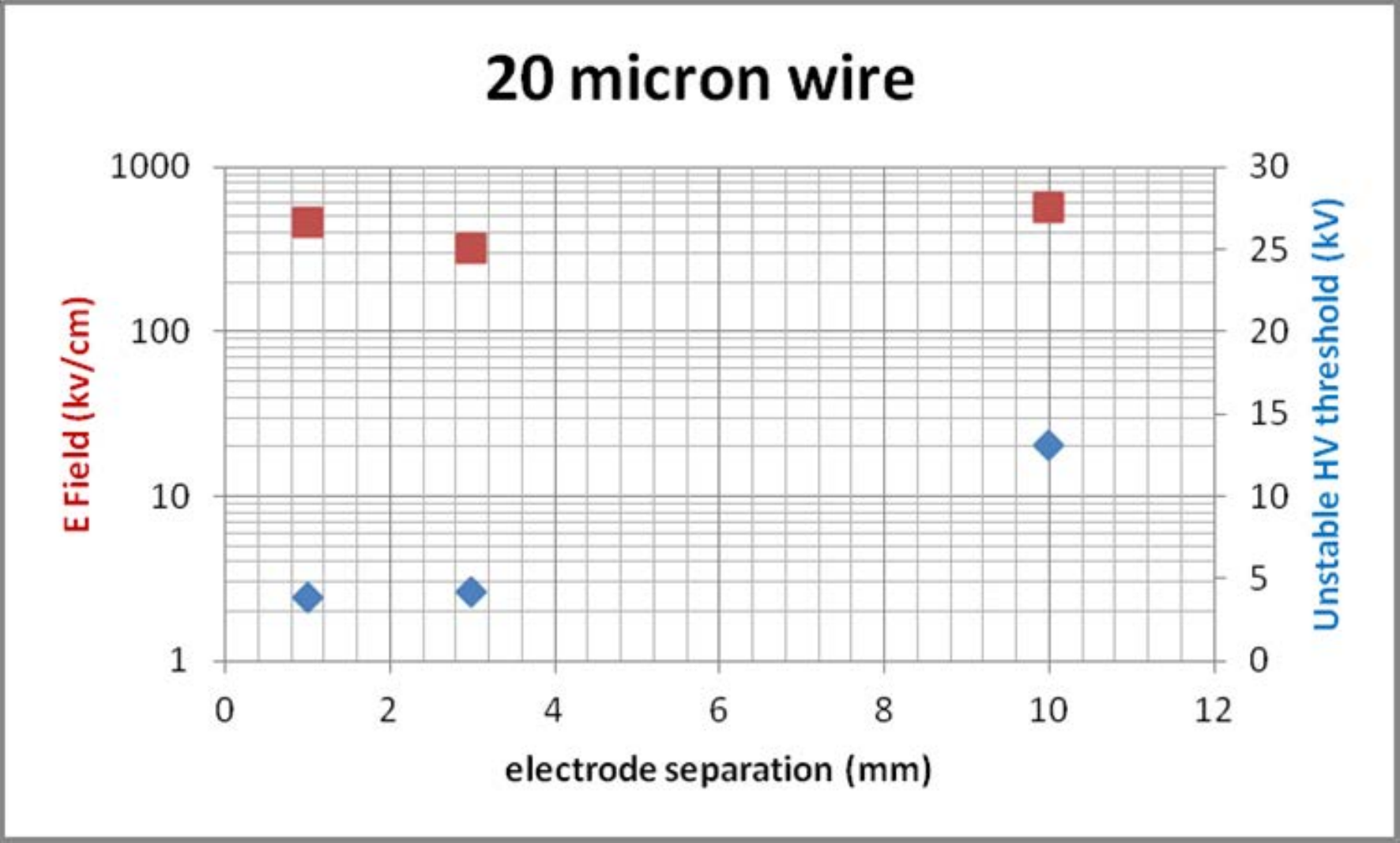}
\caption{The onset of high voltage instability for a 20 micron Au-plated tungsten wire as a function of electrode separation in the EXO-200 high voltage liquid xenon test cell. The righthand side vertical scale gives the associated surface electric field (red points) while the lefthand side scale gives the applied high voltage (blue points). }
\label{fig:20micron}
\end{center}
\end{figure}

A plot of the data taken for the 20 micron wire as a function of separation distance between the electrodes shows the general trend is shown in Fig.~\ref{fig:20micron}.  For a 1 cm separation distance, the same distance as between the cathode support ring at high voltage and the grounded copper xenon vessel in EXO-200, instability occurs at -13 kV,  which is comparable to the behavior seen in the TPC.  But there are no features as small as 20 microns, or even 100 times this size, standing off 5~mm from any high voltage surface in EXO-200.

\subsubsection{Dielectrics}

An upper electrode is designed to hold small samples of plastics and is used to test surface breakdown effects of dielectrics.  Tests of acrylic, teflon and resistive PEEK have been made, with perhaps the most significant observation being that these materials tend to electrostatically attract debris in the test cell, leading to poorer high voltage breakdown performance.

\subsubsection{Photoelectric Effects}

Ultra-high purity liquid xenon is remarkably transparent in the far UV, and admits electron transport for long distances with attenuation lengths of several meters in both cases. In addition, the electron band-structure in the liquid, as distinct from gaseous xenon, results in a reduced effective work function for immersed conductors by $\sim$0.67~eV~\cite{Schmidt}.  These facts suggest that enhanced photoelectric effects and the lack of discharge quenching could be partly responsible for high voltage breakdown phenomena in liquid xenon detectors.  As the addition of quenching agents is not an option for EXO-200, instead the effect of using high-work function conductors such as platinum ($W \ge5.6$ eV) has been investigated.  Tests of plantinum-plated SS wires and pure platinum wires, to be compared to stainless wires ($W \sim4.4$ eV), do not show any effect.  Perhaps significant effects only occur for work function values above $\sim$7 eV, which is the typical energy of the 178~nm liquid xenon scintillation light in liquid xenon.

\subsubsection{Conclusions and Future Plans}

The EXO-200 high voltage test chamber shows that liquid xenon can stably support bulk electric fields above 200 kV/cm, and local surface fields higher than 300 kV/cm.  Exotic effects in liquid xenon have not been observed so far, but more mundane high voltage breakdown issues due to suspended debris or unintended sharp features can not be eliminated.  Dielectric charge buildup effects have not been thoroughly tested, but instability effects in the lab and in EXO-200 do not show the delayed onset expected for these phenomena.  Further testing of electron emission and its control, using polarity-reversed electric fields and insulating film barriers, is planned.  In addition, simple in situ tests of the EXO-200 TPC are also under consideration.

\subsection{The LUX experiment}
{\it Contributed by C.~H.~Faham, Lawrence Berkeley National Laboratory, Berkeley, CA 94720, USA for the LUX Collaboration}
\newline

\subsubsection{Introduction}

The Large Underground Xenon experiment (LUX) is a 370~kg two-phase liquid xenon TPC that looks for interactions from galactic dark matter in the form of Weakly Interacting Massive Particles (WIMPs). The LUX detector is operating at the 4,850~ft level of the SURF in Lead, South Dakota, USA. LUX recently released its first dark matter search results from 85.3 live-days of data collection, which place the world's most stringent limits on WIMP dark matter interaction rates \cite{Akerib:2013tjd}.

\begin{figure}
\centering
\includegraphics[width=4in]{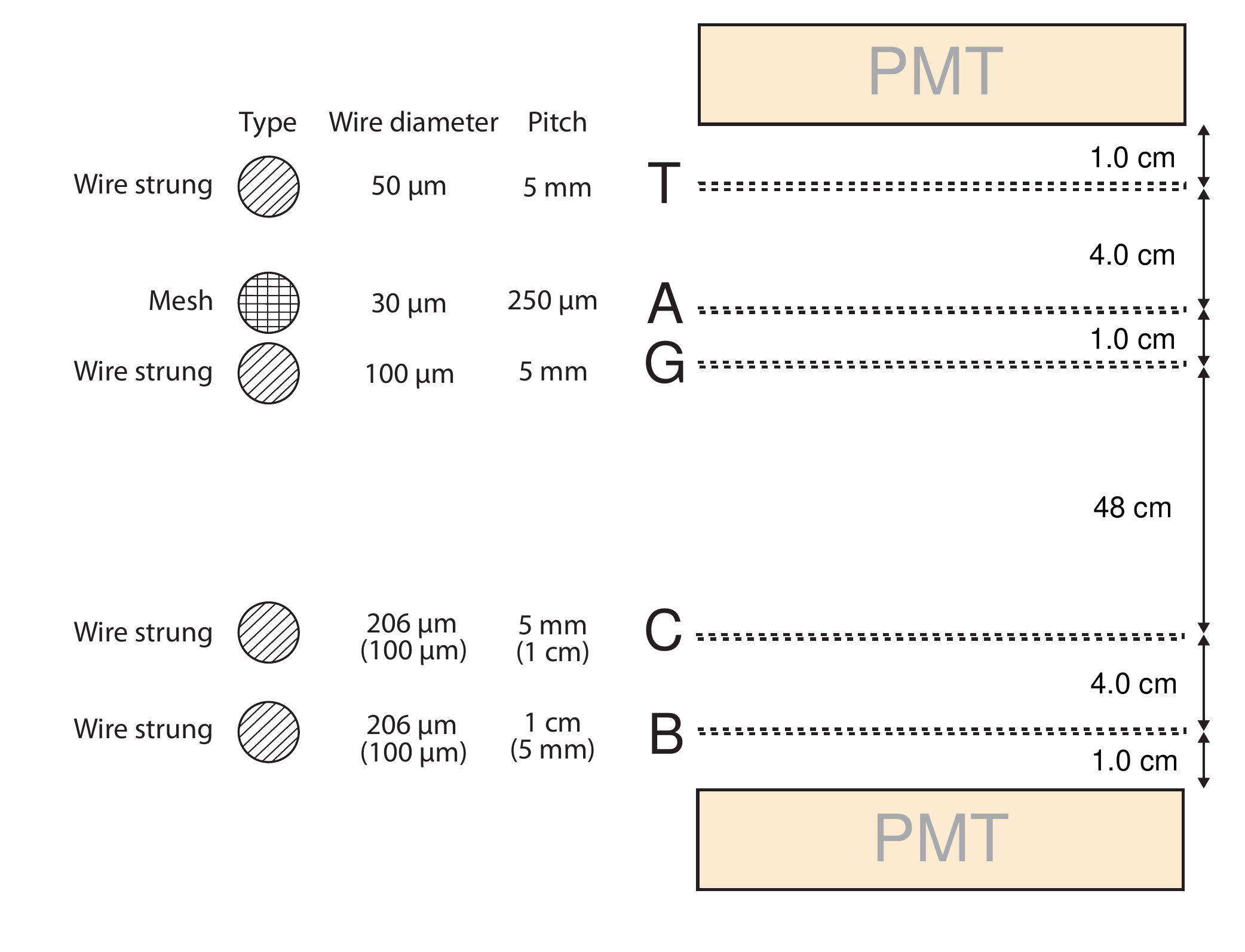}
\caption{\label{fig:lux_grid_configuration} The LUX electrostatic grid configuration: top (T), anode (A), gate (G), cathode (C) and bottom (B) grids. All grids are made from stainless steel wire. For C and B, the pitch and diameter values used during surface operations are shown in parenthesis.}
\end{figure}

The LUX active volume is enclosed by PTFE walls on the sides and is viewed by two arrays of 61 PMTs, one above the active region in the gas and one below it in the liquid \cite{Akerib:2012ys}. Electrostatic grids, shown schematically in Fig. \ref{fig:lux_grid_configuration}, are inside the active volume and produce a vertical electric field. The cathode (C) and gate (G) grids define a region of electron drift in the liquid, while the anode (A) and gate (G) grids define an electron extraction and electroluminescence (EL) production region in the gas. The liquid xenon level is placed halfway between A and G. The top (T) and bottom (B) grids shield the PMT arrays from the drift and EL fields. The grids are made from stainless steel wires. All grids are wire-strung, with the exception of A, which is a woven mesh. Fig. \ref{fig:lux_grid_configuration} shows the grid's wire pitch, wire diameter and their relative spacing.

Higher electric fields, and therefore, higher applied voltages at the grids, yield better detector performance and better WIMP sensitivity. For the A and G grids, the magnitude of the field translates into an electron extraction efficiency from the liquid into the gas phase. In order to achieve near 100\% extraction efficiency, a field of 10~kV/cm in the gas phase is required between A and G, which translates into a voltage differential of 8.5~kV. There is some evidence that a higher drift electric field between C and G yields better discrimination between the background electron recoils and signal nuclear recoils in liquid xenon. ZEPLIN-III obtained discrimination between electron and nuclear recoils at nearly 99.99\% with a drift field of 3.8~kV/cm \cite{Lebedenko:2008gb}. In comparison, XENON10 achieved 99.9\% discrimination with a drift field of 730~V/cm \cite{Angle:2007uj} and LUX achieved 99.6\% discrimination at 181~V/cm \cite{Akerib:2013tjd}.

\begin{figure}
\centering
\includegraphics[width=4in]{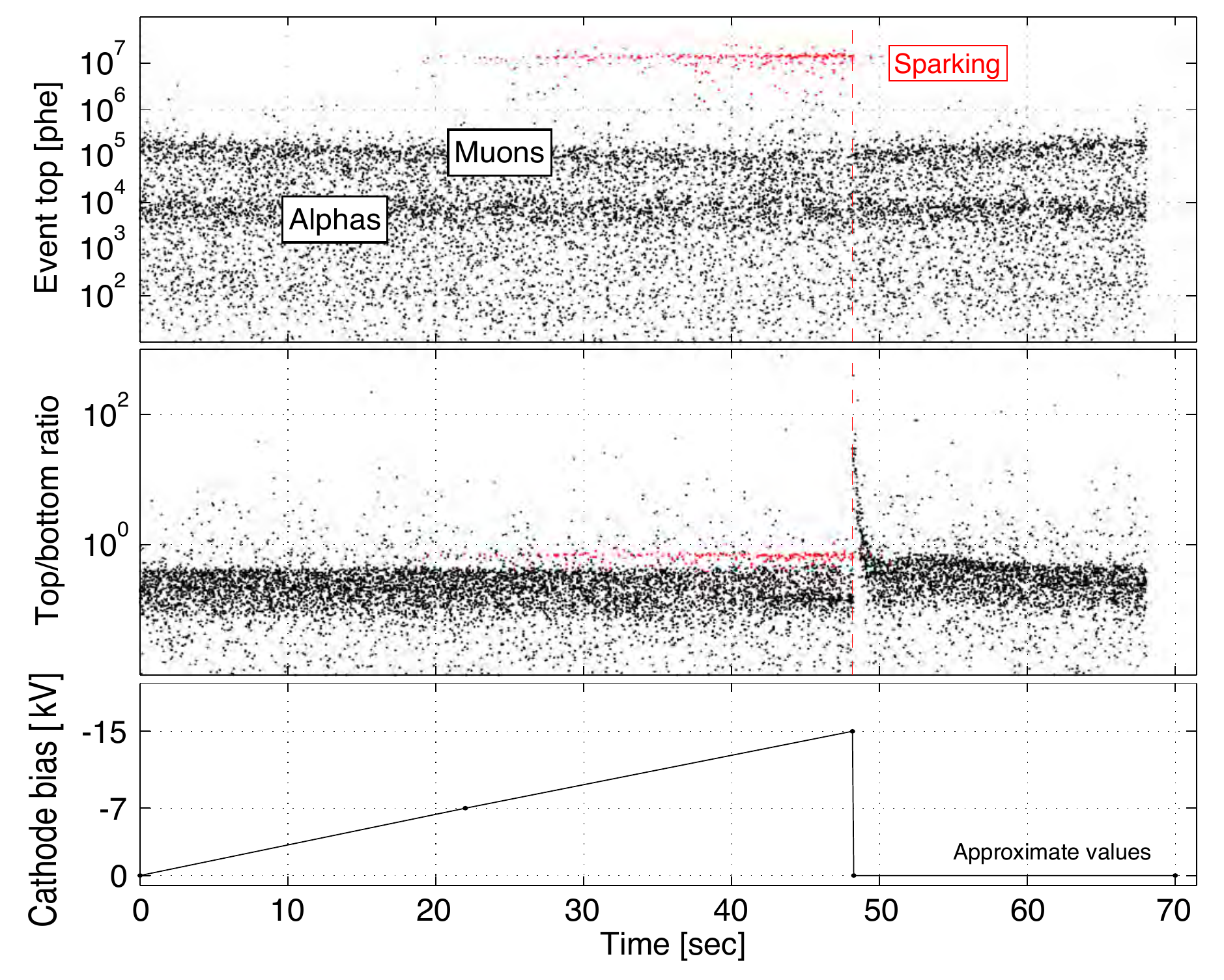}
\caption{Event area, top/bottom ratio and C grid bias as a function of time during one of the LUX high voltage bias trials in liquid xenon at the surface laboratory. For this trial, light production at the $10^7$~phe level appeared at about -7~kV and electric breakdown took place at -15~kV.}
\label{fig:sparking_lxe} 
\end{figure}

This summary describes the limitations in the C-B and A-G grid high voltage observed during surface and underground operations. There has been no evidence that the high voltage discharge is related to the high voltage feed-through or C grid cable connection. The feed-through is rated for continuous operation at -60 kV and was tested without discharge up to -100 kV \cite{Akerib:2012ys}.

\subsubsection{High Voltage Limitations During Surface Operation}

The LUX experiment was constructed and commissioned in a surface laboratory at SURF. During surface operations with liquid xenon, discharge-related optical pulses were observed when the C voltage approached $-7$~kV and  electric breakdown was observed between $-8$~kV and $-15$~kV on the C grid. Fig.~\ref{fig:sparking_lxe} shows one of the four biasing trials performed in liquid xenon. The event area, PMT top/bottom ratio and approximate C bias voltage are shown as a function of time. For this run, the PMTs were set to a gain of $1\times10^5$, which resulted in a DAQ pulse detection threshold of about 50~photoelectrons (phe). For reference, alpha particles ($10^4$~phe) from a $^{222}$Rn injection \cite{Akerib:2012ak} and muons ($10^5$~phe) are shown in the event area subplot. Light production at the level of $10^7$ phe per pulse was observed with the C biased at around -7~kV (red dots), and an electric discharge was observed at -15~kV, which is shown in the top/bottom ratio subplot as a spike. The spike is due to a reduction in the signal seen by the bottom PMTs when the B grid charges after the electric discharge from the C. Table \ref{table:surface_breakdown_summary} lists the bias trials performed and the resulting voltage for light production and electric breakdown.

\begin{figure}
\centering
\includegraphics[width=4in]{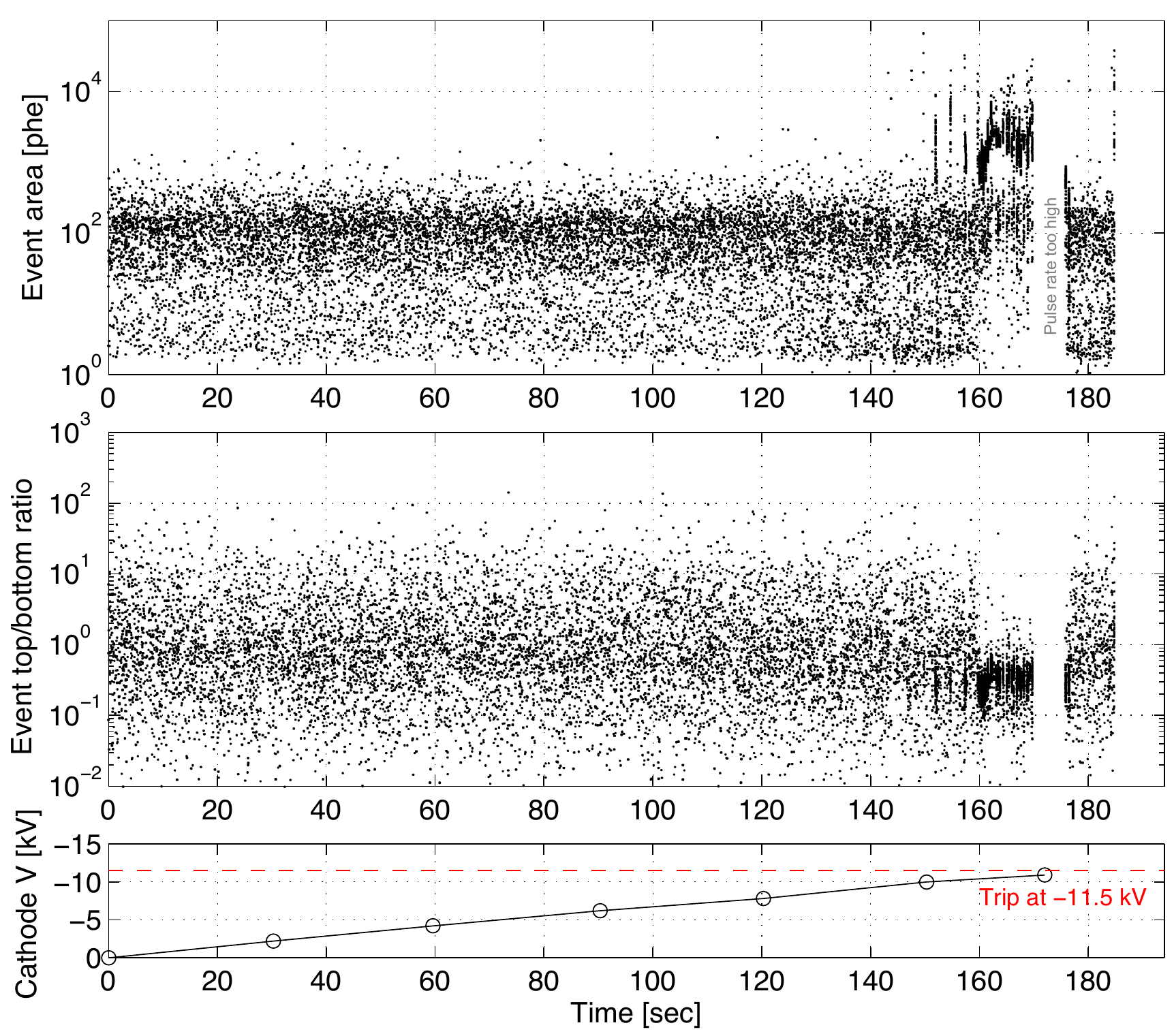}
\caption{A LUX test using a pressure of 1~bar of nitrogen gas. This test shows the light discharge appearing at -10~kV.  Electric breakdown took place at -11.5~kV. This test checks the grid high voltage performance prior to liquid xenon operations.
}
\label{fig:sparking_n2} 
\end{figure}

\begin{table}
\centering
\caption{The LUX liquid xenon and gas nitrogen trials for biasing the C grid.}
\label{table:cathode_breakdown_analysis_summary}
\begin{center}
\begin{tabular}{ | l | c | c |}
\hline
Conditions                              & Light     & Breakdown \\ \hline
liquid xenon 173~K, 1,280~torr           & -7~kV   & -10~kV   \\ \hline
liquid xenon 173~K, 1,280~torr           & none    & -8~kV      \\ \hline
liquid xenon 178~K, 1,680~torr           & -7~kV   & -14~kV   \\ \hline
liquid xenon 178~K, 1,680~torr           & -7~kV   & -15~kV    \\ \hline 
1 bar N$_2$ gas, room temp. & -10~kV & -11.5~kV \\ \hline
\end{tabular}
\end{center}
\label{table:surface_breakdown_summary}
\end{table}

When surface operations with liquid xenon were finished, the detector was evacuated, warmed to room temperature and filled with 1~bar of nitrogen gas. A C bias test was performed similar to the ones in liquid xenon, and the results are shown in Fig.~\ref{fig:sparking_n2} and listed in Table~\ref{table:surface_breakdown_summary}. Light discharges at the $10^3$~phe level were observed at around a -10~kV bias, and electric breakdown took place at a -11.5~kV bias. This result suggests that a breakdown test with 1~bar of nitrogen gas can be used to identify grid high voltage problems prior to liquid xenon commissioning.

\begin{figure}
\centering
\includegraphics[width=4in]{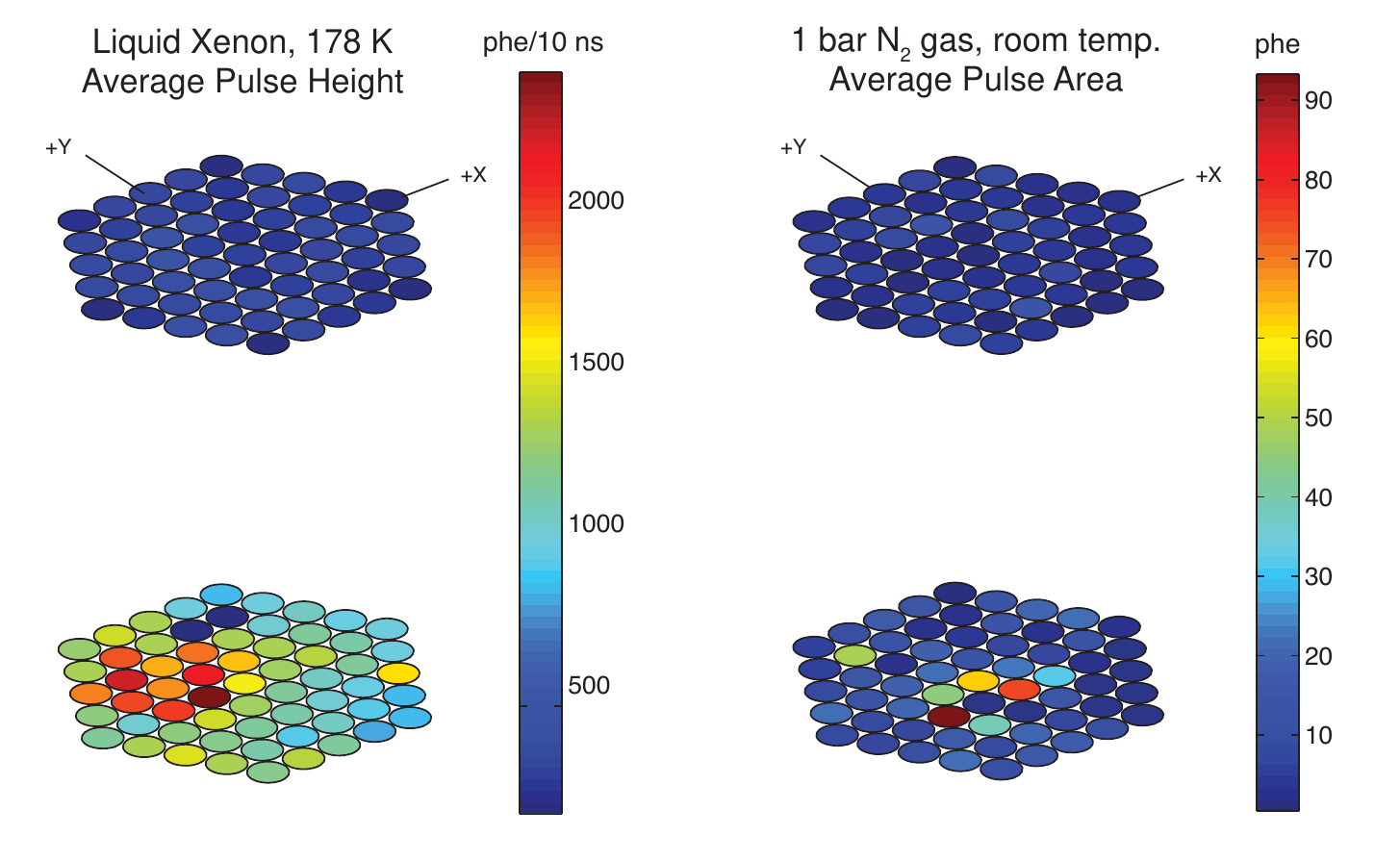}
\caption{Average hit patterns in LUX for the surface C grid breakdown in liquid xenon (left) and in nitrogen gas (right).}
\label{fig:surface_avg_hit_patterns} 
\end{figure}

The average hit patterns for the liquid xenon and nitrogen gas tests are shown in Fig. \ref{fig:surface_avg_hit_patterns}. Note that the pulse height was used for the liquid xenon data as a proxy for area in order to mitigate the effects of PMT saturation.

The highest electric field in a wire-strung grid configuration is the field at the surface of the wires. This wire surface field has a few-percent sinusoidal dependence in azimuth that can be ignored for practical purposes. The wire surface field can be approximated for wire-strung grids by an analytical formula~\cite{mcdonald:2003el}:

\begin{equation}
E_{\textrm{wire}} = (E_\uparrow - E_\downarrow) \cdot \frac{p}{\pi d},
\end{equation}
for a wire diameter $d$, grid pitch $p$ and electric field above (below) $E_\uparrow$ ($E_\downarrow$). For grounded G and B grids, a C voltage of -7~kV corresponds to -67~kV/cm at the surface of the cathode wires, and a C voltage of -15~kV corresponds to -130~kV/cm at the surface of the wires.

\subsubsection{High Voltage Limitations During Underground Operation}

After surface operations, the detector was opened and the C and B grids were replaced. The C and B wires were doubled in diameter and the C pitch was halved in order to reduce the electric field at the C wire by a factor of 4. Furthermore, the wire surface was changed from regular-finish to ultra-finish. The grids were not electropolished or burned-in prior to installation, but were cleaned in an isopropanol bath and the wires were individually inspected under a microscope. 

\begin{figure}
\centering
\includegraphics[width=4in]{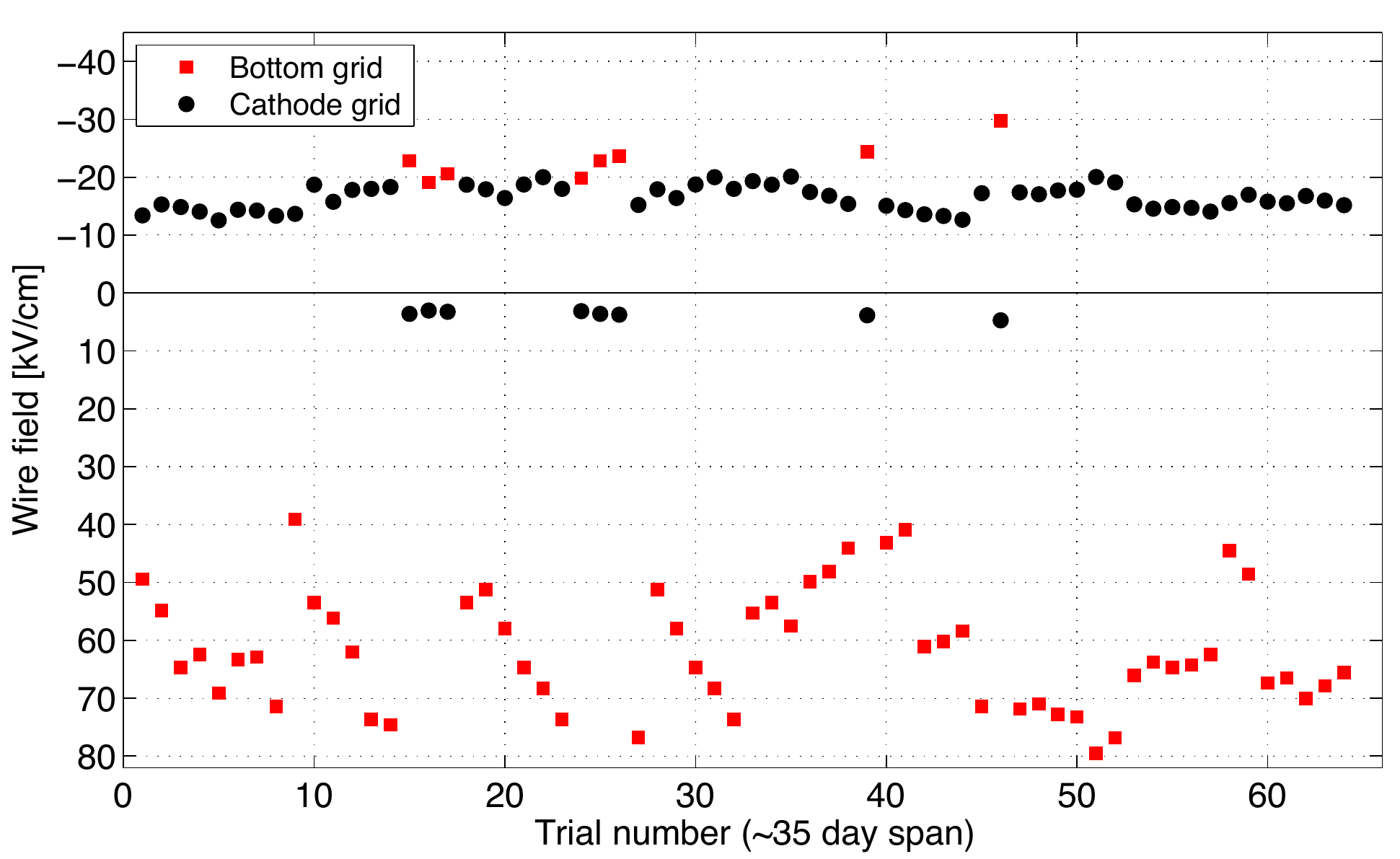}
\caption{\label{fig:BC_time_series} LUX electric field at the surface of the B or C grid wires when light discharge appeared during underground operations.}
\end{figure}

During underground commissioning, light discharge was observed at around -11~kV on the C or -3~kV on the B with the other grids grounded. The B, C wire surface fields for trials that took place during a $\sim35$~day span are shown in Fig. \ref{fig:BC_time_series}. The light discharge takes place when either the B or the C wire surface electric fields were between -15~kV/cm and -30~kV/cm. Positive electric fields at the wire that exceed 80~kV/cm do not result in light production. The electric fields achieved during underground operations are lower than what was achieved during surface operations, which points to a new source of electrical discharge such as particle debris.

\begin{figure}
\centering
\includegraphics[height=3in]{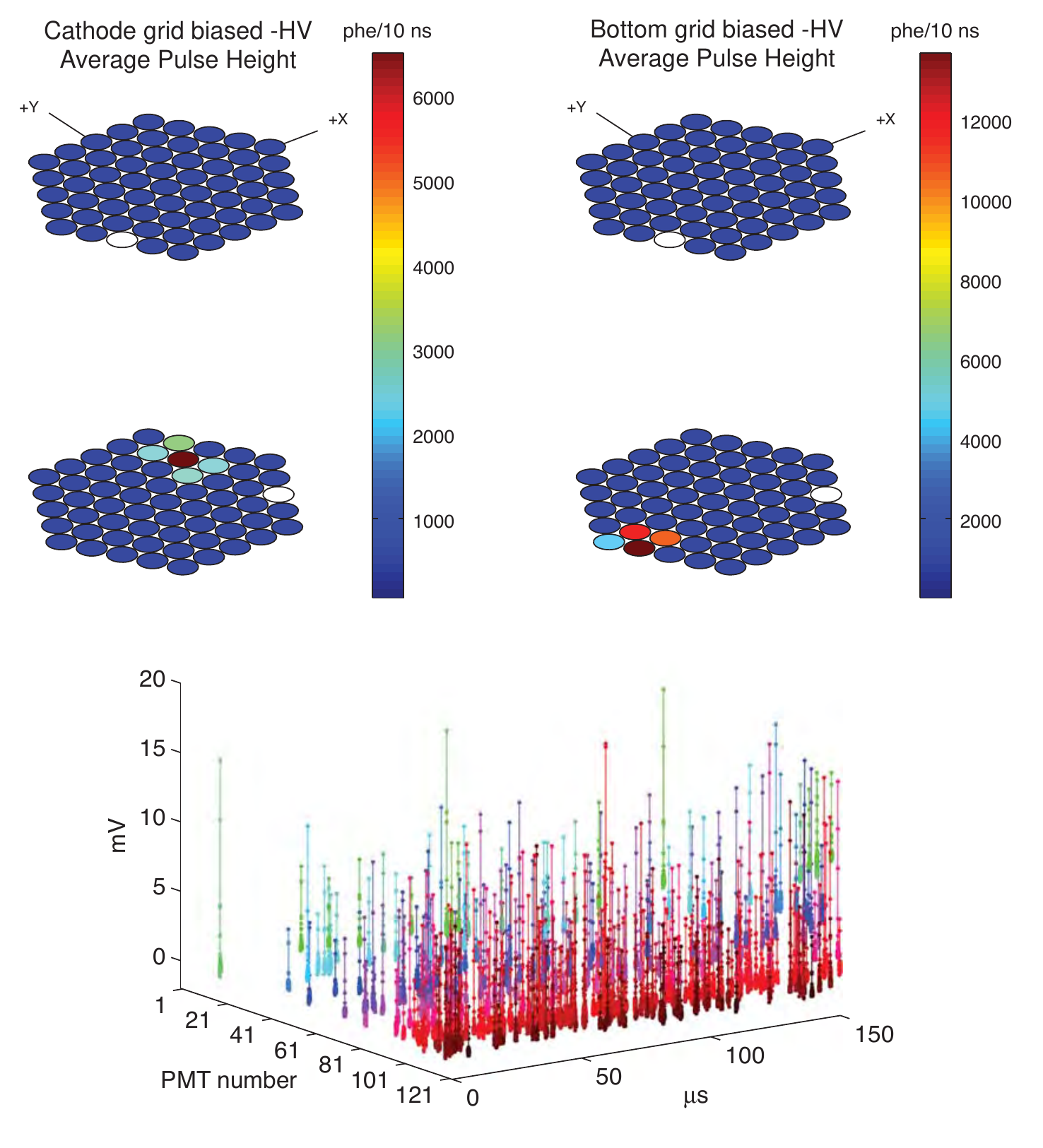}
\caption{Average hit pattern in LUX for the underground light discharge between C and B grids (top), and example of light discharge waveform (bottom). }
\label{fig:ug_avg_hit_patterns} 
\end{figure}

The average pulse height hit patterns for light discharge with the C biased to negative high voltage and B grounded, and for the reverse situation are shown in Fig.~\ref{fig:ug_avg_hit_patterns}. The light discharge waveforms are continuous sphe-level pulses, shown in the bottom subplot of Fig.~\ref{fig:ug_avg_hit_patterns}. If the voltage is increased beyond the onset of light discharge, large DAQ-saturating pulses are also produced at a lower rate.

Light discharge is also observed when the voltage difference between the A-G grids exceeds 5.5~kV, which corresponds to a field at the G grid wire surface of -53~kV/cm. The hit pattern localization is in the top array, and moves over time. This light discharge limits the electron extraction efficiency during underground operations to 65\% \cite{Akerib:2013tjd}.

\subsubsection{Tests Performed and Future Work}

Several tests were performed on the C and B grids in an attempt to reach higher voltages without light discharge production. First, a ``soak-in" of the grids took place, where the C and B grids were continuously biased for about 2~weeks below the onset of light production. The ZEPLIN-III collaboration utilized this method with modest success to progressively achieve higher cathode voltages. In LUX, however, this process yielded no long-term improvement. Liquid-phase conditioning was performed next, where the C and B grids were biased at/above the onset of light production for up to 100~minutes. No long-term improvement was reported with this procedure either.

It has been observed in the literature that bubble nucleation at the surface of a cathode grid lowers the onset of electric discharge \cite{Hanaoka:1993dc, Atrazhev:2010gc}. In order to test whether quenching bubble formation at the wires would allow higher voltage levels without light discharge, a bubble-quenching test was performed by introducing a 5~K temperature gradient in detector. No effect was observed in the voltage onset of light production with this test.

The leading hypothesis for the underground breakdown behavior is debris in the liquid xenon. Due to the high density of liquid xenon, 3~g/cm$^3$, material debris of many types, e.g. aluminum, float in it. There is evidence of conductor debris in LUX during the liquid xenon filling period, when bottom array PMTs temporarily shorted when the liquid surface crossed that level. This hypothesis will be tested by moving the liquid xenon level across a grid. There are also small PTFE pieces resting on the anode grid, a fine woven mesh, that may be responsible for the A-G behavior. Lastly, there is also the possibility of having wire surface asperities or insulator film depositions that can enhance the local field and produce light discharge, e.g. the Malter effect. Gas-phase conditioning will be performed on the A-G and C-B grids in order to see if any hypothetical contaminants can be removed.

\subsection{The XENON experiment}
\label{sec:messina}
{\it Contributed by M.~Messina, Columbia University, New York, NY 10027, USA}
\newline

\subsubsection{Introduction}

The results reported in this section have been obtained in the framework of the R\&D effort of the Columbia University group in the XENON project. The XENON detectors are dual-phase TPCs in which the liquid xenon is used as a target and detection medium for particle interactions in which ultraviolet scintillation photons and ionization electrons are generated. The latter, once extracted into the gas phase, lead to the generation of delayed scintillation photons. The generation of high voltage is one of the key technological challenges one faces when constructing such dual-phase TPC's.
 
One key of the XENON TPC is the utilization of both the prompt VUV and non-prompt ionization-induced scintillation signals to localize events inside the detector. The light is measured by PMT arrays installed on top and bottom of the field cage of the TPC. Measurement of the non-prompt signal requires an electric field to drift electrons from the interaction point through the liquid. Larger drift volumes require larger fields, and thus a larger applied voltage. Moreover, it is important to understand how the ionization electron yield depends on the field amplitude and if this has any effect on yield per unit energy. The behavior of this yield ultimately determines the minimum energy threshold for the detector, and thus has important consequences for the sensitivity of the experiments.

\subsubsection{The high voltage experience of the XENON experiments}

To generate an electric field suitable for a $\sim 1$~m drift, a voltage on the order of tens of kV is required. To accomplish this voltagefeed-through, the most common solution is to feed the voltage from outside by means of a vacuum-compatible feed through feed-through.  Although feed-throughs capable of handling very high voltage are commercially available, they cannot be used in detectors designed for a dark matter search because of their high radioactivity. Typically the insulating material is a ceramic which has a large amount of radioactive $^{40}$K and thus a major source of gamma background.  Thus, for the XENON project a custom made feed-through has been developed. It has the design of a cylindrical capacitor where the outer shell is grounded and the central electrode is biased at the desired voltage. A plastic dielectric capable of withstanding cryogenic temperatures of 98~K is placed between the electrodes. Polytetrafluoroethylene (PTFE) was selected for this purpose since it is the cleanest commercially available dielectric. The XENON10\cite{XE10} and XENON100\cite{XE100} TPCs have used such custom-made feed-throughs developed at Columbia. The XENON100 TPC is the only two-phase liquid xenonTPC which has been operated with a voltage in excess of 15~kV, and was limited by the phenomena described below, and not by the feed-through. More recently a new design for much higher voltage and with PTFE replaced by high-density Polyethylene was developed by the XENON group at UCLA. Several feed-through prototypes have been rmade, and tested in liquid nitrogen up to 100~kV and in liquid xenon up to 75~kV. Details of this testing are given in \S~\ref{sec:hanguo}.

It should be noted that the high voltage required to operate a TPC implies not only the presence of an appropriate feed-through, but also that any other components are constructed to prevent sparking when this voltage is applied. This aspect will be clarified in the next section.

\subsubsection{R\&D for the XENON experiments}

\begin{figure}[h]
\centering
\includegraphics[height=1in]{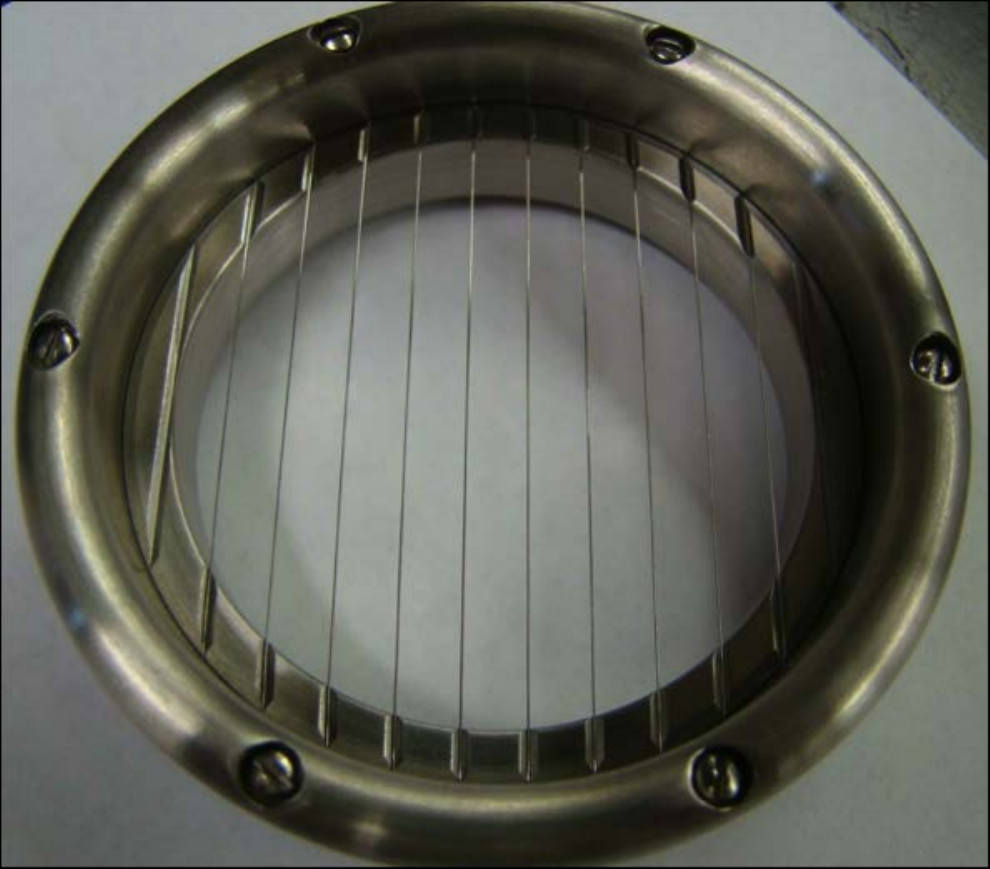}
\includegraphics[height=1in]{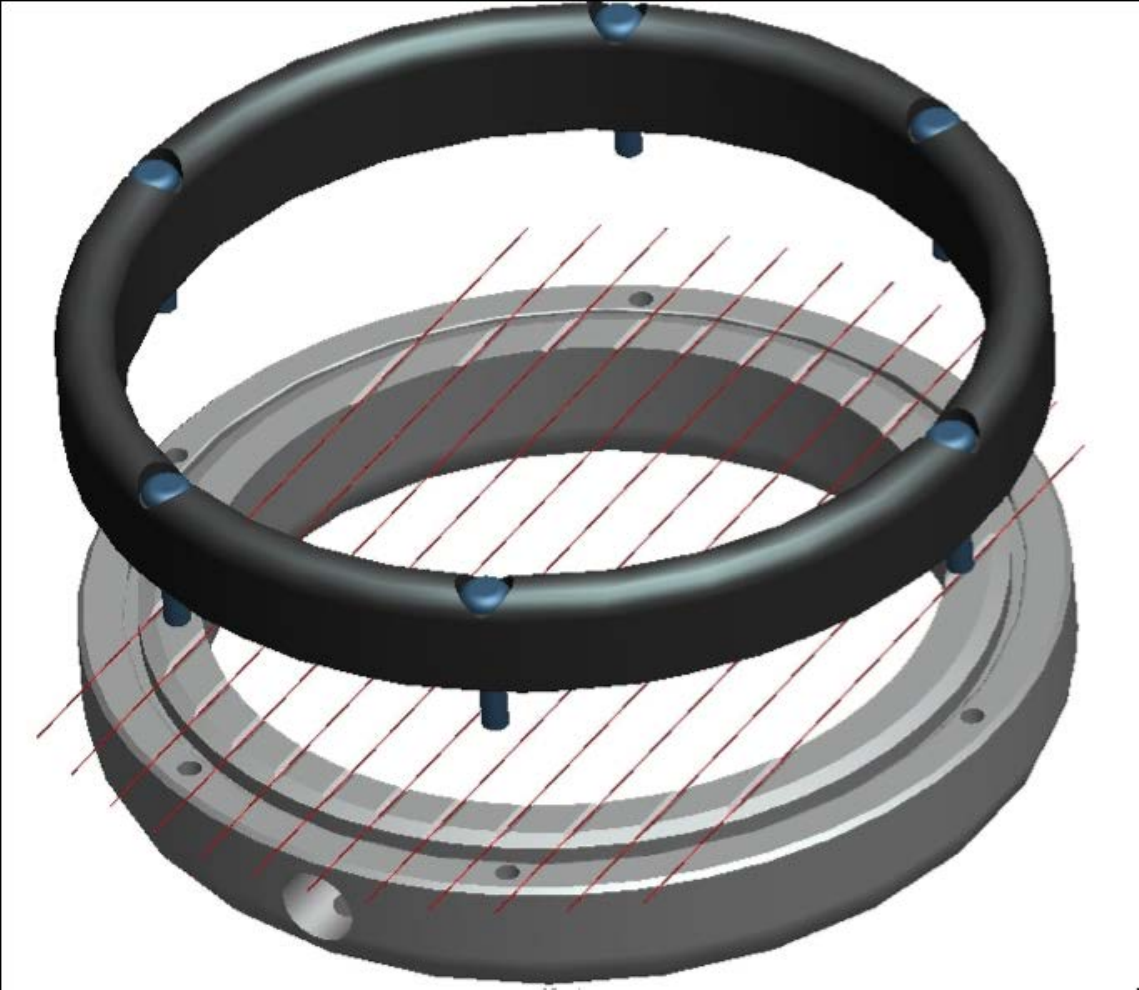}\\
\includegraphics[width=1.in]{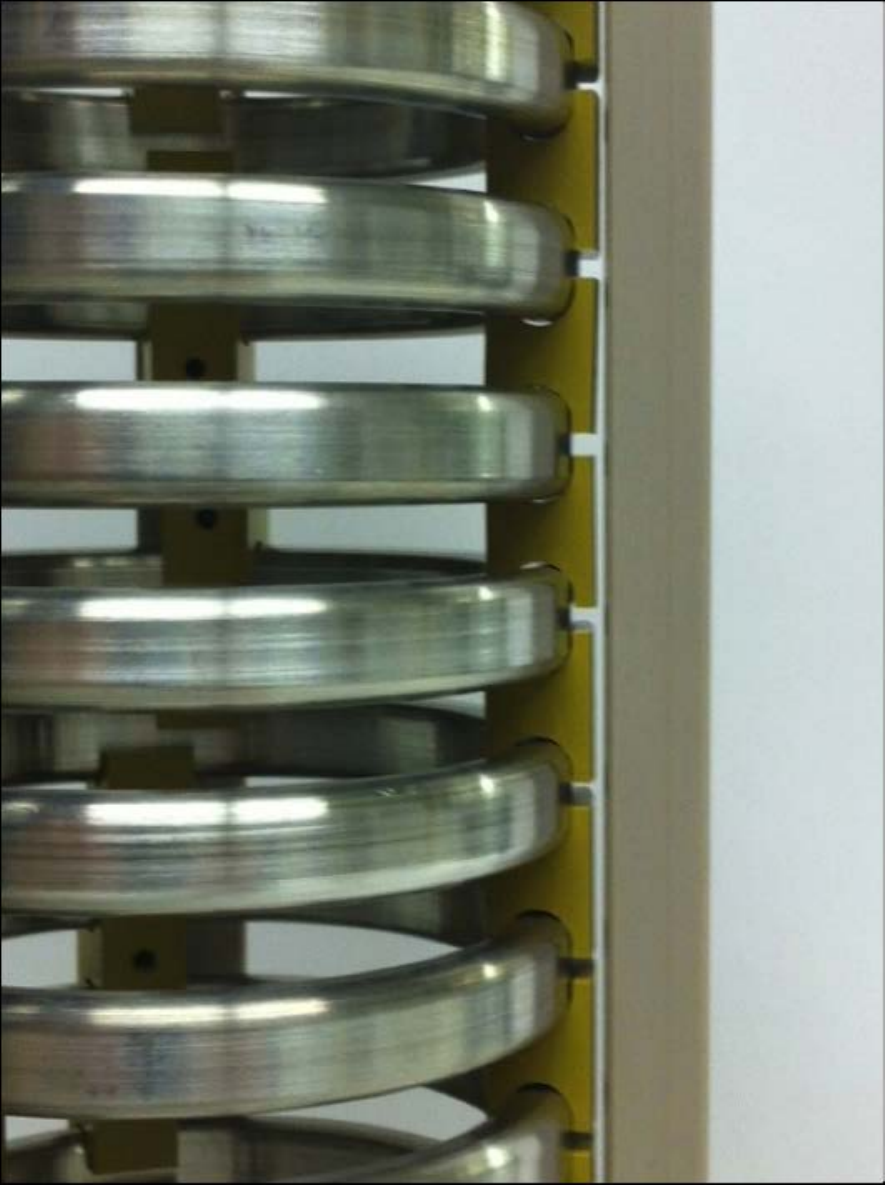}
\includegraphics[width=1.in]{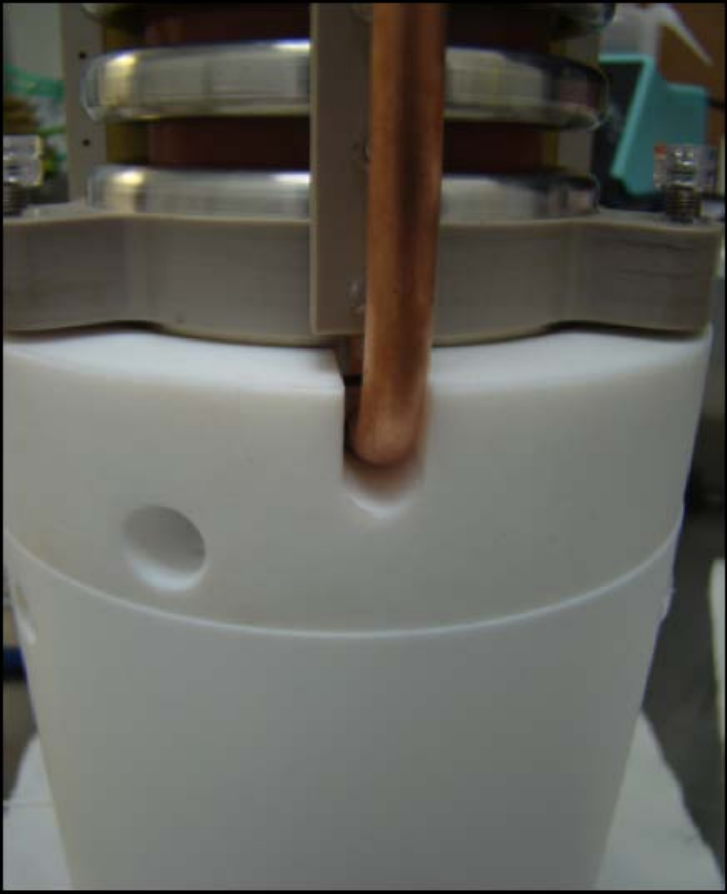}\\
\includegraphics[width=2in]{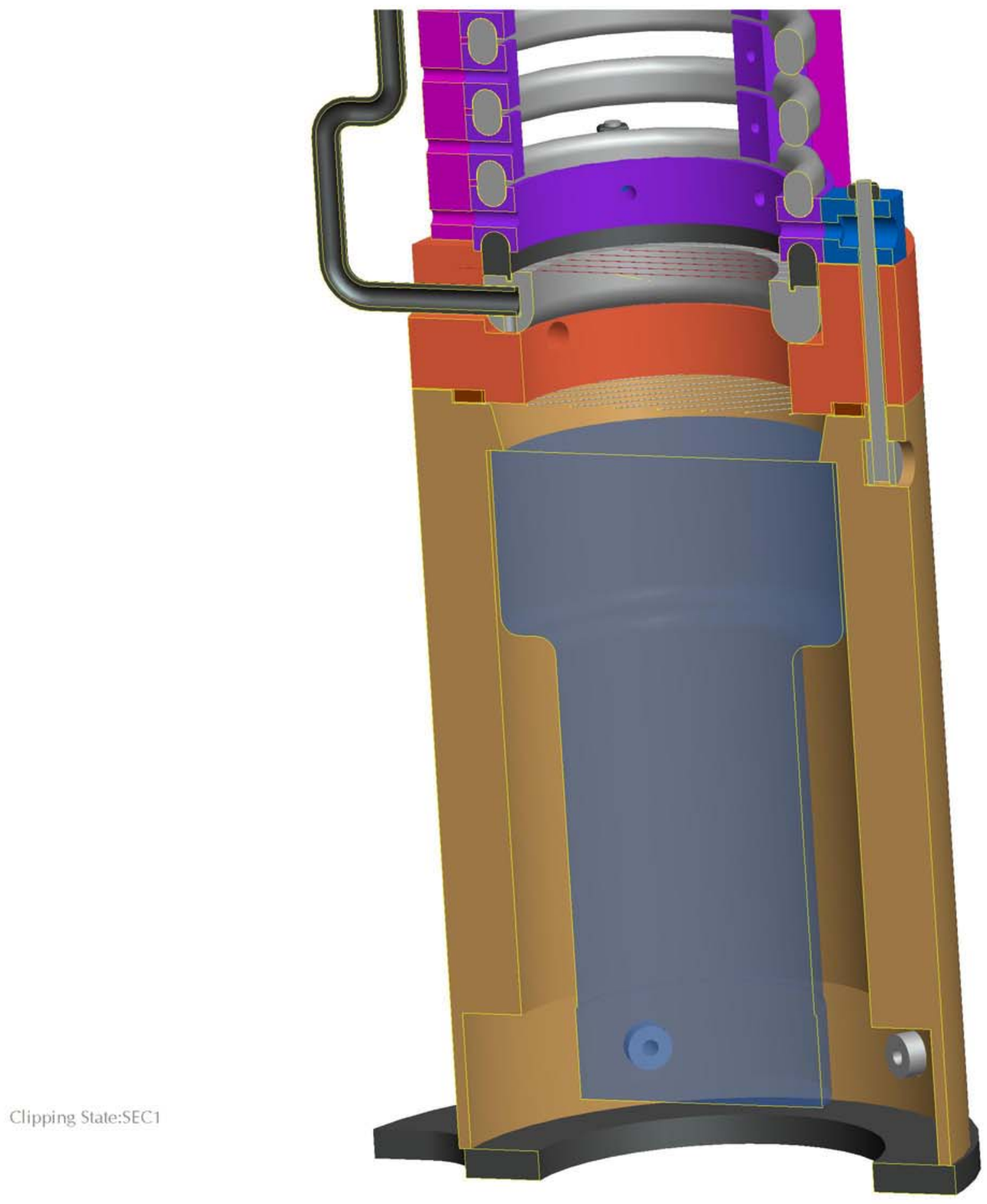}
\caption{(top) A picture and a drawing of the cathode for the 1~m TPC at Columbia University for XENON R\&D purposes . On the cathode farm the 200~$\mu$m OD wire are also visible. (middle) Detail of the external of the field cage and of the copper rod connecting the cathode and the feed-through. (bottom) A zoom of the inside of the TPC where the details of the  shaping rings, the cathode with its connection to the feed-through, the mesh protecting the bottom PMT and the PMT positioning are depicted.}
\label{electrodes}
\end{figure}

In Fig.~\ref{electrodes}, the drawings of the cathode and the cross-section of the field shaping rings are shown together with a zoom of the drawing of the inner side of the 1~m TPC installed in the Columba University R\&D facility\cite{ref2:cryodemonstartorpaper}. The flat shape of the rings prevent points where the field can reach $>100$~kV/cm.  The 0.5~cm distance between the rings prevents the field lines from escaping outside of the field cage, preserving the mean value of the electric field devoted to drift electrons.  Furthermore, the large radius of curvature of the bottom part of the cathode is designed to prevent a large field on its surface.  For the cathode, this problem is particularly important since it is the last electrode and is exposed to the ground: proximity to the ground drastically increases the probability of a spark starting on its surface. Counter to common practice, it has also been found that installing the voltage divider and associated electrical components inside the field cage helps mitigate the effects of irregularities in the resistors and/or the presence of sharp soldering edges which are common in such components, both of which can lead to discharges.

In Fig.~\ref{electrodes} , the region between the cathode and the cathode surface of the bottom PMTs is visible.  The latter is protected by a screening mesh kept at the same potential as the PMT's cathode to prevent the accumulation of ions on the PMT window's surface.  This location is also a region of high electric field, with a sizable distance of 5~cm between the cathode and the PMT, and obtained using a specially designed PTFE spacer. The surface of the spacer is corrugated such that the mobility of the ionization electrons across the field lines is greatly reduced. This corrugation is necessary to prevent the production of scintillation light by means of surface conduction between the cathode and screening mesh.

The last item in the list above is the nucleation of bubbles caused by local phase transition. In this case, the gas in the bubble is exposed to the same large field and can thus experience a breakdown much before the liquid.  High temperatures in the bubble produce even more gas, often leading to a chain reaction and eventually a discharge.  The removal of impurities from the liquid helps in this respect, since the nucleation takes place around those atoms and molecules liquefying at lower temperature than the one of the liquid.  Plating any surface at high voltage with a high work-function metal is also useful to avoid this discharge, as the plating reduces the surface emanation of extraneous molecules or atoms and increases the work function of the electrons. Unfortunately, at least in principe, the problem of ions which can start bubble formation still remains.  Interesting discussions and possible improvements were discussed in \S~\ref{sec:PEREVERZEV}, with the goal of preventing problems related to ion accumulation. 

With the care taken for all the points discussed above, the Columbia 1-meter drift TPC developed as Demonstrator for the XENON1T project, was operated with very high voltage, up to 75kV on its cathode. The sum of the knowledge and experience gathered by these studies have been included in the XENON1T TPC design.

\clearpage

\subsection{The LZ experiment}
{\it Contributed by E.~Bernard, Yale University, New Haven, CT 06520, USA for the LZ Collaboration}
\newline
\subsubsection{Introduction}

The LZ detector is a proposed two-phase liquid xenon dark matter detector.  It is the successor to the LUX experiment~\cite{Akerib:2012ys,Akerib:2013tjd}.  The detector contains a cylindrical TPC filled with 7000 kg of liquid xenon.  The TPC is viewed from the ends of the cylinder by arrays of low-background PMTs.  Elastic scattering of WIMPs by xenon nuclei produce scintillation light and ionized xenon.  A vertical electric field drifts free electrons from the interaction sites upward to the xenon surface, where they produce a delayed electroluminescence signal in the gas region of the detector.  The pattern of electroluminescence light sensed by the upper PMT array gives the horizontal position of the interaction, while the depth of the interaction within the liquid xenon is determined by time projection.  The detector is surrounded by large tanks of liquid scintillator that operate as an active shield to veto gamma and neutron events.  The detector and liquid scintillator tanks are also immersed in a large water tank that provides further gamma and neutron shielding.  

The detector is designed to operate with a cathode voltage of up to -200 kV, with nominal operation at -100 kV.  The electric field within the TPC is graded by a set of field rings connected in series by resistors as a linear voltage divider.  At a cathode voltage of -200 kV, the rings establish a drift field of 1430 V/cm across the 140 cm drift length.  The region between the cathode and the grounded lower PMT array is only 15 cm long and the corresponding field is 13.3 kV/cm.  This ``reverse field region" is similarly graded with field rings. The field rings are made of carbon-loaded conductive plastic.  These are captured by interlocking teflon rings that form the reflective interior of the TPC.  The surfaces of each teflon ring terminate at the conductive plastic, leaving a slight gap of xenon between each insulating segment.  The teflon rings also contain internal passages for the thick-film surface-mount resistors that connect adjacent conductive rings.  In both the TPC and reverse field regions the rings repeat with a pitch of 2.5 cm.

To avoid unwanted electroluminescence, the LZ detector limits fields to a maximum of 50~kV/cm within the liquid xenon.  This limit necessitates placing a gap of 4~cm between the field rings and the wall of the cryostat that contains the detector.  The lower third of the cryostat wall is bulged outward from the field cage to increase the gap to 8~cm where the voltage is highest.  The xenon in this gap is observed by PMTs to provide a coincidence signal that is used to veto gamma rays that Compton scatter within the TPC and subsequently interact in the gap.

Differential contraction shortens the height of the plastic field cage relative to the titanium cryostat.  The TPC is supported from below, so that the motion of the TPC due to thermal contraction is accommodated at the top of the TPC, where the electric fields are smallest.

\subsubsection{Cathode high voltage delivery in LZ}

\begin{figure}
\begin{center}
\includegraphics[width=4.5in]{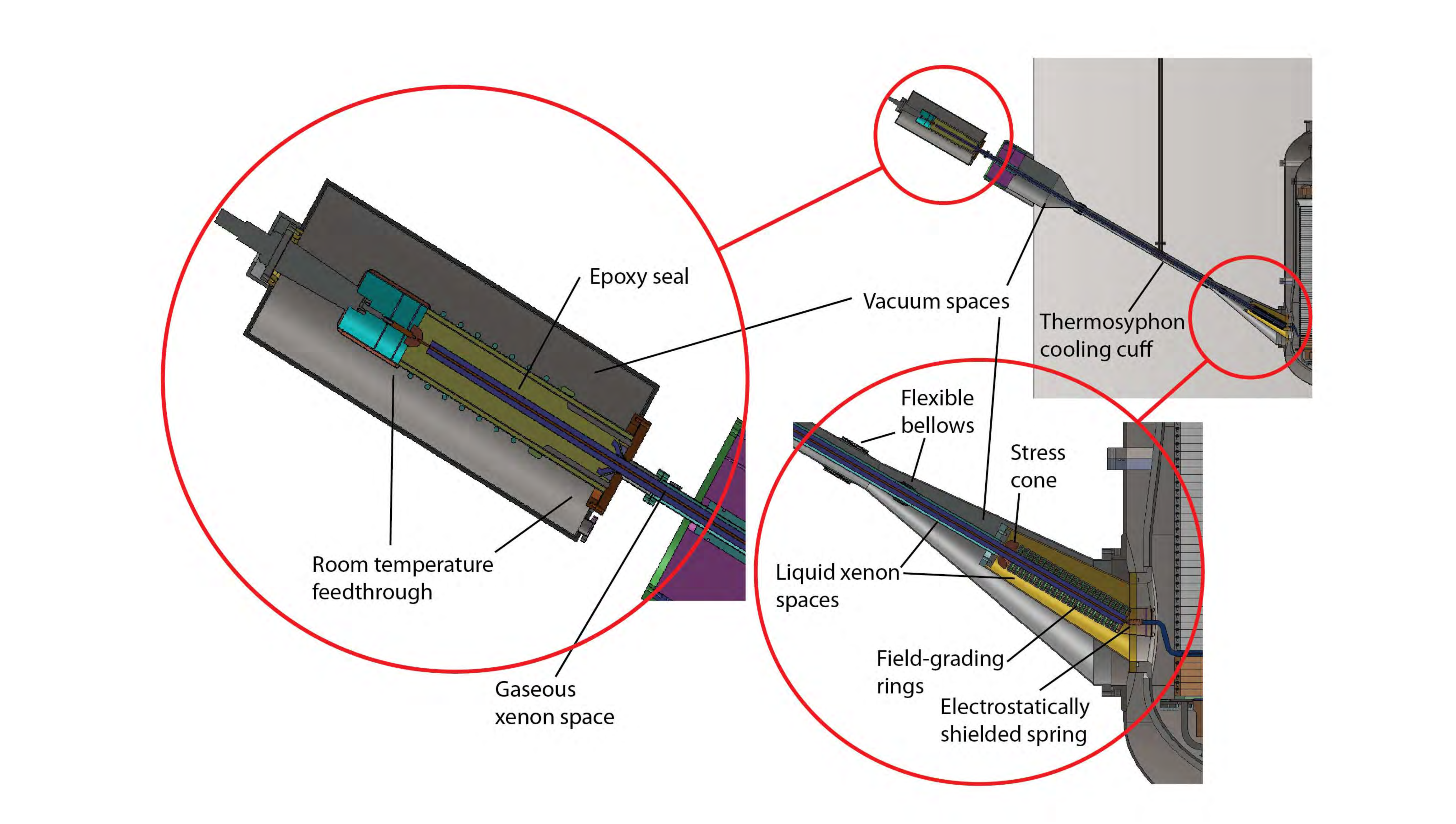}
\caption{Features of the high voltage delivery to the cathode of the LZ detector.}
\label{fig:CHV}
\end{center}
\end{figure}

High voltage is provided to the cathode of the LZ detector through a side arm of the inner cryostat, as shown in Fig.~\ref{fig:CHV}.  This design avoids routing the large high voltage cable along the side of the TPC or the reverse field regions. Such a routing would require a substantial reduction in the diameter of the TPC due to the 50~kV/cm field constraint.  The high voltage arm extends outward from the inner vessel at an angle of 30 degrees above the horizontal.  The arm is an extension of the xenon space of the inner cryostat and is surrounded by insulating vacuum.  In normal operation, liquid xenon fills a length of approximately 2.8 meters of the arm, rising to a level equal to that in the detector.  The upper end of the arm penetrates the water tank wall.

The high voltage arm contains a length of model 2077 high voltage cable by Dielectric Sciences.  This cable is rated to 300 kV DC and consists of a 5 mm diameter copper core surrounded by 3.3 cm diameter polyethylene insulation.  This cable is covered by a tinned copper braid.  At the upper end of the high voltage arm the cable is terminated by a custom epoxy feed-through that seals to the cable polyethylene and a conflat flange.  The design of this feed-through is described in the next section.  The feed-through forms a high voltage connection between the cable, located within the xenon space, to a terminal, located in an high-vacuum can.  A vacuum high voltage feed-through produced by Parker Medical connects the high-vacuum space to a conventional x-ray cable and high voltage power supply.  The feed-through operates at approximately room temperature.  

From the outside of the detector, the 2077 cable penetrates from the high vacuum space into gaseous xenon in the TPC volume.  The cable passes through the surface of the liquid xenon as it descends from the feed-through down the high voltage arm toward the detector.  The cable is supported in the arm until it is within 40 cm of the cathode.  At this point the grounded braid of the cable is terminated in a flared structure, the stress cone, that minimizes the field at the cable surface.  The bare cable continues and the center conductor connects to a flexible spring that connects to the cathode.  The spring is shrouded in a curved shield connected to the cathode to prevent high fields at the small features of the spring.  The bare polyethylene between the stress cone and the spring is surrounded by 16 rings made of carbon-doped conductive plastic.  The rings are connected by resistors to the cathode and grounded braid to produce a linear voltage grading along the length of the cable.  The large diameter of the rings also serves to reduce the field in the xenon around the cable.  The rings make electrical contact to the cable surface through canted coil springs.

\subsubsection{The LZ epoxy high voltage feed-through}

The LZ feed-through is based on a design used successfully in the LUX experiment.  This style of feed-through terminates a high voltage cable in a mass of epoxy that is vacuum sealed to a conflat flange.  The grounding braid of the cable is removed from a length of the cable and this length is encased in Stycast 2850 epoxy plastic.  The epoxy is filled with alumina and has a dielectric strength of 15 kV/mm.  The cable is terminated at a copper sphere, half of which is embedded in the epoxy, leaving the exposed half to act as a terminal.   The epoxy shrinks slightly as it cures, providing symmetric pressure to the polyethyelene of the cable and forming a helium-leak-tight vacuum seal.  The epoxy and cable are cured within a thin fiberglass laminate tube.  The fiberglass tube is also concentric about a SS tube and sealed to it by epoxy.  The tube is welded to a standard conflat flange.  In this way a hermetic seal is made between the cable and the conflat flange.  The point of the cable at which the grounding braid ends is tightly contained within a smooth stainless steel tube that is gradually curved away from the cable.  This design minimizes the field enhancement caused by the abrupt termination of the grounding braid.  All of the spaces that contain high fields are filled with epoxy, eliminating the possibility of gas breakdown.

This style of feed-through was developed for the LUX detector and sealed a Heinzinger HVC100 cable to a 6 inch conflat flange.  Two such feed-throughs were successfully tested to $-100$~kV by terminating the distant end of the cable in silicone transformer oil.  For the LZ detector, the epoxy seals the much larger Dielectric Sciences 2077 cable to an 8 inch conflat flange.

\begin{figure}[t]
\begin{center}
\includegraphics[width=2.5in,keepaspectratio]{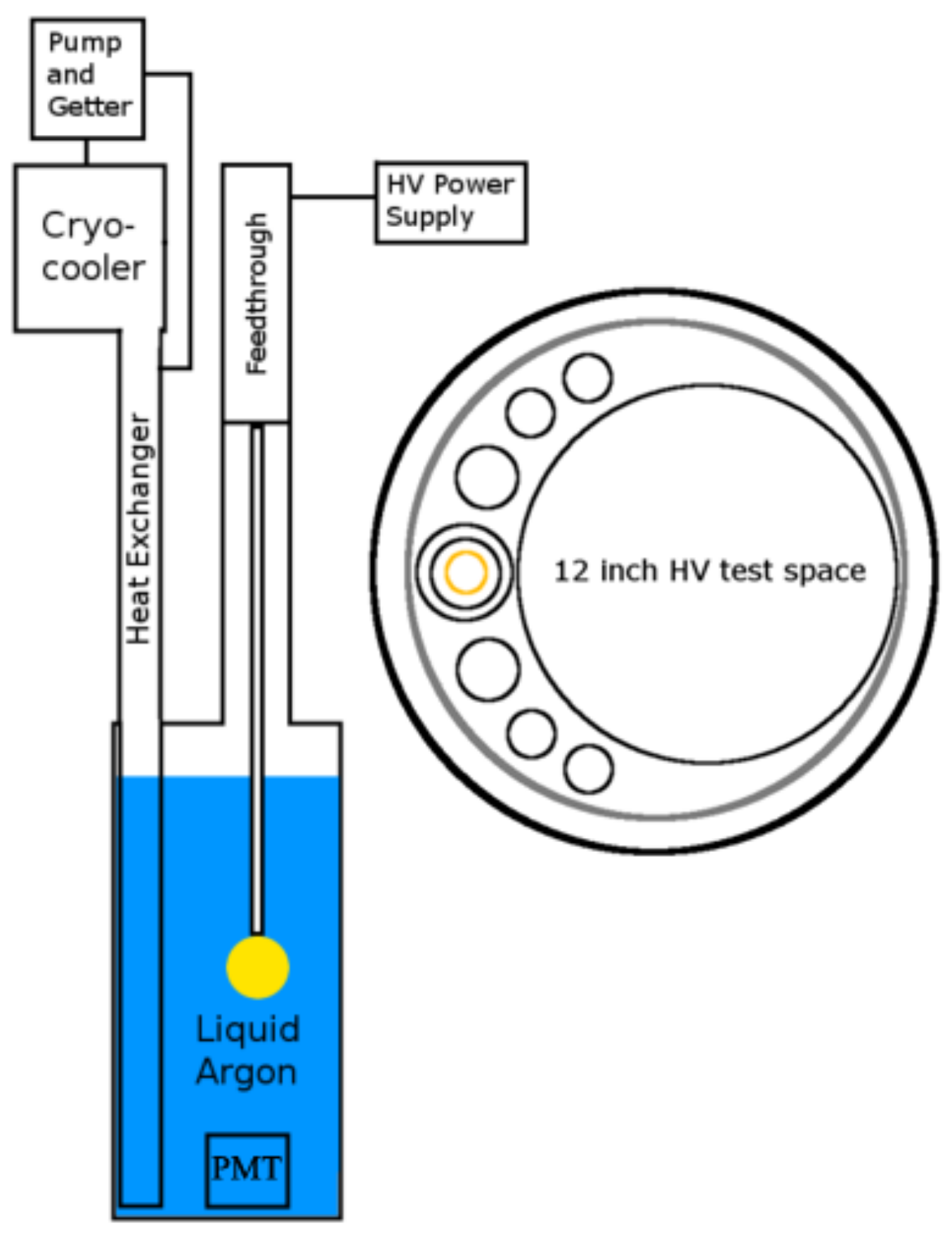}
\caption{Block diagram (left) and top flange (right) of the liquid argon high voltage test system for the LZ experiment.}
\label{fig:AR_HV} 
\end{center}
\end{figure}

\subsubsection{The liquid argon high voltage test system for LZ}

Although the LZ detector will be a liquid xenon TPC, a liquid argon high voltage test system has been constructed at Yale University to test the feed-through design in a less expensive medium. The feed-through seals the DS 2077 cable to an 8 inch conflat flange.  The bare length of cable surrounded by epoxy is 180~cm long so that fields are very low outside of the epoxy.  This design allows the feed-through to operate with only air insulation.  The outside of the feed-through is surrounded by corona rings and the high voltage end is terminated in a 22 inch diameter sphere.  The sphere is connected to an air insulated power supply produced by Glassman High Voltage.  This feed-through is designed to hold -300 kV; it has been tested to -200 kV for 20 days without any sign of aging or breakdown. 

The feed-through and power supply sit on a metal grating.  The cable, enclosed in a vacuum bellows, emerges from the bottom of the feed-through and passes through the metal grating.  The bellows seals to the top of a 180~cm tall dewar of liquid argon, and the cable hangs into the liquid argon from above and terminates at a 4 inch sphere.  The sphere sits 30 cm above a TPB-coated quartz plate that shifts the UV light predominantly emitted by discharges in liquid argon to 425 nm.  This opening is observed by an 8 inch phototube that sits immediately beneath the quartz plate.  

This system provides an open cylindrical bore 12 inches in diameter and 45 inches long in which test structures can be observed while supplied with up to -200 kV.  This system will be used to determine if the LZ high voltage structures can operate at full voltage without producing electroluminescence.  With the possible addition of an electron lifetime monitor, this system could also explore the relationship between electroluminescence threshold and electron lifetime.

\subsubsection{LZ cathode wire R\&D}
\label{ref:tomas}
{\it Contributed by A.~Tom\'as, Blackett Laboratory, Imperial College London, UK in collaboration with\\
H.~Ara\'ujo, Blackett Laboratory, Imperial College London, UK\\
A.~Bailey, Blackett Laboratory, Imperial College London, UK\\
T.~Sumner, Blackett Laboratory, Imperial College London, UK} 
\newline

The ability of a cathode electrode, usually a wire grid, to withstand the high electric field experienced at the surface of individual wires is under study at Imperial College London. Nominally these fields are much lower than the onset fields for electroluminescence or charge multiplication in the liquid, the former being 400--700~kV/cm for liquid xenon~\cite{Derenzo74,Masuda79}. The fact remains that most practical cathode electrodes in double phase detectors have been limited to surface fields of 40--65~kV/cm\cite{Howard04,Lebedenko09,Burenkov09,Akimov12a,Akimov12b,Akerib13}, although the XED test chamber operated with substantially higher values of up to 220~kV/cm~\cite{Shutt07a}. The cold emission of electrons might be a key phenomenon to understand the origin of these limitations -- as well as a useful diagnostic -- or an undesirable secondary effect. Single electron emission, possibly related with cathode breakdown, is a serious issue to achieve the low energy thresholds, near or below keV energies, which are required in the hunt for low mass WIMPs.

Electron emission from metal surfaces can be boosted in two different ways: $i)$ a local enhancement of the electric field at a particular point on the surface due to wire imperfections such as tips, impurities, crystal structure defects, and domain edges,  and $ii)$ the presence of a thin insulator film covering the conductor surface such as xenon ice, deposition of impurities, oxidation and other dynamic effects as described in \S~\ref{sec:PEREVERZEV} that will result in a lower effective work function. The presence of single electron events has been reported and studied by several experiments~\cite{Edwards08,Sorensen08,Burenkov09,Santos11,Aprile13}, although these were mostly found to be correlated with larger detected interactions -- at least below the onset of breakdown.

At Imperial College the R\&D activity focusses on a small chamber to test cathodes made from a single wire; in this way high voltage breakdown issues can be avoided as the applied voltage will be $\sim$10~kV and, simultaneously, challenging electric fields can be applied to the wire surface, i.e. $\sim$400~kV/cm for 100~$\mu$m diameter wire. The upper surface of the wire will be observed with a PMT looking down from the gas phase, which will also detect the VUV electroluminescence (EL) from any charge released from it and then drifted up and emitted into the vapor. Any electron emission from the cathode can, if accompanied by prompt light, be reconstructed to the cathode depth by electron drift time with single electron sensitivity allowing the measurement of minute electron emission currents, well before macroscopic breakdown sets in.

\begin{figure}[h]
\centering
\includegraphics[width=4in]{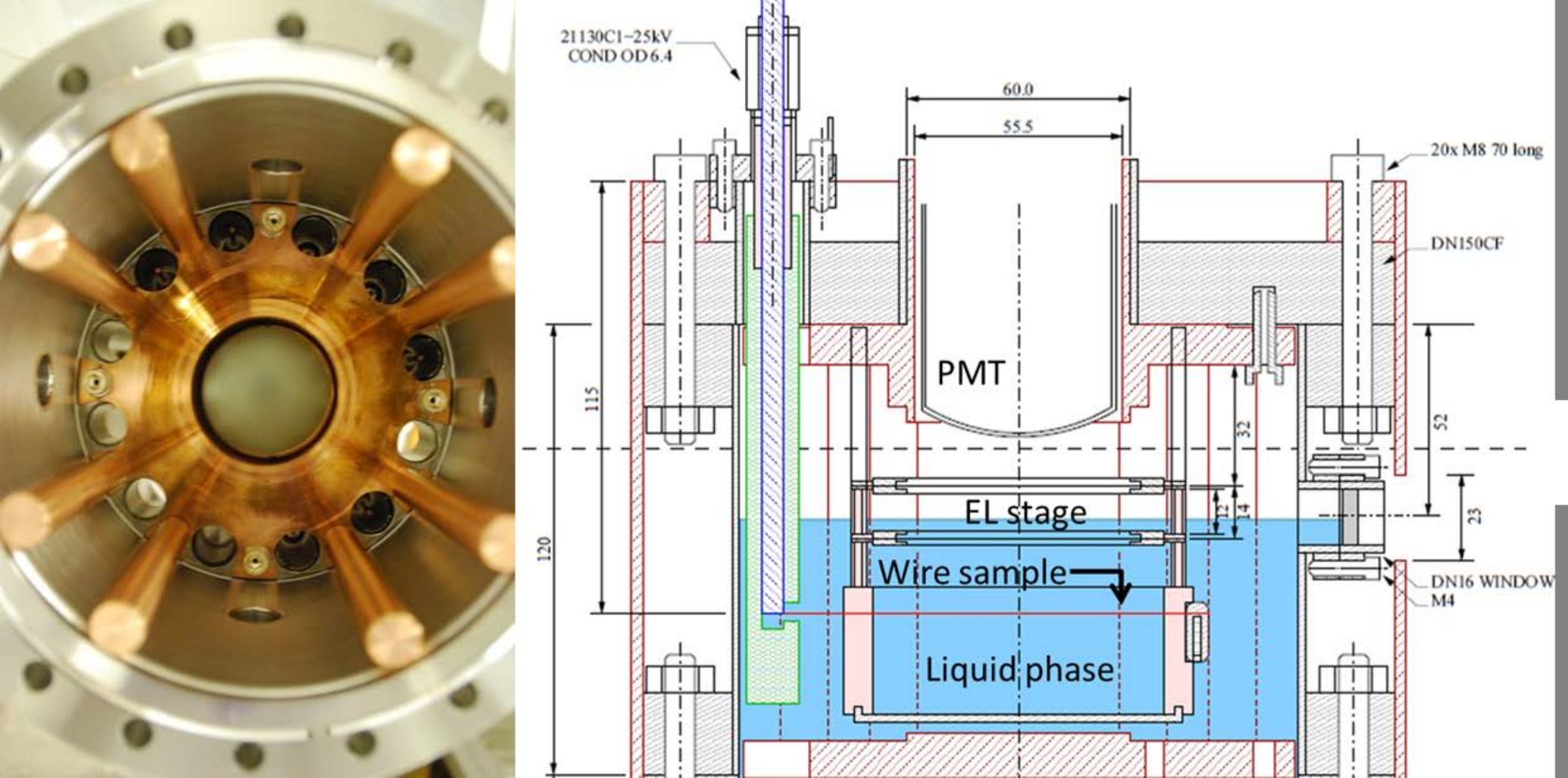}
\caption{Left: Interior of the cathode wire test chamber at Imperial College London for LZ R\&D purposes. The upper part is shown with the PMT, available ports, copper structure for heat distribution and 4 view-ports. Right: A sketch of the chamber.}
\label{fig:setupHandling}
\end{figure}

The test chamber shown in Fig.~\ref{fig:setupHandling}) will be operated with 1.3 litres (4~kg) of liquid xenon in equilibrium with vapor; the gaseous phase where EL is generated is typically 5~mm thick and maintained near 1.6~bar. The drift distance from the cathode is approximately 25~mm. This double-phase TPC is viewed by a 2-inch quartz-window PMT. In addition, four view-ports allow direct inspection. The subsystems required to operate the chamber were those developed for the ZEPLIN-III experiment at the Boulby Underground Laboratory~\cite{Akimov07,Akimov12b}; the gas handling system is built from UHV stainless steel components, including a 2.5~m$^3$ dump vessel (safety), a pump-driven recirculation loop through a SAES hot getter, a mass spectrometer, and an the lifetime monitor. System DAQ and slow control were also inherited from ZEPLIN-III. Pulse finding and data reduction software \cite{Neves11} will allow full exploitation of the 2-ns-sampled waveforms.

The chamber is cooled by means of a 1-inch cold finger immersed in liquid nitrogen, providing $\sim$10~W of cooling power with an autonomy of $\sim$7~h between liquid nitrogen fills and very simple and reliable operation.

The main aim of this research program is to investigate high voltage delivery into liquid xenon unambiguously limited by the electrode wires themselves. Then, correlation with wire material, diameter, finish, and environmental conditions such as liquid purity, pressure, and conditioning history will be studied. The possible emission of electrons or/and light from the cathode preceding instability will be recorded; in particular, the possibility of measuring very small currents thanks to single electron sensitivity is a key feature of the work. Tests to address the real mechanisms leading to cold electron emission will be carried out. The electrode may be modified later so as to test small grids or plastic materials. The system is scheduled to begin operation early in 2014.


\subsection{The NEXT experiment}
\label{sec:monrabal}
{\it Contributed by F.~Monrabal, Instituto de Fisica Corpuscular (IFIC), CSIC \& Universitat de Valencia, 46980 Paterna, Valencia, Spain in collaboration with \\
C. Sofka, Texas A\& M University College Station, Texas 77843-4242, USA, \\
J.J. Gomez-Cadenas, Instituto de Fisica Corpuscular (IFIC), CSIC \& Universitat de Valencia, 46980 Paterna, Valencia, Spain}
\newline

The Neutrino Experiment with a Xenon TPC (NEXT) \cite{Alvarez:2012sma} is a search for $\beta\beta 0 \nu$  in $^{136}$Xe using a 100-150 kg high-pressure gaseous xenon (HPGXe) time projection chamber. To amplify the ionization signal the detector uses EL. Electroluminescence is a linear process with a large gain that allows to obtain both good energy resolution and event topological information for background rejection. Compared with the other two Xenon experiment, EXO and KamLAND-ZEN, NEXT displays a much better energy resolution, which can be as good as 0.5\% at $Q_{\beta\beta}$, compared with $\sim$4\%  by EXO and 10\% by KamLAND-ZEN, and the ability to use the topological information of the event to characterize the signal. 

\subsubsection{NEXT-100 High Voltage Requirements}

The electroluminescence threshold in Xenon is at $\sim$0.8kV/cm/bar, and the NEXT detector will operate well above that threshold. Prototypes have operated stably in a window of 2 to 3 kV/cm /bar.With those requirements, the NEXT-100 gate as seen in Fig.~\ref{fig:DEMOFieldCage}) should be able to operate up to 25 kV.

Electron recombination is smaller in gas than in liquid Xenon and then, there is no necessity of using very high fields in the drifting region to preserve the charges and the energy resolution. The electric field planned for use in the NEXT-100 detector is 300 V/cm. In order to accommodate 100 kg of gas Xenon at 15 bar, a drift volume of 1.2 meters is required. This drift distance feed-throughimplies a voltage on the cathode 40kV above the voltage in the gate of the electroluminescence region. The high voltage feed-through is then required to hold up to 60-70 kV.

\subsubsection{NEXT-DEMO }

\begin{figure}
\centering
\includegraphics[height=3in]{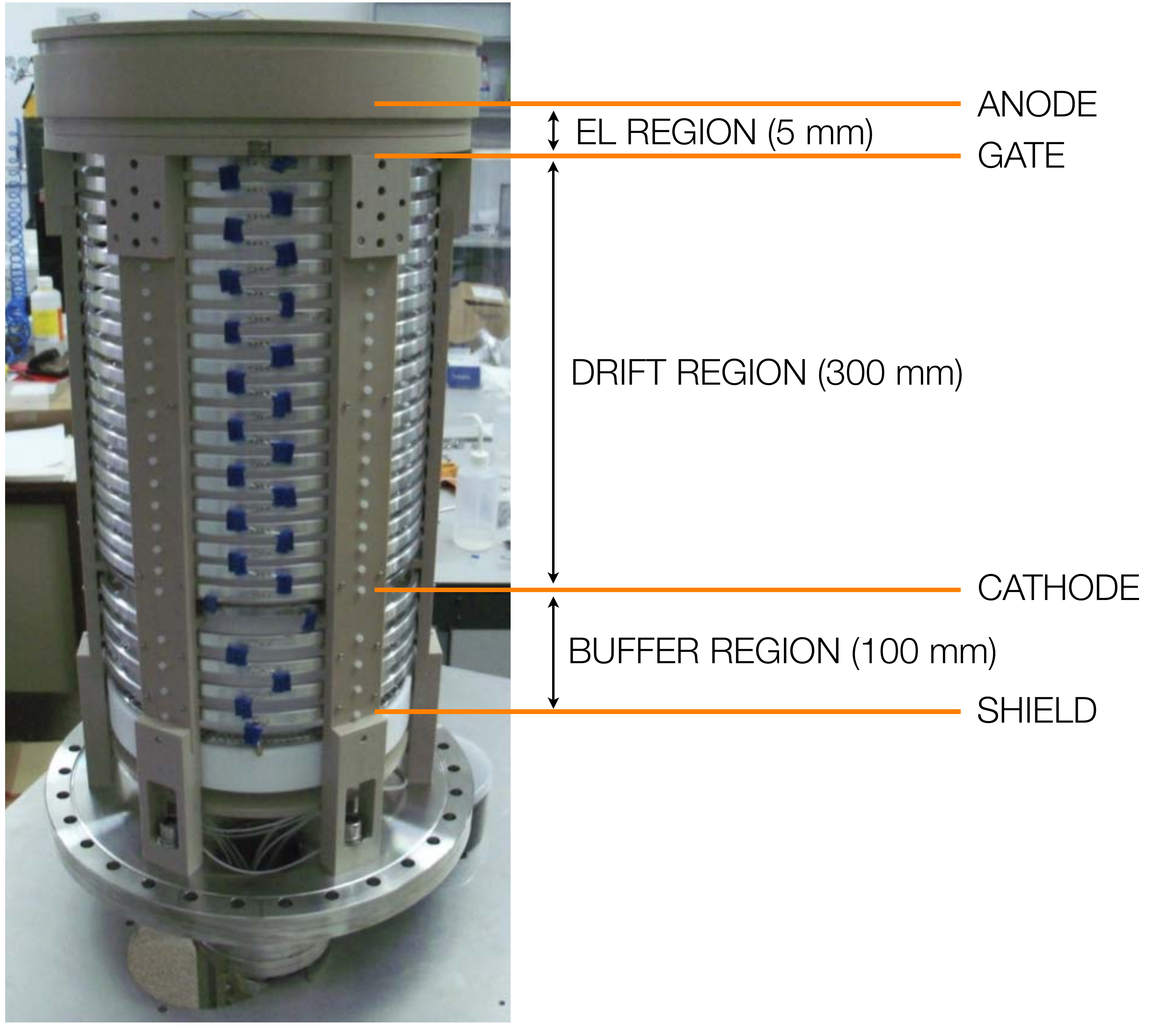}
\caption{External view of the NEXT-DEMO TPC field cage mounted on one end-cap. The approximate positions of the different regions of the TPC are indicated.}
\label{fig:DEMOFieldCage} 
\end{figure}

A large prototype, known as NEXT-DEMO, is operating at IFIC in Valencia, Spain to demonstrate the full NEXT-100 concept. It has  two planes of photodetectors, one at the cathode and one at the anode, a 30~cm long drift region and a field cage capable of withstanding large voltages. The design operative pressure is 10 bar. The vessel is a stainless steel cylinder, 60\,cm long and 30\,cm diameter. The fiducial volume is defined by a hexagonal light-tube made of PTFE reflector panels, 160\,mm across the diagonal and 300\, mm drift region. The electroluminescence region is made of two parallel grids separated by 5 mm. The maximum designed drift field is 1 kV/cm and the maximum electroluminescence field is 40 kV/cm. Figure~\ref{fig:DEMOFieldCage} shows the field cage of NEXT-DEMO. 

Recent results have demonstrated the energy resolution capabilities of the technology, showing energy resolution that extrapolates to 0.87\% FWHM at $Q_{\beta\beta}$ \cite{Alvarez:2012nd}. The tracking capabilities of the detector operating with a tracking plane of SiPMs behind the EL region has also been shown and first electrons from radioactive sources reconstructed successfully  \cite{Alvarez:2013gxa}.

\begin{figure}[h]
\centering
\includegraphics[height=2in]{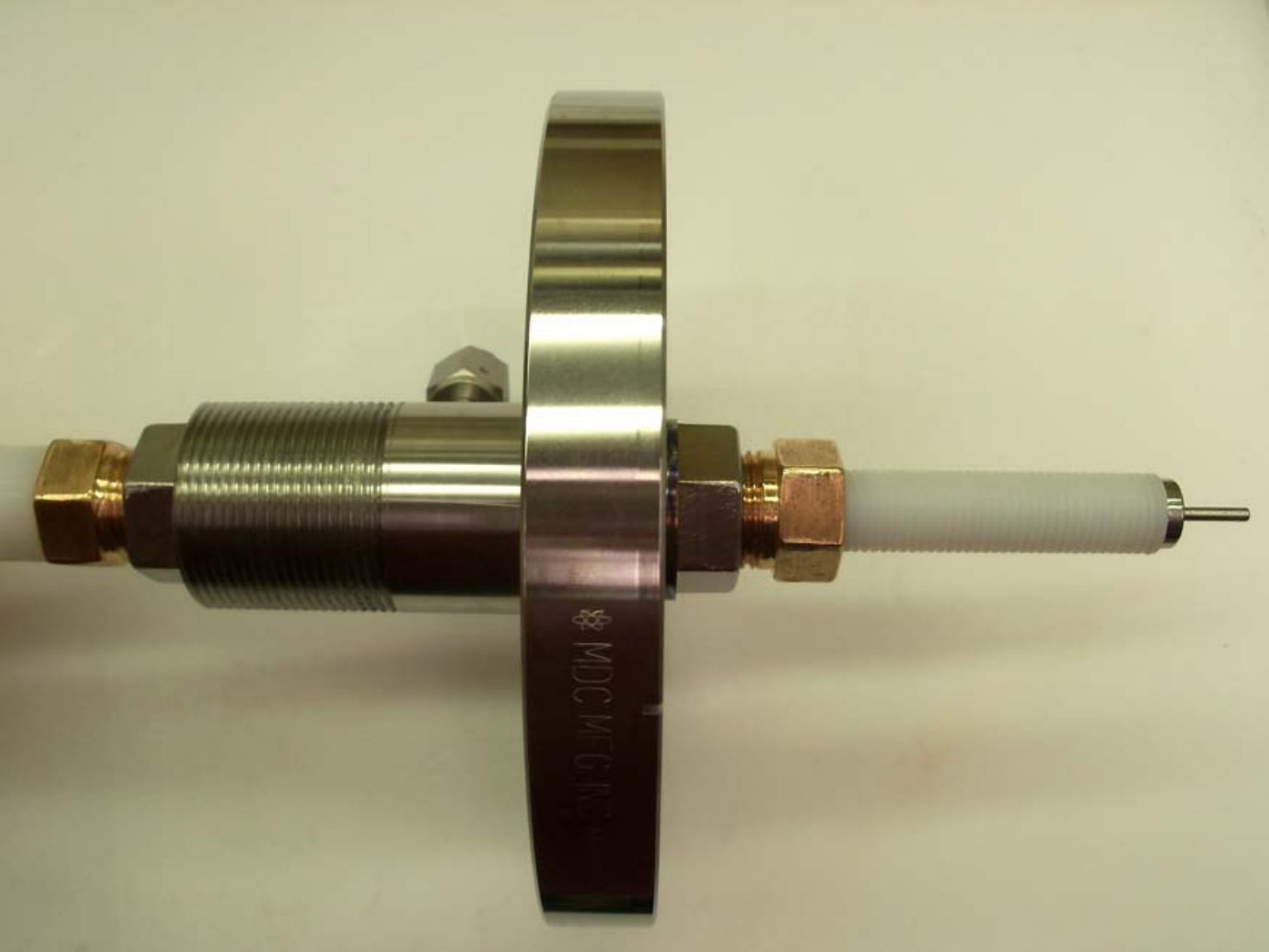}
\caption{The NEXT-DEMO high-voltage feed-through, designed and built by Texas A\&M}
\label{fig:feedthroughNEXTDEMO} 
\end{figure}

\subsubsection{NEXT-DEMO feed-through}

\begin{figure}
\centering
\includegraphics[height=2in]{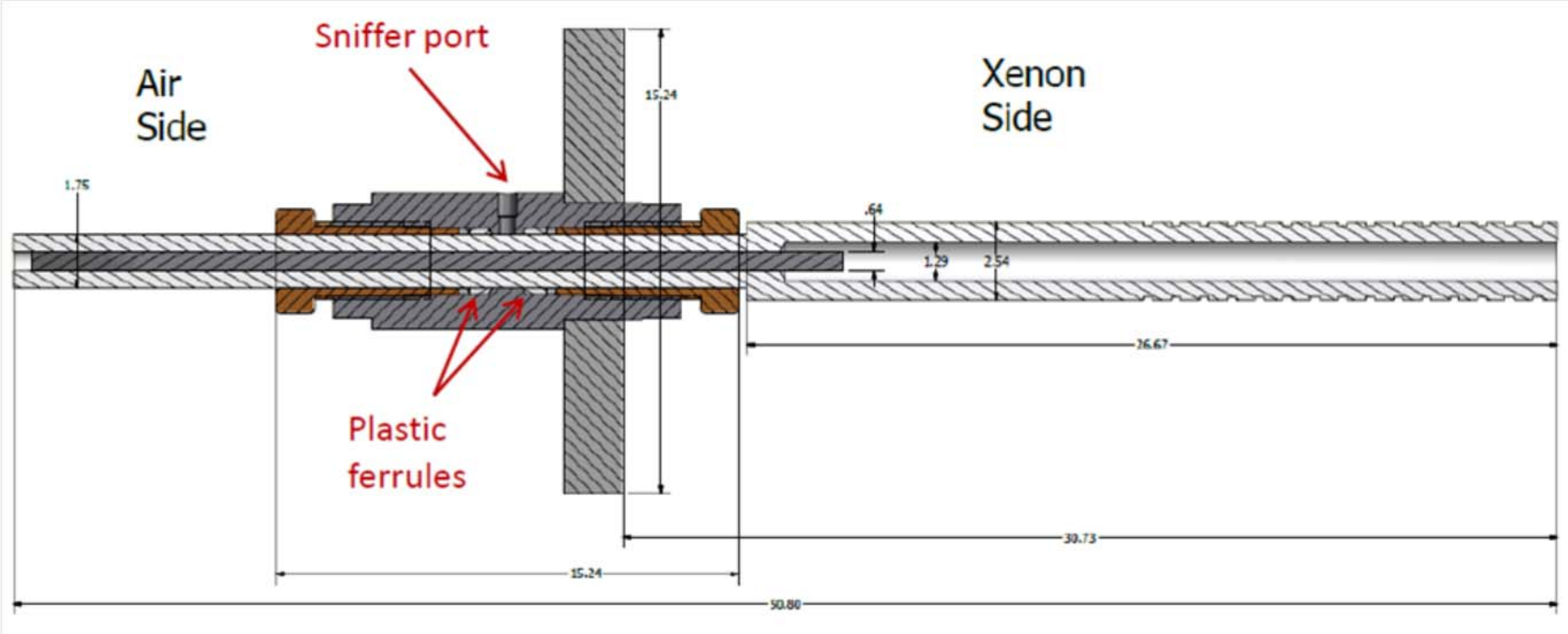}
\caption{Cathode high-voltage feed-through for NEXT designed for operation up to 100 kV.}
\label{fig:feedthroughInternal} 
\end{figure}

The high voltage feed-through used in the DEMO prototype was designed and constructed at Texas A\&M and is shown in Fig.~\ref{fig:feedthroughNEXTDEMO}.  The cathode high voltage feed-through for NEXT will be constructed using a compression seal approach, as illustrated in figure \ref{fig:feedthroughInternal}. A metal rod is pressed into a plastic tube (Polyethylene, which has a high dielectric strength), and is then clamped using SS backing rings and plastic ferrules from both the pressure side and air side. A sniffer port is placed between the seals to assure that xenon is not leaking. The feed-through will be attached to a flange located on the energy plane end-cap. A shielded cable will be connected to the feed-through and placed through the PMT support plate. The unshielded portion of the cable, with an additional resistive coating, will then run along the inside of the buffer fields rings and mate with the cathode via a spring loaded junction. This approach, with the exception of the resistive coating, has been used in NEXT-DEMO, where a cathode voltage of 45 kV has been achieved. A smaller prototype was tested to 100~kV in vacuum and 70~kV in nitrogen at 3~bar. It has been demonstrated to be xenon leak tight at 10 bar and 10-7 mbar vacuum

\begin{figure}[Htb]
\centering
\includegraphics[height=2in]{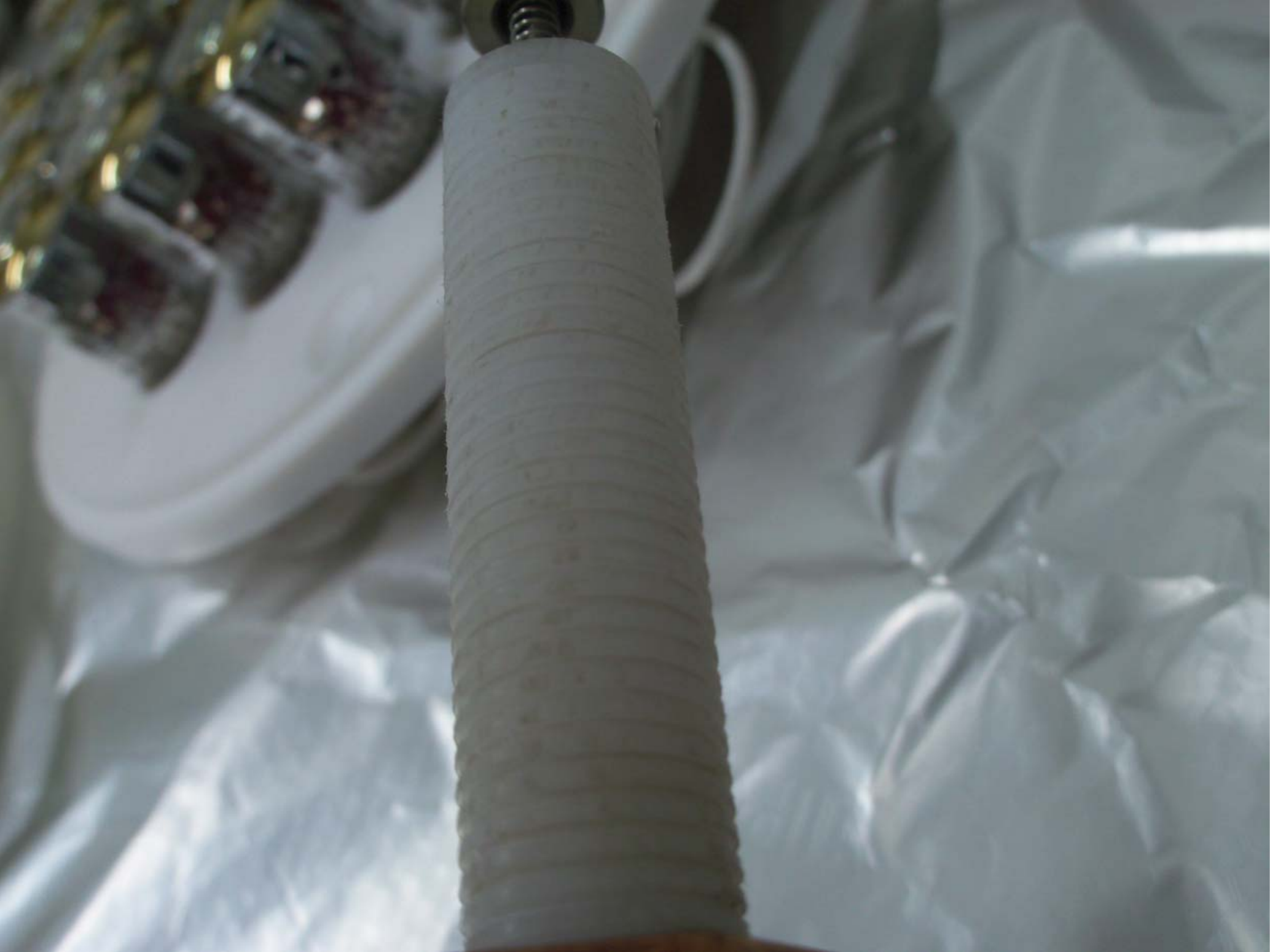}
\caption{Deail of the carbon tracks produced in the polyethylene of the cathode feed-through during the tests in NEXT-DEMO.}
\label{fig:feedthroughtracks} 
\end{figure}

\begin{figure}[Htb]
\centering
\includegraphics[height=2in]{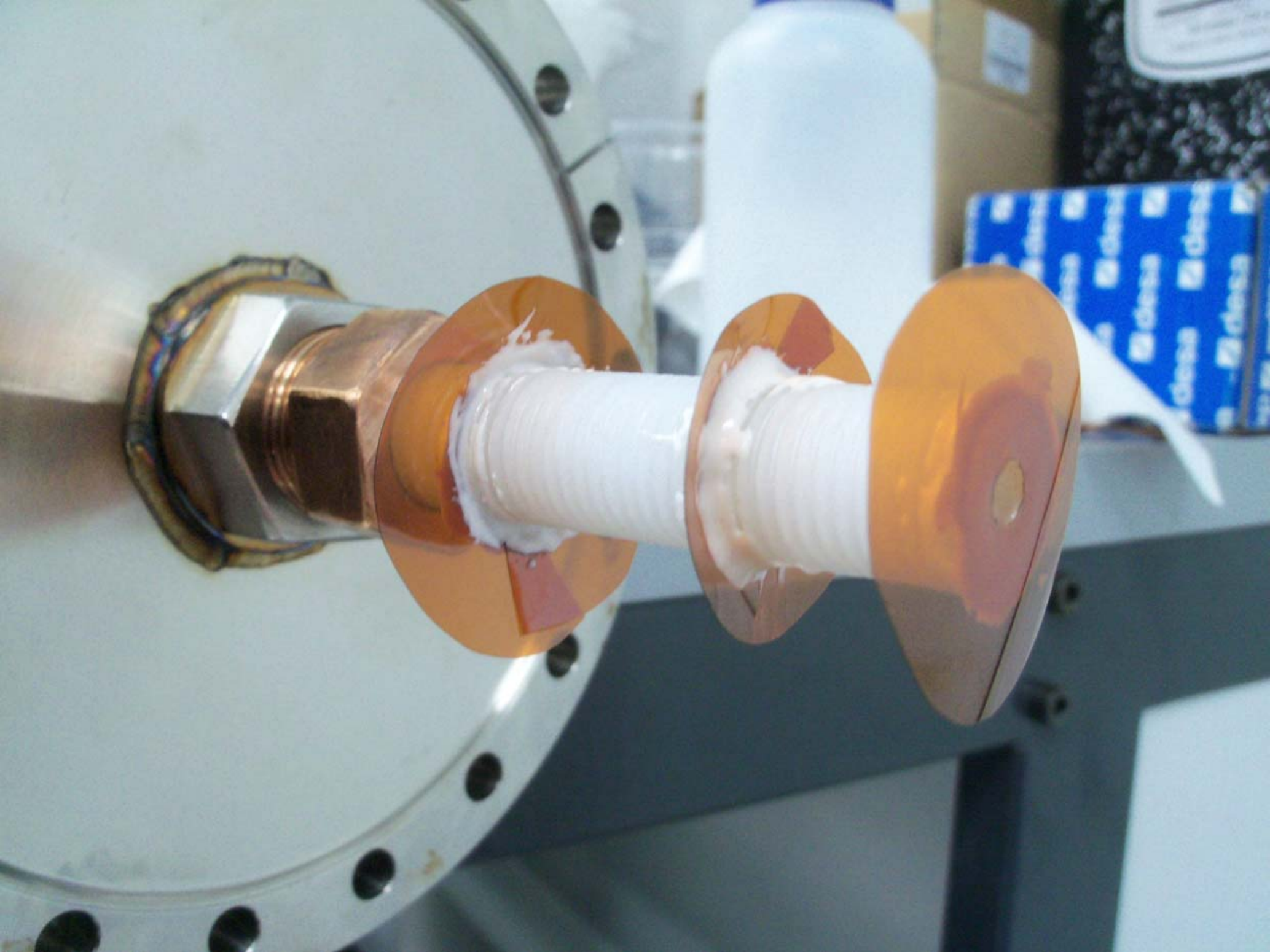}
\caption{Kapton layers preventing charges moving up in the feed-through in NEXT-DEMO.}
\label{fig:feedthroughkapton} 
\end{figure}

\subsubsection{NEXT-DEMO High Voltage Tests}

A series of tests with the feed-through has been performed in the NEXT-DEMO prototype. The tests check the behavior of the feed-through in a xenon atmosphere at 10 bar. During the tests the field cage was not inside the detector to allow for a direct visual localization of the sparks produced, and also to prevent any damage in the field cage or the EL mesh. The test demonstrates that both cathode and anode feed-through could operate at the required voltage. However, some carbon tracks did appear in the polyethylene of the feed-through as seen in Fig.~\ref{fig:feedthroughtracks}. In order to increase the stability of the feed-throughs, a series of kapton layers were glued to the polyethylene using vacuum epoxy in order to inhibit the movement of free charges along the dielectric surface of the plastic as shown in Fig.~\ref{fig:feedthroughkapton}.  With this new configuration NEXT-DEMO detector has been operating in stable conditions the last two years. 

\subsubsection{Conclusions}
The NEXT-DEMO detector has been operating for 2 years without major problems. The voltages needed for NEXT-100 detector can be achieved using a very similar design to the one used for the prototype. The NEW detector will placed and operated in Laboratorio Subterrneo de Canfranc (LSC), and the final design for the NEXT-100 feed-through will be tested there.



\section{Lessons Learned and Summary}

With proper care it is indeed possible to achieve very high electric fields in the liquid noble gases. The fundamental dielectric strengths of liquid argon, liquid xenon, and liquid helium have been measured and found to be greater than 100~kV/cm under good conditions (see Figs.~\ref{fig:results},~\ref{breakdown},~\ref{zeller},~and~\ref{fig:20micron}).  The ICARUS liquid argon TPC is perhaps the best-known example of a working experiment that has achieved stable high voltage performance at -150~kV over a long period of time, and consequently the ICARUS feed-through design has been adopted by many subsequent experiments. 

Nevertheless a number of experiments have not achieved their intended high voltage specification, and, in many cases, the onset of high voltage discharge is far below that expected from the dielectric strength of the liquid medium. This indicates that these detectors are probably not limited by the fundamental properties of the liquid {\it per se}, but are instead limited by practical considerations such as the construction methodology, the basic design of the high voltage system, or the quality of the conducting and dielectric surfaces. Indeed, a common element among the successful experiments is that the high voltage system was treated as a major focus of the design from the beginning. A careful simulation of the electric field for a given electrode and insulator geometry followed by bench-top tests seems to be a basic requirement for success, with particular attention paid to the regions with the very highest fields.

There is general agreement that in the presence of a high electric field the interface between dielectric surfaces and the liquid must be treated with care. Charge that accumulates on surfaces could be problematic if it tends to move along the surface in the presence of a tangential component of the electric field. In the past a common practice was to avoid all such surfaces in the design phase, protecting dielectrics with field rings or replacing them with materials of finite resistivity to allow charge to be removed in a controlled manner. However, for many contemporary applications, exposed dielectrics are unavoidable due to the demanding purity and reflectivity requirements, so this issue must be managed in some other way. A common technique is to corrugate the dielectric surface so that charge that accumulates becomes trapped and is unable to move across the entire surface. An electric field simulation plays a major design role here.

Among those experiments that have experienced high voltage difficulties, it seems likely that a variety of effects are at work. Working hypotheses include gas bubbling, surface asperities, the presence of particulates, and the Malter effect.  In liquid helium, gas bubbling is believed to be a leading cause of high voltage discharge (\S~\ref{sec:snsbreakdown}), and tests in liquid argon provide supporting evidence (\S~\ref{sec:test}). It has been suggested that to mitigate bubbling difficulties, one should design the high voltage system such that it performs well in a purely gaseous environment. The purity of the liquid medium has also received attention due to the evidence that extremely pure liquids have a lower dielectric strength than those that are less pure, as shown in Fig.~\ref{zeller}.

High voltage performance has long been a central focus in liquid argon TPCs and in liquid helium neutron EDM experiments. The growth of liquid xenon technology in the last decade has brought renewed attention from that community as well. Since bench-top tests and working experiments such as ICARUS have demonstrated that the liquid nobles are capable of withstanding the electric fields that modern experiments require, it seems that robust performance should be routinely achievable with appropriate engineering, design, testing, and construction controls. This experience bodes well for future experimental efforts that employ these technologies.

\section{Acknowledgements}

The speakers at this workshop represented many colleagues from both experimental collaborations and local university groups.  The work they presented was supported by a variety of funding agencies.  The following statements of acknowledgement were supplied by the speakers and have not been edited.

The work presented in \S~\ref{sec:strauss} was on behalf of the Albert Einstein Center, Laboratory of High Energy Physics of the University of Bern.  The work presented in \S~\ref{sec:hanguo} is supported by the following research grants: NSF-PHY:1004060, 1242545, and 1104720; URA (Fermilab): 604071; BNL: 238854; LANL:240213-1. It was also supported by the Physics and Astronomy Machine Shop at UCLA supervised by Harry Lockart. The work in \S~\ref{sec:monrabal} was supported by the Ministerio de Economia y Competitividad of Spain under grants CONSOLIDER-Ingenio 2010 CSD2008-0037 (CUP) and FPA2009-13697-C04-04; the Director, Office of Science, Office of Basic Energy Sciences, of the US Department of Energy under contract number DE-AC02-05CH11231; and the Portuguese FCT and FEDER through the program COMPETE, project PTDC/FIS/103860/2008. 

\bibliographystyle{JHEP}
\bibliography{2013_hvnl_workshop_summary}

\providecommand{\href}[2]{#2}\begingroup\raggedright\begin{thebibliography}{10}

\bibitem{Malter}
L.~Malter, {\it Anomalous secondary electron emission},  {\em Phys. Rev.} {\bf
  49} (1936) 879.

\bibitem{Zorin}
I.~Chirikov-Zorin and O.~Pukhov, {\it On sensitivity of gas-discharge detectors
  to light},  {\em Nucl. Instrum. Meth. A} {\bf 371} (1996) 375--379.

\bibitem{Cieslikowski}
P.~L. D.~Cieslikowski, A.J.~Dahm, {\it Investigation of thin helium films with
  surface-bound electrons},  {\em Phys. Rev. Lett.} {\bf 58} (1987) 1751--1754.

\bibitem{Brush}
J.~A.~V. L.W.~Brush, R. D.~Diehl, {\it Progress in the measurement and modeling
  of physisorbed layers},  {\em Rev. Mod. Phys.} {\bf 79} (2007) 1381-- 1454.

\bibitem{Forster}
F.~R. F.~Forster, S.~Hufner, {\it Rare gases on noble-metal surfaces: An
  angle-resolved photoemission study with high energy resolution},  {\em J.
  Phys. Chem. B} {\bf 108} (2004) 14692.

\bibitem{Huuckstadt}
C.~Huuckstadt et~al., {\it Work function studies of rare-gas/noble metal
  adsorption systems using a kelvin probe},  {\em Phys. Rev. B} {\bf 73} (2006)
  075409.

\bibitem{LLNLAr}
in preparation.

\bibitem{Sorensen}
P.~Sorensen et~al., {\it Lowering the low-energy threshold of xenon detectors},
   \href{http://xxx.lanl.gov/abs/1011.6439}{{\tt arXiv:1011.6439}}.

\bibitem{jv1}
J.~Va`vra, {\it Volume resistivity of some insulators at cold temperature and
  electrodless tpcs},  \href{http://xxx.lanl.gov/abs/SLAC-PUB-15838}{{\tt
  SLAC-PUB-15838}}.

\bibitem{jv2}
J.~Va`vra, {\it {Review of Wire Chamber Aging}},  {\em Nucl. Instrum. Meth. A}
  {\bf 252} (1986) 547.

\bibitem{ja}
J.~Allison et~al., {\it {AN ELECTRODELESS DRIFT CHAMBER}},  {\em Nucl. Instrum.
  Meth.} {\bf 201} (1982) 341.

\bibitem{ITO07}
T.~M. Ito, {\it {Plans for a Neutron EDM Experiment at SNS}},  {\em J. Phys.
  Conf. Ser.} {\bf 69} (2007) 012037.

\bibitem{GOL94}
R.~Golub and S.~K. Lamoreaux, {\it {Neutron electric dipole moment, ultracold
  neutrons and polarized He-3}},  {\em Phys. Rep.} {\bf 237} (1994).

\bibitem{BAK06}
C.~A. Baker et~al., {\it {An Improved experimental limit on the electric dipole
  moment of the neutron}},  {\em Phys. Rev. Lett.} {\bf 97} (2006) 131801.

\bibitem{WEB56}
K.~H. Weber and H.~S. Endicott {\em Trans. Am. Inst. Elec. Eng.} {\bf 75}
  (1956) 371.

\bibitem{GOL86}
R.~Golub {\em Sov. Phys. Tech. Phys.} {\bf 31} (1986) 945.

\bibitem{nEDMnote}
In the actual nEDM experiment, the achievable field was further lowered due to
  other factors.

\bibitem{KOF60}
M.~J. Kofoid {\em Trans. Am. Inst. Elec. Eng.} {\bf 79} (1960) 999.

\bibitem{COB58}
J.~D. Cobine, {\em Gaseous Conductors: Theory and Engineering Applications}.
\newblock McGraw Hill, 1958.

\bibitem{futureargon1}
A.~Ereditato and A.~Rubbia, {\it {Ideas for future liquid Argon detectors}},
  {\em Nucl. Phys. Proc. Suppl.} {\bf 139} (2005) 301.

\bibitem{futureargon2}
A.~Ereditato and A.~Rubbia, {\it {Conceptual design of a scalable multi-kton
  superconducting magnetized liquid Argon TPC}},  {\em Nucl. Phys. Proc.
  Suppl.} {\bf 154} (2006) 163.

\bibitem{futureargon3}
A.~Ereditato and A.~Rubbia, {\it {Conceptual design of a scalable multi-kton
  superconducting magnetized liquid Argon TPC}},  {\em Nucl. Phys. Proc.
  Suppl.} {\bf 155} (2006) 233.

\bibitem{nitrogen1}
A.~Ereditato et~al., {\it {Study of ionization signals in a liquid Argon TPC
  doped with Nitrogen}},  {\em JINST} {\bf 3} (2008) P10002.

\bibitem{nitrogen2}
A.~Ereditato et~al., {\it {Ionization signals from electrons and
  alpha-particles in mixtures of liquid Argon and Nitrogen: Perspectives on
  protons for Gamma Resonant Nuclear Absorption applications}},  {\em JINST}
  {\bf 5} (2010) P10009.

\bibitem{masterzeller}
M.~Zeller, {\it Development of a tpc filled with a mixture of liquid argon and
  nitrogen for g-ray resonant absorption applications},  Master's thesis,
  Universitat Bern, 2009.

\bibitem{uvlaser1}
B.~Rossi et~al., {\it {A Prototype liquid Argon Time Projection Chamber for the
  study of UV laser multi-photonic ionization}},  {\em JINST} {\bf 4} (2009)
  P07011.

\bibitem{uvlaser2}
I.~Badhrees et~al., {\it {Measurement of the two-photon absorption
  cross-section of liquid argon with a time projection chamber}},  {\em New. J.
  Phys.} {\bf 12} (2010) 113024.

\bibitem{gapd}
I.~Kreslo et~al., {\it {Pulse-shape discrimination of scintillation from alpha
  and beta particles with liquid scintillator and Geiger-mode multipixel
  avalanche diodes}},  {\em JINST} {\bf 6} (2011) P07009.

\bibitem{icarus}
{\bf ICARUS} Collaboration, S.~Amerio et~al., {\it {Design, construction and
  tests of the ICARUS T600 detector}},  {\em Nucl. Instrum. Meth. A} {\bf 527}
  (2004) 329.

\bibitem{30kvhv}
A.~Ereditato et~al., {\it {30 kV coaxial vacuum-tight feedthrough for operation
  at cryogenic temperatures}},  {\em JINST} {\bf 5} (2010) T11002.

\bibitem{argontuberef}
A.~Ereditato et~al., {\it Design and operation of argontube: a 5~m long drift
  liquid argon tpc},  {\em JINST} {\bf 8} (2013) P07002.

\bibitem{microboone}
MicroBooNE collaboration, \emph{MicroBooNE technical design review}, {\tt
  http://www-microboone.fnal.gov/}.

\bibitem{swan1}
D.~Swan, {\it Electron attachment processes in liquid argon containing oxygen
  or nitrogen impurity},  {\em Proc. Phys. Soc.} {\bf 82} (1963) 74.

\bibitem{swan2}
D.~Swan, {\it Drift velocities of electrons in liquid argon, and the influence
  of molecular impurities},  {\em Proc. Phys. Soc.} {\bf 83} (1964) 659.

\bibitem{swan3}
D.~Swan and T.~Lewis, {\it The influence of cathode and anode surfaces on the
  electric strength of liquid argon},  {\em Proc. Phys. Soc.} {\bf 78} (1961)
  448.

\bibitem{phdzeller}
M.~Zeller, {\em Advances in liquid Argon TPC's for particle detectors}.
\newblock PhD thesis, {Universit\"at Bern}, 2013.

\bibitem{bib1}
{\bf LBNE} Collaboration, C.~Adams et~al., {\it {Scientific Opportunities with
  the Long-Baseline Neutrino Experiment}},
  \href{http://xxx.lanl.gov/abs/1307.7335}{{\tt arXiv:1307.7335}}.

\bibitem{bib2}
{\bf CAPTAIN} Collaboration, H.~Berns et~al., {\it {The CAPTAIN Detector and
  Physics Program}},  \href{http://xxx.lanl.gov/abs/1309.1740}{{\tt
  arXiv:1309.1740}}.

\bibitem{bib3}
D. Alton and others, (2010) \\{\tt
  http://lartpc-docdb.fnal.gov/0005/000581/001/\\DarkSide50\_DOE\_Project\_Narrative\_FNAL.pdf}.

\bibitem{bib4}
L.~G. Christophorou, {\it Insulating gases},  {\em Nucl. Instrum. Meth. A} {\bf
  268} (1988) 424--433.

\bibitem{Stahl:2012}
A.~Stahl et~al., {\it {Expression of Interest for a very long baseline neutrino
  oscillation experiment (LBNO)}},
  \href{http://xxx.lanl.gov/abs/CERN-SPSC-2012-021}{{\tt CERN-SPSC-2012-021}}.

\bibitem{Cantini:2013}
C.~Cantini et~al., {\it {Long-term operation of a double phase LAr LEM Time
  Projection Chamber with a simplified anode and extraction-grid design}},
  \href{http://xxx.lanl.gov/abs/1312.6487}{{\tt 1312.6487}}.

\bibitem{Badertscher:2013}
A.~Badertscher et~al., {\it {First operation and performance of a 200 lt double
  phase LAr LEM-TPC with a 40x76 cm$^2$ readout}},  {\em JINST} {\bf 8} (2013)
  P04012.

\bibitem{Badertscher:2011a}
A.~Badertscher et~al., {\it {First operation of a double phase LAr Large
  Electron Multiplier Time Projection Chamber with a two-dimensional projective
  readout anode}},  {\em Nucl. Instrum. Meth. A} {\bf 641} (2011) 48.

\bibitem{Badertscher:2011b}
A.~Badertscher et~al., {\it {A tagged low-momentum kaon test-beam exposure with
  a 250L LAr TPC (J-PARC T32)}},  {\em J. Phys. Conf. Ser.} {\bf 308} (2011)
  012016.

\bibitem{Badertscher:2010}
A.~Badertscher et~al., {\it {Operation of a double-phase pure argon Large
  Electron Multiplier Time Projection Chamber: Comparison of single and double
  phase operation}},  {\em Nucl. Instrum. Meth. A} {\bf 617} (2010) 188.

\bibitem{Badertscher:2008}
A.~Badertscher et~al., {\it {Construction and operation of a Double Phase LAr
  Large Electron Multiplier Time Projection Chamber}},
  \href{http://xxx.lanl.gov/abs/0811.3384}{{\tt 0811.3384}}.

\bibitem{LAGUNA_LBNO}
LAGUNA-LBNO FP7 Design study, EC Grant Agreement no 284518,
  \url{http://laguna-science.eu}.

\bibitem{Heinzinger}
Heinzinger PNC 100000. Heinzinger electronic GmbH, Rosenheim, Germany.
  \url{http://www.heinzinger.com}.

\bibitem{Rogowski:1923}
W.~Rogowski, {\it {The electrical strength at the edge of flat plate
  capacitors: A contribution to the theory of spark gaps and feedthroughs}},
  {\em Arch.Elektrotech.} {\bf 12} (1923) 1.

\bibitem{COMSOL}
COMSOL Multiphysics: \url{http://www.comsol.com}.

\bibitem{EXOdet}
M.~Auger et~al., {\it {The EXO-200 detector, part I: Detector design and
  construction}},  {\em JINST} {\bf 7} (2012) P05010.

\bibitem{SAES}
SAES, {\tt http://www.saespuregas.com/}.

\bibitem{MAXWELL}
ANSYS/Ansoft, \\{\tt http://www.ansoft.com/products/em/maxwell/}.

\bibitem{Schmidt}
W.~Tauchert and W.F.~Schmidt, Z. Naturforsch. 30a, (1975) 1085-1086. See also
  Liquid State Electronics of Dielectric Liquids', W.F.~Schmidt, CRC Press,
  1997.

\bibitem{Akerib:2013tjd}
{\bf LUX} Collaboration, D.~S. Akerib et~al., {\it {First results from the LUX
  dark matter experiment at the Sanford Underground Research Facility}},
  \href{http://xxx.lanl.gov/abs/1310.8214}{{\tt arXiv:1310.8214}}.

\bibitem{Akerib:2012ys}
{\bf LUX Collaboration} Collaboration, D.~S. Akerib et~al., {\it {The Large
  Underground Xenon (LUX) Experiment}},  {\em Nucl. Instrum. Meth. A} {\bf 704}
  (2013) 111.

\bibitem{Lebedenko:2008gb}
{\bf ZEPLIN-III} Collaboration, V.~N. Lebedenko et~al., {\it {Result from the
  First Science Run of the ZEPLIN-III Dark Matter Search Experiment}},  {\em
  Phys. Rev. D} {\bf 80} (2009) 052010.

\bibitem{Angle:2007uj}
{\bf XENON} Collaboration, J.~Angle et~al., {\it {First Results from the
  XENON10 Dark Matter Experiment at the Gran Sasso National Laboratory}},  {\em
  Phys. Rev. Lett.} {\bf 100} (2008) 021303.

\bibitem{Akerib:2012ak}
{\bf LUX} Collaboration, D.~S. Akerib et~al., {\it {Technical Results from the
  Surface Run of the LUX Dark Matter Experiment}},  {\em Astropart. Phys.} {\bf
  45} (2013) 34.

\bibitem{mcdonald:2003el}
K.~McDonald, 'Notes on Electrostatic Wire Grids,' \\ {\tt
  www.hep.princeton.edu/$\sim$mcdonald/examples/grids.pdf}.

\bibitem{Hanaoka:1993dc}
R.~Hanaoka et~al., {\it Pre-breakdown current in liquid nitrogen under dc
  nonuniform field},  {\em Nucl.\ Instrum.\ Meth.\ A} {\bf 327} (1993) 107.

\bibitem{Atrazhev:2010gc}
V.~Atrazhev et~al., {\it Mechanisms of impulse breakdown in liquid: The role of
  joule heating and formation of gas cavities},  {\em IEEE Transactions on
  Plasma Science} {\bf 38} (2010), no.~10.

\bibitem{XE10}
J.~Angle et~al., {\it {First Results from the XENON10 Dark Matter Experiment at
  the Gran Sasso National Laboratory}},  {\em Phys. Rev. Lett.} {\bf 100}
  (2008) 021303.

\bibitem{XE100}
E.~Aprile et~al., {\it {Dark Matter Results from 225 Live Days of XENON100
  Data}},  {\em Phys. Rev. Lett.} {\bf 109} (2012) 181301.

\bibitem{ref2:cryodemonstartorpaper}
E.~Aprile et~al., {\it {Performance of a cryogenic system prototype for the
  XENON1T Detector}},  {\em JINST} {\bf 7} (2012) P10001.

\bibitem{Derenzo74}
S.~E. Derenzo, T.~S. Mast, H.~Zaklad, and R.~A. Muller, {\it Electron avalanche
  in liquid xenon},  {\em {Phys. Rev. A}} {\bf 9} (1974) 2582--2591.

\bibitem{Masuda79}
K.~{Masuda}, S.~{Takasu}, T.~{Doke}, T.~{Takahashi}, A.~{Nakamoto},
  S.~{Kubota}, and E.~{Shibamura}, {\it A liquid xenon proportional
  scintillation counter},  {\em {Nucl. Instrum. Meth.}} {\bf 160} (1979)
  247--253.

\bibitem{Howard04}
A.~S. {Howard}, {\it {Dark matter searches by the Boulby Collaboration and
  liquid xenon prototype development}},  {\em {Phys. At. Nucl.}} {\bf 67}
  (2004) 2032--2040.

\bibitem{Lebedenko09}
V.~N. Lebedenko, H.~M. Ara{\'u}jo, E.~J. Barnes, A.~Bewick, R.~Cashmore,
  V.~Chepel, et~al., {\it Results from the first science run of the
  {ZEPLIN-III} dark matter search experiment},  {\em {Phys. Rev. D}} {\bf 80}
  (2009) 052010.

\bibitem{Burenkov09}
A.~A. {Burenkov}, D.~Y. {Akimov}, Y.~L. {Grishkin}, A.~G. {Kovalenko}, V.~N.
  {Lebedenko}, V.~N. {Solovov}, et~al., {\it Detection of a single electron in
  xenon-based electroluminescent detectors},  {\em {Phys. At. Nuc.}} {\bf 72}
  (2009) 653--661.

\bibitem{Akimov12a}
V.~N. Lebedenko, H.~M. Ara{\'u}jo, E.~J. Barnes, A.~Bewick, R.~Cashmore,
  V.~Chepel, et~al., {\it Results from the first science run of the
  {ZEPLIN-III} dark matter search experiment},  {\em {Phys. Rev. D}} {\bf 80}
  (2009) 052010.

\bibitem{Akimov12b}
D.~Akimov, I.~Aleksandrov, V.~Belov, A.~Bolozdynya, A.~Burenkov, Y.~Efremenko,
  et~al., {\it Measurement of single-electron noise in a liquid-xenon emission
  detector},  {\em {Instrum. Exp. Tech.}} {\bf 55} (2012) 423--428.

\bibitem{Akerib13}
{\bf LUX} Collaboration, D.~S. Akerib et~al., {\it {First results from the LUX
  dark matter experiment at the Sanford Underground Research Facility}},  {\em
  ArXiv e-prints} (Oct., 2013) [\href{http://xxx.lanl.gov/abs/1310.8214}{{\tt
  arXiv:1310.8214}}].

\bibitem{Shutt07a}
T.~Shutt, C.~E. Dahl, J.~Kwong, A.~Bolozdynya, and P.~Brusov, {\it Performance
  and fundamental processes at low energy in a two-phase liquid xenon dark
  matter detector},  {\em {Nucl. Instrum. Meth. A}} {\bf 579} (2007) 451--453.

\bibitem{Edwards08}
B.~{Edwards}, H.~M. {Ara{\'u}jo}, V.~{Chepel}, D.~{Cline}, T.~{Durkin},
  J.~{Gao}, et~al., {\it Measurement of single electron emission in two-phase
  xenon},  {\em Astropart.Phys.} {\bf 30} (2008) 54--57.

\bibitem{Sorensen08}
P.~F. Sorensen, {\em A Position-Sensitive Liquid Xenon Time-Projection Chamber
  for Direct Detection of Dark Matter: The XENON10 Experiment}.
\newblock PhD thesis, Brown University, 2008.

\bibitem{Santos11}
E.~Santos, B.~Edwards, V.~Chepel, H.~Ara{\'u}jo, D.~Akimov, E.~Barnes, et~al.,
  {\it Single electron emission in two-phase xenon with application to the
  detection of coherent neutrino-nucleus scattering},  {\em {JHEP}} {\bf 2011}
  (2011) 1--20.

\bibitem{Aprile13}
E.~Aprile, M.~Alfonsi, K.~Arisaka, et~al., {\it {Observation and applications
  of single-electron charge signals in the XENON100 experiment}},  {\em ArXiv
  e-prints} (2013) [\href{http://xxx.lanl.gov/abs/1311.1088}{{\tt
  arXiv:1311.1088}}].

\bibitem{Akimov07}
{\bf ZEPLIN-III} Collaboration, D.~Y. Akimov, G.~J. Alner, H.~M. Ara{\'u}jo,
  A.~Bewick, C.~Bungau, {A.~A.~Burenkov}, et~al., {\it The {ZEPLIN}-{III} dark
  matter detector: Instrument design, manufacture and commissioning},  {\em
  {Astropart. Phys.}} {\bf 27} (2007) 46--60.

\bibitem{Neves11}
F.~Neves, D.~Y. Akimov, H.~M. Ara{\'u}jo, E.~J. Barnes, V.~A. Belov, A.~A.
  Burenkov, et~al., {\it {ZE3RA}: the {ZEPLIN-III} reduction and analysis
  package},  {\em {JINST}} {\bf 6} (2011) P11004.

\bibitem{Alvarez:2012sma}
{\bf NEXT} Collaboration, V.~Alvarez et~al., {\it {NEXT-100 Technical Design
  Report (TDR): Executive Summary}},  {\em JINST} {\bf 7} (2012) T06001.

\bibitem{Alvarez:2012nd}
{\bf NEXT} Collaboration, V.~Alvarez et~al., {\it Initial results of next-demo,
  a large-scale prototype of the next-100 experiment},
  \href{http://xxx.lanl.gov/abs/1211.4838}{{\tt arXiv:1211.4838}}.

\bibitem{Alvarez:2013gxa}
{\bf NEXT} Collaboration, V.~Alvarez et~al., {\it {Operation and first results
  of the NEXT-DEMO prototype using a silicon photomultiplier tracking array}},
  {\em JINST} {\bf 8} (2013) P09011,
  [\href{http://xxx.lanl.gov/abs/1306.0471}{{\tt arXiv:1306.0471}}].

\end{thebibliography}\endgroup

\end{document}